\newtheorem{theorem}{Theorem}
\newtheorem{proposition}{Proposition}
\newtheorem{assumption}{Assumption}
\newtheorem{p-assumption}{Preference Assumption}
\newtheorem{gr-assumption}{Satisfaction Assumption}
\newtheorem{e-assumption}{Monotonic Assumption}
\newtheorem{e-assumption-p}{Parametric Assumption}
\newtheorem{cr-assumption}{Coherent-risk Assumption}

\newtheorem{lemma}{Lemma}
\newtheorem{corollary}{Corollary}

\documentclass[12pt]{article}

\usepackage{amsmath}
\usepackage{amssymb}
\usepackage{amsfonts}

\usepackage{graphicx}
\newcommand{\qed}{{\mbox{} \hspace*{\fill}{\vrule height5pt width4pt
depth0pt}}\\}

\def\M{\hspace*{0.75em}}

\begin{document}

\title{Game-theoretic Modeling of Players' Ambiguities on External Factors}
\author{Jian Yang\\
Department of Management Science and Information Systems\\
Business School, Rutgers University\\
Newark, NJ 07102\\
Email: jyang@business.rutgers.edu}
\date{February 2017}
\maketitle

\begin{abstract}
We propose a game-theoretic framework that incorporates both incomplete information and general ambiguity attitudes on factors external to all players. Our starting point is players' preferences on payoff-distribution vectors, essentially mappings from states of the world to distributions of payoffs to be received by players. There are two ways in which equilibria for this preference game can be defined. When the preferences possess ever more features, we can gradually add ever more structures to the game. These include real-valued utility-like functions over payoff-distribution vectors,  sets of probabilistic priors over states of the world, and eventually the traditional expected-utility framework involving one single prior. We establish equilibrium existence results, show the upper hemi-continuity of equilibrium sets over changing ambiguity attitudes, and uncover relations between the two versions of equilibria. 
Some attention is paid to the enterprising game, in which players exhibit ambiguity seeking attitudes while betting optimistically on the favorable resolution of ambiguities. The two solution concepts are unified at this game's pure equilibria, whose existence is guaranteed when strategic complementarities are present. The current framework can be applied to settings like auctions involving ambiguity on competitors' assessments of item worths.

\vspace*{.1in}
\end{abstract}

\newpage

\section{Introduction}\label{introduction}

\subsection{Motivation}

In the traditional expected-utility approach to games involving incomplete information, such as Harsanyi's \cite{H67-8}, players' types are used to model private messages they receive about the uncertain external environments. After observing his own type $t_n$, it is customary that player $n$ will form a probabilistic understanding $p_{n,t_n}\equiv (p_{n,t_n|t_{-n}})_{t_{-n}\in T_{-n}}$ about other players' types $t_{-n}\in T_{-n}$. Presented with others' strategies, he will strive to maximize his own expected utility, where expectation is taken with the aforementioned assessment $p_{n,t_n}$. To precisely describe the actual uncertainty, however, it often takes more than even the entire collection $t\equiv (t_n)_{n\in N}$ of all players' types. This is where states of the world $\omega$ will come into play.

A more detailed model might allow player $n$, after observing his own type $t_n$, to predict that the actual $\omega$ will come from some subset $\Omega_{n,t_n}$ of states of the world. Of course, it would be necessary that $\Omega_{n,t_n}$ and $\Omega_{n,t'_n}$ be non-overlapping when $t_n\neq t'_n$, and that $\bigcup_{t_n\in T_n}\Omega_{n,t_n}$ for every player $n$ be some common space $\Omega$. Indeed, such is essentially the information structure introduced by Aumann \cite{A76}. However, there seems to be no urgency in exploiting the ``$\omega$-level'' detail in the traditional approach, due to the ``integration'' of its effects. 

Such is not necessarily the case when players' ambiguities have to be taken into account. Suppose all players adopt potentially random actions based on their types. When the actual $\omega\in \Omega$ is eventually revealed, all players' types will be known: there is a unique $t\equiv (t_n)_{n\in N}$ such that $\omega\in \Omega_t\equiv \bigcap_{n\in N}\Omega_{n,t_n}$. Then, every player will know the probabilistic distribution over his payoffs that he is supposed to experience. Of course, post decision, he will eventually experience just one single payoff. Now during the play, right after receiving his own type $t_n$ but before knowing anything about others' types $t_{-n}$ let alone the true $\omega$, player $n$ should anticipate one payoff distribution say $\pi(\omega)$ per $\omega\in \Omega_{n,t_n}\equiv \bigcup_{t_{-n}\in T_{-n}}\Omega_{t_n,t_{-n}}$. He will certainly want to make the vector $\pi\equiv (\pi(\omega)|\omega\in\Omega_{n,t_n})$ as likable to himself as possible. Each payoff-distribution vector $\pi$ is essentially an act first proposed by Anscombe and Aumann \cite{AA63}.

A natural apparatus to express the ``$(n,t_n)$''-player's taste is a strict preference relationship $\succ_{n,t_n}$ on all payoff-distribution vectors. With it, one can understand $\pi\succ_{n,t_n}\pi'$ as the player-type pair liking $\pi$ better than $\pi'$ but not the other way around. The traditional expected-utility approach basically uses what we shall call a real-valued satisfaction function $s_{n,t_n}$ on payoff-distribution vectors $\pi$ to facilitate each $(n,t_n)$-player's preference relation:
\begin{equation}\label{stfn}
\pi\succ_{n,t_n}\pi'\;\;\;\mbox{ if and only if }\;\;\;s_{n,t_n}(\pi)>s_{n,t_n}(\pi');
\end{equation}
moreover, $s_{n,t_n}$ is specially built from one single probabilistic prior $\rho_{n,t_n}$ on the state space $\Omega_{n,t_n}$ and a utility function $u_{n,t_n}$ on payoffs, in the fashion of
\begin{equation}\label{ref-tradition}
s_{n,t_n}(\pi)=s^0_{n,t_n}(\pi,\rho_{n,t_n})=\int_{\Omega_{n,t_n}}\left[\int u_{n,t_n}\cdot d[\pi(\omega)]\right]\cdot \rho_{n,t_n}(d\omega).
\end{equation}
This, at least in formality, amounts to $s_{n,t_n}(\pi)=\int u_{n,t_n}\cdot d[\int_{\Omega_{n,t_n}}\pi(\omega)\cdot \rho_{n,t_n}(d\omega)]$ as well.

So it is as though the various $\pi(\omega)$-components are first averaged over using weights provided by the $\rho_{n,t_n}$-distribution; then the integrated payoff distribution or lottery is combined with the utility function $u_{n,t_n}$ to generate the satisfaction level over the given payoff-distribution vector $\pi$. The linear treatment of the lottery on payoffs through the utility function $u_{n,t_n}$ can be legitimized by von Neumann and Morgenstern's \cite{VNM44} axioms; the appropriateness of using a single prior $\rho_{n,t_n}$ to assemble the lottery can be reasoned using Savage's \cite{S54} arguments. Both features have been used in the traditional modeling of not only incomplete-information but also normal-form games; see, e.g., Harsanyi \cite{H67-8} and Nash \cite{N50}\cite{N51}. 
Of course, there is no need to single out the latter because they can be treated as special incomplete-information games in which every player has only one type.

Both linear treatments of payoff distributions and the latter's formations out of payoff-distribution vectors through the uses of single probabilistic priors have been criticized. First, Allais \cite{A53} challenged the notion that people use linear functionals over payoff distributions to reach decisions. Empirical studies in support of this contention can be found, e.g., in Camerer and Ho \cite{CH94} and Wu and Gonzalez \cite{WG96}. 

Then, Ellsberg \cite{E61} argued that decision makers (DMs) often do not even know the probabilities to be assigned to different states of the world. For instance, there are probably not enough data to estimate the chance of a new financial crisis to occur within the next two years; also, there has no precedent to be relied on to assess probabilities concerning the climate change due to human actitivies. Hence, to many situations the single-prior assumption on uncertain factors can be ill suited. Starting from Schmeidler \cite{S89}, researchers resorted to tools like Choquet integration and capacities, i.e., non-additive probabilities, to help with single-agent decision making involving general  ambiguity attitudes; see, e.g., Gilboa and Marinacci \cite{GM13}. Under axioms associated with ambiguity aversion, Gilboa and Schmeidler \cite{GS89} legitimized the worst-prior form to be taken by a DM. 
In this form,
\begin{equation}\label{s-form}
s_{n,t_n}(\pi)=\inf_{\rho\in P_{n,t_n}}s^0_{n,t_n}(\pi,\rho),
\end{equation}
where $P_{n,t_n}$ is a set of prior distributions on the space $\Omega_{n,t_n}$ and $s^0_{n,t_n}$ is defined in~(\ref{ref-tradition}).

\subsection{Main Contributions}

In the current strategic setting involving incomplete information, failure to account for players' diverse ambiguity attitudes could lead to weird predictions or dangerous prescriptions. In auctions, especially those involving works of art, offshore oilfields, or electromagnetic spectra, participants often do not know for sure the actual worths to themselves of the item being auctioned; very likely, they are also uncertain about the distributions their competitors assign to the item's worths; in addition, some may fear losing the object more than they regret about overpaying for it. How can a model capture these features then? The prevalent auction theory takes the traditional approach to incomplete-information games; hence, it is not able to model bidders' unconventional ambiguity attitudes.

We make an attempt at overcoming the above shortcoming by defining a more general preference game from the mere preference relations $\succ_{n,t_n}$, without imposing any structural requirement. Incidentally, a certain set of requirements led Anscombe and Aumann \cite{AA63} to the simultaneous emergence of both the utility function $u_{n,t_n}$ and the probabilistic prior $\rho_{n,t_n}$ of the traditional form~(\ref{ref-tradition}). But without help from any such structure, behavioral equilibria can already be defined. There is admittedly a growing game-theoretic literature on ambiguity considerations. Against this backdrop, this work still makes substantial contributions. 

First, we propose a game-theoretic framework starting from players' preference relations on payoff-distribution vectors. This enables the incorporation of players' diverse ambiguity attitudes on external factors. Our emphasis here is not the consideration of preferences itself. In various strategic settings, this has been done by, e.g., Schmeidler \cite{S69}, Mas-Colell \cite{M74}, Shafer and Sonnenschein \cite{SS75}, Khan and Sun \cite{KS90}, and Grant, Meneghel, and Tourky \cite{GMT16}. It is preferences on {\em payoff-distribution vectors} that we want to stress. We believe such preferences provide more flexibility than those on actions, action distributions, integrated payoff distributions, or payoff vectors. This can probably be attested to by the inclusion of various existing models in the current framework. On the flip side, while based on players' anticipations of their own payoffs rather than other players' strategies, our preferences are more natural and parsimonious choices for model primitives.

Second, we give definitions to two prominent types of behavioral equilibria and establish their existence and continuity in various circumstances. Previously unknown relations between the two equilibrium notions are uncovered as well.
The first, {\em action}-based interpretation leaves every player in control of his action whilst maintaining a long-term commitment to his portion of a behavioral equilibrium. The second, {\em distribution}-based interpretation ties every player's action to the outcome of a random device in a fashion consistent to his portion of an equilibrium. It will soon be clear that an action-based equilibrium assigns weights only to actions that leave no room for improvement by any other pure action; whereas, a distribution-based equilibrium leaves no room for improvement by any other distribution of actions. Since there are ``more'' action distributions than pure actions to compete against, distribution-based equilibria are in general ``harder to come by'' than action-based ones.

Third, we step into the less traveled ambiguity-seeking territory and make interesting findings. Since Ellsberg's \cite{E61} pioneering work, most attention has been paid to ambiguity aversion as an alternative attitude to ambiguity neutrality. However, 
experiments involving human subjects showed that ambiguity seeking could be equally prevalent; see, e.g., Curley and Yates \cite{CY89} and Charness, Karni, and Levin \cite{CKL13}. 
We also believe that optimistic assessments of uncertain gains is part of what drive people to participate in auctions, embark on exploratory journeys, and start new firms.
Thus, the case opposite to that assuming~(\ref{s-form}) is equally if not more interesting. 
We call the corresponding game ``enterprising'' because each
\begin{equation}\label{radical-form}
s_{n,t_n}(\pi)=\sup_{\rho\in P_{n,t_n}}s^0_{n,t_n}(\pi,\rho),
\end{equation}
so that players make optimistic bets on favorable resolutions of their ambiguities. 

The action-distribution distinction turns out to be irrelevant for the enterprising game's pure equilibria, given that they exist. When equipped with strategic complementarity features, the game can be shown to possess not only pure equilibria, but also those with monotone trends with respect to players' types as well as external conditions. These results can be considered as extensions of those achieved for the traditional counterpart as laid out in van Zandt and Vives \cite{VZV07}. As normal-form games are incomplete-information games with singleton type spaces, the results also generalize those that appeared in traditional supermodular games studied by Topkis \cite{T79}, Milgrom and Roberts \cite{MR90}, and Vives \cite{V90}.

Finally, our findings are applicable to settings like 
auctions involving ambiguity on bidders' assessments of item worths and competitive pricing involving uncertain demand curves.

\subsection{Outline of Results}

Under mild conditions, we show that action-based equilibria always exist. When the preferences $\succ_{n,t_n}$ connote ambiguity aversion, distribution-based ones will come into being as well; see Theorem~\ref{t-existence}. 
Both sets of equilibria are upper hemi-continuous in players' ambiguity attitudes; see Theorem~\ref{t-dep}. When the preferences are representable by real-valued functions $s_{n,t_n}$ satisfying~(\ref{stfn}), our game is specialized to the so-called satisfaction kind. For this game, action-based equilibria will exist in general and so will distribution-based equilibria when the $s_{n,t_n}$'s are quasi-concave. 
When there is a set $P_{n,t_n}$ of prior distributions on the state space $\Omega_{n,t_n}$, so that each $s_{n,t_n}$ takes Gilboa and Schmeidler's \cite{GS89} form~(\ref{s-form}), we shall obtain the so-called alarmists' game. In it, players express aversions to ambiguities. Due to concavity of the $s_{n,t_n}$'s, the game has both action- and distribution-based equilibria. 

For the preference game, rudimentary understandings on relations between the action- and distribution-based equilibria can be formed. Our message will become considerably sharper for the satisfaction game. For it, we can conclude that distribution-based equilibria will be action-based ones when players are ambiguity-seeking and the two types will be identical when players are ambiguity-neutral; see Theorem~\ref{ep-satisfaction}. Relatedly, as might have been suspected, the distinction between the two versions of equilibria will cease to matter for the traditional expected-utility game; see Theorem~\ref{ep-ea}. Our derivation relies on concepts like continuous kernels and their integrations, as well as intermediate results like Lemma~\ref{meaningful} that might be of value elsewhere. When we focus on pure equilibria, we again confirm the earlier ``comparative rarity'' observation by showing that any pure distribution-based equilibrium must also be a pure action-based one; see Theorem~\ref{pure-pa}.

Our attention then shifts to the enterprising game in which players demonstrate ambiguity-seeking traits. As a special satisfaction game with convex $s_{n,t_n}$ functions, any distribution-based equilibrium of this game is necessarily an action-based one.
When confined to pure strategies, we also have the equivalence between the two types of equilibria; see Theorem~\ref{chedan}.
One technical result involved in its proof is Lemma~\ref{l-okla}. It is an extension of a well known finite-dimensional property, stating that the maximum of a convex function over a convex region in $\Re^d$ for some $d=1,2,...$ can always be achieved at extreme points. 

Of special interest is the case
where (i) each $\Omega_t=\{t\}\times\tilde{\Omega}$ for some common state space $\tilde{\Omega}$ and (ii) all ambiguities of an $(n,t_n)$-player are on $\tilde{\Omega}$ rather than other players' types $t_{-n}$. This reflects the situation where players can form subjective probabilities $p_{n,t_n|t_{-n}}$ on their opponents' types much like in the traditional game, but with extra ambiguities on other external factors. Borrowing ideas from works dealing with subjects like lattices and submodularity, including Milgrom and Shannon \cite{MS94}, Zhou \cite{Z94}, Topkis \cite{T98}, and Yang and Qi \cite{YQ13}, we can extend the traditional analysis of games possessing strategic complementarities and obtain the existence of monotone pure equilibria as well as their monotone comparative statics properties; see Theorems~\ref{big-t} and~\ref{big-t-p}. An enabling technical result is Lemma~\ref{l-import} on the preservation of increasing differences under maximization, much like a well known one about the preservation of supermodularity under maximization.

In the following, we discuss existing game-theoretic literature with ambiguity considerations in Section~\ref{literature}, and give a general formulation in Section~\ref{formulation}. The existence and continuity of the two types of equilibria are derived in Section~\ref{analysis}. We next delve into various special cases in Section~\ref{cases}, 
and establish relationships between the two equilibrium concepts in Section~\ref{relation}. The framework's suitability to auctions is discussed in Section~\ref{application}; finally, our conclusion is reached in Section~\ref{conclusion}. A special enterprising game whose features lead to monotone pure equilibria are detailed in Appendix~\ref{app-dc}.

\section{Literature Survey}\label{literature}

Normal-form games incorporating general ambiguity attitudes have been studied. Dow and Werlang \cite{DW94} used convex capacities to model players' beliefs about opponents' behaviors and arrived at equilibrium belief profiles. Eichberger and Kelsey \cite{EK00} extended the study to situations involving $n\geq 3$ players and identified players' confidence degrees for equilibrium parametrization purposes. Marinacci \cite{M00}, on the other hand, gave more flexible definitions to players' vaguenesses on their beliefs, which could then be used in comparative statics studies. Klibanoff \cite{K96} and Lo \cite{L96} adopted Gilboa and Schmeidler's \cite{GS89} notion of ambiguity aversion and used convex and closed sets of probabilistic priors on products of other players' mixed strategies, reducible to those on their pure actions,
as the basis on which players make decisions. 
Epstein \cite{E97} let players be ambiguous about opponents' pure strategies as well as their ambiguity attitudes, and studied the iterated elimination of dominated strategies.

Players in the above games were allowed to have qualms about opponents' behaviors. We, like some studies of incomplete-information games involving general ambiguity attitudes, focus on the complementary situation where players have vagueness about 
factors external to all of them.
We argue for merits of the ambiguity-on-external-factor rather than ambiguity-on-opponent-behavior consideration as follows.
First, as shown momentarily, mixed strategies chosen by players are often enforceable. Second, uncertainties about the state of the world can pose a much bigger problem than those about other players' behaviors. Think of a Stag Hunt game where each participant has only to choose between {\em cooperate} and {\em defect}, and yet there are millions of combinations in numbers, sizes, and speeds of the stags and hares on the hunting ground, as well as other factors like temperature and wind. Third, no longer having to model players' behaviors through non-probabilistic means, we can apply conventional tools built on countably additive probabilities to our analysis. 
Uncertainty about opponents' types will still indirectly lead to uncertainty about their preferences as well as behaviors. Incidentally, Bade \cite{Bd11} introduced behavioral ambiguity by allowing players to base their actions on uncontrollable factors that are not observable by opponents.

Regarding the verifiability of mixed strategies pronounced by players, there seem to be at least two plausible solutions.
First, when the game merely reflects one encounter in many repeated interactions, we propose that each player takes a frequentist approach to the compliance of his proclaimed mixed strategy. Thus, he is at almost total control of his own action in each play, but has to maintain agreed-upon frequencies to various actions in the long run. Second, it is possible that every player is given a random number generator whose output is private knowledge in-game but 
public knowledge post-game, and the player has to act according to an agreed-upon mapping from the random device's output to his action. We call the first perspective action-based because players get to choose their actual actions and the second distribution-based because players need to decide on distributions of actions before they start to act. 
Equilibrium concepts in Dow and Werlang \cite{DW94} 
and Marinacci \cite{M00} are of the action-based variety; whereas, those in Klibanoff \cite{K96} and Lo \cite{L96} are of the distribution-based variety. 
Also, Kajii and Ui \cite{KU05} called the first kind ``equilibria in beliefs'' and the second kind ``mixed equilibria''.


Among works on incomplete-information games involving general ambiguity attitudes, we note that Epstein and Wang \cite{EW96}
used preference relations over acts to express ambiguity attitudes, and also allowed ambiguities over opponents' preferences. This setup gave rise to infinite sequences of preferences over preferences, much like Mertens and Zamir's \cite{MZ85} sequences of beliefs over beliefs. Under reasonable assumptions about allowable preferences, authors justified the emergence of those largest necessary type spaces that contain players' personal characteristics. Ahn \cite{A07} and Di Tillio \cite{D08} worked along a similar line, with the former modeling ambiguities using sets of beliefs and the latter imposing less restrictions on preferences but more on payoff and state spaces. 


In our current study, exogenous factors $\omega$ within the state space $\Omega$ contain no information on either opponents' behaviors or their ambiguity attitudes. Each of player $n$'s types $t_n$ is a private message he receives about the actual external factor. The $(n,t_n)$-player is uncertain which $\omega\in \Omega_{n,t_n}$ has been realized. Consequently, he is uncertain about the actual opponent-type profile $t_{-n}\in T_{-n}$ because he does not know which $\Omega_{t_n,t_{-n}}\subseteq \Omega_{n,t_n}$ the $\omega$ is in. The various preference relations $\succ_{n,t_n}$ on
``$|\Omega_{n,t_n}|$-dimensional'' payoff-distribution vectors reflect players' ambiguity attitudes which might be under the sway of messages they receive. Like their behaviors which as we have emphasized are observable and enforceable, players' preferences are assumed to be commonly known. 

In real life, these entities might simply translate into a few choices on players' personalities. For instance, it is possible that one player be labeled ``mildly conservative all the time'' while another ``fairly reckless when knowing that the stake is high''. 
Just allowing ambiguities on factors external to all players is applicable enough to a wide variety of practically relevant cases. Take, for example, an auction of the exploration right to an offshore oil field. All bidders, being major petrochemical firms, probably know each other well through past interactions. The major uncertainty then stems from the potential of the field itself, which is also responsible for all the private readings delivered to the bidding oil majors.

Some works, like ours, dealt with equilibrium existence issues. Kajii and Ui \cite{KU05} effectively studied the alarmists' game, albeit with finite action and state spaces. They showed the existence of both action- and distribution-based equilibria; see, respectively, their Propositions 2 and 1.
Moreover, Azrieli and Teper \cite{AT11} treated what might be considered a special satisfaction game. In its interim version, payoff-distribution vectors $\pi$ are first turned into  expected-utility vectors $\int u_{n,t_n}\cdot d\pi\equiv (\int u_{n,t_n}\cdot d[\pi(\omega)]|\omega\in\Omega_{n,t_n})$ using utility functions $u_{n,t_n}$. The latter are then assessed using functionals say $j_{n,t_n}$. So for this game,
\begin{equation}\label{atat}
s_{n,t_n}(\pi)=j_{n,t_n}\left(\int u_{n,t_n}\cdot d\pi\right).
\end{equation}
Authors showed that quasi-concavity of the $j_{n,t_n}$'s would lead to the existence of distribution-based equilibria; see their Definition 2 which allows a player to maximize his action distribution instead of letting him fill the distribution's support with optimal pure actions. Recently, Grant, Meneghel, and Tourky \cite{GMT16} proposed the Savage game in which players possess preferences over strategy profiles made up of all players' deterministic action plans. They identified sufficient conditions for pure-equilibrium existence. We think that empowering players with abilities to directly rank strategy profiles might have overstated actual players' sophisticated-ness, and have shifted too much burden from the game's analysis to its setup. Our setup where players have preferences over payoff- rather than strategy-related entities, besides being likely more realistic and parsimonious, avoids the issue pertaining to the observability of other players' strategies. Also, with randomization of actions intrinsic in our equilibria, we need no special requirements on preferences to achieve existence results.

Riedel and Sass \cite{RS13} allowed players to adjust ambiguities about the random devices used in action generations. Their resultant Ellsberg equilibria generalized Nash equilibria. While dealing with ambiguous mechanism design, Bose and Renou \cite{BR14} assumed that each $\Omega_t=\{t\}$ and that players adopt~(\ref{s-form}) as their attitudes.
Ambiguities have also been considered in auctions; see, e.g., Lo \cite{L98} and Bose, Ozdenoren, and Pape \cite{BOP06}. Meanwhile, the otherwise traditional model of Milgrom and Weber \cite{MW82} allowed bidders to waver on worths of the item.

\section{General Formulation}\label{formulation}

\subsection{Game Primitives}\label{no-need}

Given space $X$ with metric $d_X$, we use ${\cal B}(X)$ for its Borel $\sigma$-field and ${\cal P}(X)$ for the space of probabilities defined on the mesurable space $(X,{\cal B}(X))$. The space ${\cal P}(X)$ is endowed with the Prokhorov metric $\psi_X$, which also induces weak convergence. It will be separable when $X$ is. When the latter is compact, ${\cal P}(X)$ will be so too. Given metric spaces $X$ and $Y$, let ${\cal C}(X,Y)$ be the space of continuous mappings from $X$ to $Y$. Its members will be uniformly continuous when $X$ is compact; they will further be bounded when $Y$ is the real line $\Re$.

We let the finite $N\equiv \{1,...,\bar n\}$ be the set of players. Each player $n\in N$ is associated with a finite type space $T_n\equiv\{1,...,\bar t_n\}$. 
For convenience, we call player $n$ with type $t_n$ the $(n,t_n)$-player. This player's action comes from some metric space $A_{n,t_n}$. Let $T\equiv \prod_{n\in N}T_n$ be the space of type profiles and for each such profile $t\equiv (t_n)_{n\in N}\in T$, let $A_t\equiv \prod_{n\in N} A_{n,t_n}$ be the space of allowable action profiles under $t$.

Suppose metric space $\Omega$ hosts states of the world. Given $n\in N$, let $(\Omega_{n,t_n})_{t_n\in T_n}$ be a partition of $\Omega$, with each $\Omega_{n,t_n}$ containing all states of the world that correspond to player $n$'s type $t_n$. Even when player $n$ knows his type to be a certain $t_n$, he should anticipate the state of the world $\omega$ to potentially come from anywhere in $\Omega_{n,t_n}$. Given $t\equiv(t_n)_{n\in N}\in T$, use $\Omega_t\equiv \bigcap_{n\in N}\Omega_{n,t_n}$ for the set hosting all states of the world that correspond to each player $n$ his type $t_n$.
Spaces concerning the states of the world or in our own words, external factors, can be ``averaged away'' from the modeler's view if the traditional approach is taken. Here, with general ambiguity attitudes being considered, they will not.


After introducing players, types, actions, and states of the world, we now turn to payoffs. For $n\in N$ and $t\in T$, let there be Borel-measurable functions $r_{n,t}\equiv r_{n,t_n,t_{-n}}$ from $A_t\times \Omega_t$ to some metric space $R_{n,t_n}$, so that each $r_{n,t_n,t_{-n}}(a,\omega)$ stands for the payoff to player $n$ under type profile $t\equiv (t_n,t_{-n})$, pure action profile $a\equiv (a_n)_{n\in N}\in A_t$, and state of the world $\omega\in \Omega_t$. Note that the payoff spaces $R_{n,t_n}$ do not have to be one-dimensional real sets. When nothing escapes a player's notice, for instance, we could let $R_{n,t_n}=A_{n,t_n}\times \prod_{m\neq n}(\bigcup_{t_m\in T_m}A_{m,t_m})\times\Omega_{n,t_n}$, and let each $r_{n,t_n,t_{-n}}$ be the identity map on $A_t\times\Omega_t\equiv A_{n,t_n}\times\prod_{m\neq n}A_{m,t_m}\times[\Omega_{n,t_n}\cap (\bigcap_{m\neq n}\Omega_{m,t_m})]$ being treated as from $A_t\times\Omega_t$ to the current superset  $R_{n,t_n}$. Even for this extreme case, the abilities to rank opponents' actions as implied by our preferences pose much lower requirements than those of Grant, Meneghel, and Tourky \cite{GMT16}, which amounted to being able to rank opponents' strategies. 

Each player $n$, when seeing his type $t_n$, will be able to tell that the actual realization $\omega$ is in $\Omega_{n,t_n}$; however, nothing else, including opponents' types, can be determined. During the game's play where opponents mete out their behavioral strategies while the player himself may or may not randomize on actions, he will face choices on payoff-distribution vectors of the form $\pi\equiv (\pi(\omega)|\omega\in\Omega_{n,t_n})$, where each component $\pi(\omega)$ is a member of ${\cal P}(R_{n,t_n})$.

It will become clear that Harsanyi's \cite{H67-8} approach and by specialization, Nash's \cite{N50}\cite{N51} as well, represent a very special view on how players should rank the payoff-distribution vectors $\pi$. Here, we make generalizations. First, we just associate each $(n,t_n)$-player with a strict preference relation $\succ_{n,t_n}$ on the space $({\cal P}(R_{n,t_n}))^{\Omega_{n,t_n}}$ of such vectors. The relation is merely required to be irreflexive and transitive, to the effect that\\
\indent\M (I) $\pi\not \succ_{n,t_n}\pi$ for any $\pi\in ({\cal P}(R_{n,t_n}))^{\Omega_{n,t_n}}$;\\
\indent\M (II) $\pi\succ_{n,t_n} \pi''$ whenever $\pi\succ_{n,t_n} \pi'$ and $\pi'\succ_{n,t_n} \pi''$.\\
For example, it might be that $\Omega_{n,t_n}=\{\mbox{hot day},\mbox{cold day}\}$ and $R_{n,t_n}=\{\mbox{ice cream},\mbox{beef stew}\}$. Then, one $\succ_{n,t_n}$ might dictate that ``ice cream when it is hot and beef stew when it is cold'' is strictly preferred to ``either type of food with a 50\% chance on either type of a day'', which is in turn strictly preferred to ``beef stew when it is hot and ice cream when it is cold''.

We focus on the preference game $\Gamma\equiv (N,(T_n)_{n\in N},(A_{n,t_n})_{n\in N,t_n\in T_n},\Omega,(\Omega_{n,t_n})_{n\in N,t_n\in T_n}$, $(R_{n,t_n})_{n\in N,t_n\in T_n},(r_{n,t})_{n\in N,t\in T},(\succ_{n,t_n})_{n\in N,t_n\in T_n})$. To recap, $N$ is the set of players, each $T_n$ is player $n$'s type space, and each $A_{n,t_n}$ is the action space of the $(n,t_n)$-player; also, $\Omega$ is the space for states of the world and each $\Omega_{n,t_n}$ contains states of the world that lead to player $n$'s type $t_n$; finally, each $R_{n,t_n}$ is the payoff space of the $(n,t_n)$-player, each $r_{n,t}\equiv r_{n,t_n,t_{-n}}$ is that player's payoff function from $A_t\times \Omega_t$ to $R_{n,t_n}$ when opponents' type profile happens to be $t_{-n}$, and each $\succ_{n,t_n}$ is the preference relation adopted by the $(n,t)$-player.

\subsection{Payoff-distribution Vectors}

Every player $n$'s behavioral strategy $\delta_n$ can be understood as the vector $(\delta_{n,t_n})_{t_n\in T_n}$, where each component $\delta_{n,t_n}\in \Delta_{n,t_n}\equiv {\cal P}(A_{n,t_n})$ is a probability distribution over actions in $A_{n,t_n}$. That is, $\delta_n\in\Delta_n\equiv \prod_{t_n\in T_n}\Delta_{n,t_n}$. This way, $\delta_n$ offers a plan for player $n$ on what potentially randomized action to take under each type realization $t_n$.

Let $\Delta\equiv\prod_{n\in N}\Delta_n$ be the space of all behavioral-strategy profiles covering all players. To the $(n,t_n)$-player, opponents' type profile $t_{-n}\equiv (t_m)_{m\neq n}$ may be anything from $T_{-n}\equiv \prod_{m\neq n}T_m$ and their behavioral-strategy profile $\delta_{-n}\equiv (\delta_m)_{m\neq n}$ may be anything from $\Delta_{-n}\equiv \prod_{m\neq n}\Delta_m$. At each fixed $t_{-n}\in T_{-n}$, it is the $\delta_{-n,t_{-n}}\equiv (\delta_{m,t_m})_{m\neq n}$-portion of $\delta_{-n}$ that will materialize, and the state of the world $\omega$ must be from $\Omega_{t_n,t_{-n}}$. Note that $(\Omega_{t_n,t_{-n}})_{t_{-n}\in T_{-n}}$ forms a partition of $\Omega_{n,t_n}$. In the following, we will take the liberty to use notation like $a(\mbox{ or }\delta)_{-n,t_{-n}}\equiv (a(\mbox{ or }\delta)_{m,t_m})_{m\neq n}$ and $A(\mbox{ or }\Delta)_{-n,t_{-n}}\equiv \prod_{m\neq n}A(\mbox{ or }\Delta)_{m,t_m}$.

As noted, there are two ways in which payoff-distribution vectors can be formed during the play of a game where  players' random actions are verifiable and yet ambiguities exist on external factors. The action-based case will emerge when each player has almost a free reign on the actions to take except with the long-term goal of playing out a given randomized strategy; meanwhile, the distribution-based case will arise when players use exogenously generated random numbers to map out their chosen random strategies.

From the action-based perspective, the $(n,t_n)$-player is in total command of his own action whilst anticipating random actions from other players. Then, under his pure action $a_{n,t_n}\in A_{n,t_n}$, opponent behavioral-strategy profile $\delta_{-n}\equiv (\delta_{m,t_m})_{m\neq n,t_m\in T_m}\in \Delta_{-n}$, opponent type profile $t_{-n}\in T_{-n}$, and state $\omega\in \Omega_{t_n,t_{-n}}$, the player will expect the payoff distribution
\begin{equation}\label{mechanism}
\pi^{\mbox a}_{n,t_n,t_{-n}}(a_{n,t_n},\delta_{-n,t_{-n}},\omega)=
\left(\prod_{m\neq n}\delta_{m,t_m}\right)\cdot (r_{n,t_n,t_{-n}}(a_{n,t_n},\cdot,\omega))^{-1}\in {\cal P}(R_{n,t_n}).
\end{equation}
For any $R'_{n,t_n}\in {\cal B}(R_{n,t_n})$, the probability $[\pi^{\mbox a}_{n,t_n,t_{-n}}(a_{n,t_n},\delta_{-n,t_{-n}},\omega)](R'_{n,t_n})$ equals
\begin{equation}\label{aha}\begin{array}{l}
\int_{A_{-n,t_{-n}}}{\bf 1}(\{a_{-n,t_{-n}}\mbox{ with }r_{n,t_n,t_{-n}}(a_{n,t_n},a_{-n,t_{-n}},\omega)\in R'_{n,t_n}\})\cdot [\prod_{m\neq n} \delta_{m,t_m}](da_{-n,t_{-n}})\\
\;\;\;\;\;\;\;\;\;\;\;\;=(\prod_{m\neq n} \delta_{m,t_m})(\{a_{-n,t_{-n}}\in A_{-n,t_{-n}}|r_{n,t_n,t_{-n}}(a_{n,t_n},a_{-n,t_{-n}},\omega)\in R'_{n,t_n}\}),
\end{array}\end{equation}
where ${\bf 1}(\cdot)$ stands for the indicator function. The above reflects how opponents' random actions result with the current player's random payoff distribution. Assumptions to be made in Section~\ref{legit} will ensure that all operations in this and the ensuring Section~\ref{equili} are legitimate. Had player $n$ known his opponents' type profile $t_{-n}\in T_{-n}$, he would have anticipated the ``$|\Omega_{t_n,t_{-n}}|$-dimensional'' vector
\begin{equation}\label{buttt}
\pi^{\mbox a}_{n,t_n,t_{-n}}(a_{n,t_n},\delta_{-n,t_{-n}})\equiv \left(\pi^{\mbox a}_{n,t_n,t_{-n}}(a_{n,t_n},\delta_{-n,t_{-n}},\omega)|\omega\in \Omega_{t_n,t_{-n}}\right).
\end{equation}
However, the player is unaware of opponents' actual type profile. So he should contemplate on the ``$|\Omega_{n,t_n}|$-dimensional'' vector that is patched up from the vectors defined in~(\ref{buttt}):
\begin{equation}\label{vec-p}\begin{array}{ll}
\left(\pi^{\mbox a}_{n,t_n,t_{-n}}(a_{n,t_n},\delta_{-n,t_{-n}})\right)_{t_{-n}\in T_{-n}}&= \left((\pi^{\mbox a}_{n,t_n,t_{-n}}(a_{n,t_n},\delta_{-n,t_{-n}},\omega)|\omega\in \Omega_{t_n,t_{-n}})\right)_{t_{-n}\in T_{-n}}\\
&= \left(\pi^{\mbox a}_{n,t_n,t_{-n}}(a_{n,t_n},\delta_{-n,t_{-n}},\omega)|\omega\in \Omega_{n,t_n}\right),
\end{array}\end{equation}
where the second equality comes from $\Omega_{n,t_n}=\bigcup_{t_{-n}\in T_{-n}}\Omega_{t_n,t_{-n}}$. The resulting payoff-distribution vector $\pi^{\mbox a}_{n,t_n}(a_{n,t_n},\delta_{-n})$ is a member of $({\cal P}(R_{n,t_n}))^{\bigcup_{t_{-n}\in T_{-n}}\Omega_{t_n,t_{-n}}}\equiv({\cal P}(R_{n,t_n}))^{\Omega_{n,t_n}}$.

From the distribution-based perspective, the $(n,t_n)$-player is in control of his behavioral strategy. When he is committed to some $\delta_{n,t_n}\in \Delta_{n,t_n}$ while other players have adopted behavioral-strategy profile $\delta_{-n}\in \Delta_{-n}$, the player should, under opponent type profile $t_{-n}\in T_{-n}$ and state of the world $\omega\in\Omega_{t_n,t_{-n}}$, anticipate the additionally mixed distribution
\begin{equation}\label{oklaa}
\pi^{\mbox d}_{n,t_n,t_{-n}}(\delta_{n,t_n},\delta_{-n,t_{-n}},\omega)=\left(\delta_{n,t_n}\times \prod_{m\neq n}\delta_{m,t_m}\right)\cdot (r_{n,t_n,t_{-n}}(\cdot,\cdot,\omega))^{-1}\in {\cal P}(R_{n,t_n}).
\end{equation}
In view of~(\ref{mechanism}) and~(\ref{aha}), the above~(\ref{oklaa}) could also be understood as
\begin{equation}\label{oklab}
\pi^{\mbox d}_{n,t_n,t_{-n}}(\delta_{n,t_n},\delta_{-n,t_{-n}},\omega)=\int_{A_{n,t_n}}\pi^{\mbox a}_{n,t_n,t_{-n}}(a_{n,t_n},\delta_{-n,t_{-n}},\omega)\cdot\delta_{n,t_n}(da_{n,t_n}),
\end{equation}
in the sense that, for any $R'_{n,t_n}\in {\cal B}(R_{n,t_n})$,
\begin{equation}\label{oklac}
[\pi^{\mbox d}_{n,t_n,t_{-n}}(\delta_{n,t_n},\delta_{-n,t_{-n}},\omega)](R'_{n,t_n})=\int_{A_{n,t_n}}[\pi^{\mbox a}_{n,t_n,t_{-n}}(a_{n,t_n},\delta_{-n,t_{-n}},\omega)](R'_{n,t_n})\cdot\delta_{n,t_n}(da_{n,t_n}).
\end{equation}
This just means that the payoff distribution under the $(n,t_n)$-player's action distribution  $\delta_{n,t_n}$ is a mixture of the payoff distributions under the player's pure actions. Now, let vector $\pi^{\mbox d}_{n,t_n}(\delta_{n,t_n},\delta_{-n})$ be equated to
\begin{equation}\label{vec-a}\begin{array}{ll}
\left(\pi^{\mbox d}_{n,t_n,t_{-n}}(\delta_{n,t_n},\delta_{-n,t_{-n}})\right)_{t_{-n}\in T_{-n}}&\equiv \left((\pi^{\mbox d}_{n,t_n,t_{-n}}(\delta_{n,t_n},\delta_{-n,t_{-n}},\omega)|\omega\in \Omega_{t_n,t_{-n}})\right)_{t_{-n}\in T_{-n}}\\
&= \left(\pi^{\mbox d}_{n,t_n,t_{-n}}(\delta_{n,t_n},\delta_{-n,t_{-n}},\omega)|\omega\in \Omega_{n,t_n}\right),
\end{array}\end{equation}
i.e., the vector of all potential payoff distributions under opponent type profiles $t_{-n}\in T_{-n}$ and states of the world $\omega\in\Omega_{t_n,t_{-n}}$. It is again a member of $({\cal P}(R_{n,t_n}))^{\Omega_{n,t_n}}$.

\subsection{Equilibrium Definitions}\label{equili}

The two perspectives lead to two equilibrium notions that can be different under general ambiguity attitudes. In the action-based case, each player $n$ should respond to any opponent strategy profile $\delta_{-n}$ by choosing action distribution $\delta_{n,t_n}$ for each type realization $t_n$ that gives no chance to any action $a_{n,t_n}$ whose corresponding vector $\pi^{\mbox a}_{n,t_n}(a_{n,t_n},\delta_{-n})$ defined at~(\ref{vec-p}) could be less preferential than that of any other action. In the distribution-based case, the player should choose action distribution $\delta_{n,t_n}$ at each $t_n$ so that the corresponding vector $\pi^{\mbox d}_{n,t_n}(\delta_{n,t_n},\delta_{-n})$ defined at~(\ref{vec-a}) is not less preferential than that of any other distribution.

Start from the action-based perspective. For each player $n\in N$, type $t_n\in T_n$, and opponent strategy profile $\delta_{-n}\in \Delta_{-n}$, let $\hat A^{\mbox a}_{n,t_n}(\delta_{-n})$ be the set of actions $a_{n,t_n}$ that render the vector $\pi^{\mbox a}_{n,t_n}(a_{n,t_n},\delta_{-n})$ as defined in~(\ref{vec-p}) $\succ_{n,t_n}$-maximal:
\begin{equation}\label{maximal0}
\hat A^{\mbox a}_{n,t_n}(\delta_{-n})
=\left\{a_{n,t_n}\in A_{n,t_n}|\pi^{\mbox a}_{n,t_n}(a'_{n,t_n},\delta_{-n})\not \succ_{n,t_n}\pi^{\mbox a}_{n,t_n}(a_{n,t_n},\delta_{-n})\;\;\forall a'_{n,t_n}\in A_{n,t_n}\right\}.
\end{equation}
For any $n\in N$ and $t_n\in T_n$, let best-response correspondence $\hat B^{\mbox a}_{n,t_n}:\Delta_{-n}\rightrightarrows \Delta_{n,t_n}$ be such that, for any opponent strategy profile $\delta_{-n}\in \Delta_{-n}$,
\begin{equation}\label{b-def}
\hat B^{\mbox a}_{n,t_n}(\delta_{-n})=\left\{\delta_{n,t_n}\in \Delta_{n,t_n}|\delta_{n,t_n}(\hat A^{\mbox a}_{n,t_n}(\delta_{-n}))=1\right\},
\end{equation}
where $\hat A^{\mbox a}_{n,t_n}(\delta_{-n})$ has just been defined in~(\ref{maximal0}). Thus, $\delta_{n,t_n}$ will be considered one of the $(n,t_n)$-player's best responses to $\delta_{-n}$ when its support is made up of those $a_{n,t_n}$'s that render $\pi^{\mbox a}_{n,t_n}(a_{n,t_n},\delta_{-n})$ $\succ_{n,t_n}$-maximal among all  $\pi^{\mbox a}_{n,t_n}(a'_{n,t_n},\delta_{-n})$'s.

Now define correspondence $\hat B^{\mbox a}:\Delta\rightrightarrows \Delta$ from strategy profiles to themselves, so that
\begin{equation}\label{zia-p}
\delta'\in \hat B^{\mbox a}(\delta)\;\;\mbox{ if and only if }\;\;\delta'_{n,t_n}\in \hat B^{\mbox a}_{n,t_n}(\delta_{-n})\mbox{ for any }n\in N\mbox{ and }t_n\in T_n.
\end{equation}
A behavioral-strategy profile $\delta\equiv(\delta_{n,t_n})_{n\in N,t_n\in T_n}\in \Delta\equiv \prod_{n\in N}\prod_{t_n\in T_n}\Delta_{n,t_n}$ will be considered an action-based equilibrium of $\Gamma$ if $\delta\in \hat B^{\mbox a}(\delta)$. For convenience, we use ${\cal E}^{\mbox a}\subseteq \Delta$ to denote the set of all such equilibria.

Let us move on to the distribution-based perspective. For any player $n\in N$ and type $t_n\in T_n$, define best-response correspondence $\hat B^{\mbox d}_{n,t_n}:\Delta_{-n}\rightrightarrows \Delta_{n,t_n}$ so that, for any opponent strategy profile $\delta_{-n}\in \Delta_{-n}$,
\begin{equation}\label{maximal-k0}
\hat B^{\mbox d}_{n,t_n}(\delta_{-n})=\left\{\delta_{n,t_n}\in \Delta_{n,t_n}|\pi^{\mbox d}_{n,t_n}(\delta'_{n,t_n},\delta_{-n})\not \succ_{n,t_n}\pi^{\mbox d}_{n,t_n}(\delta_{n,t_n},\delta_{-n})\;\;\forall \delta'_{n,t_n}\in \Delta_{n,t_n}\right\}.
\end{equation}
Here, a $\delta_{n,t_n}$ will be considered one of the $(n,t_n)$-player's best responses to $\delta_{-n}$ when $\pi^{\mbox d}_{n,t_n}(\delta_{n,t_n},\delta_{-n})$ is $\succ_{n,t_n}$-maximal among all $\pi^{\mbox d}_{n,t_n}(\delta'_{n,t_n},\delta_{-n})$'s.

Now define correspondence $\hat B^{\mbox d}:\Delta\rightrightarrows \Delta$ from strategy profiles to themselves, so that
\begin{equation}\label{zia-a}
\delta'\in \hat B^{\mbox d}(\delta)\;\;\mbox{ if and only if }\;\;\delta'_{n,t_n}\in \hat B^{\mbox d}_{n,t_n}(\delta_{-n})\mbox{ for any }n\in N\mbox{ and }t_n\in T_n.
\end{equation}
A behavioral-strategy profile $\delta\in \Delta$ will be considered a distribution-based equilibrium of $\Gamma$ if $\delta\in \hat B^{\mbox d}(\delta)$. For convenience, we use ${\cal E}^{\mbox d}\subseteq \Delta$ to denote the set of all such equilibria.

When every type space $T_n$ is a singleton, the game $\Gamma$ will be normal-form. Then, all the state spaces $\Omega_t$ and $\Omega_{n,t_n}$ will be equatable to $\Omega$; hence, all payoff-distribution vectors will be of the same length. Certainly, no separate treatment is needed for this special case.

\section{Existence and Continuity of Equilibria}\label{analysis}

\subsection{Compactness and Continuity}\label{legit}

Let us provide conditions under which the equilibrium sets ${\cal E}^{\mbox a}$ and ${\cal E}^{\mbox d}$ will be nonempty. We first make the following assumptions related to compactness and continuity.

\begin{assumption}\label{compact-action}
For any $n\in N$ and $t_n\in T_n$, the action space $A_{n,t_n}$ is compact.
\end{assumption}

\begin{assumption}\label{compact-state}
The state space $\Omega$ is compact and for any type profile $t\equiv (t_n)_{n\in N}\in T\equiv \prod_{n\in N}T_n$, the subset $\Omega_t\equiv\bigcap_{n\in N}\Omega_{n,t_n}$ is closed and hence compact.
\end{assumption}

\begin{assumption}\label{compact-payoff}
For any $n\in N$ and $t_n\in T_n$, the payoff space $R_{n,t_n}$ is compact.
\end{assumption}

\begin{assumption}\label{continuity}
For any player $n\in N$ and type profile $t\in T$, the payoff function $r_{n,t}$ from $A_t\times \Omega_t\equiv \prod_{n\in N}A_{n,t_n}\times \bigcap_{n\in N}\Omega_{n,t_n}$ to $R_{n,t_n}$ is continuous.
\end{assumption}


Assumptions~\ref{compact-action} to~\ref{continuity} are all quite routine. The only exception might be Assumption~\ref{compact-state}, which requires that all the disjoint spaces $\Omega_t$ be closed and hence ``mutually distinguishable''. But this is trivially satisfied by the case where $\Omega_t=\{t\}\times\tilde\Omega$ for some common space $\tilde\Omega$.

By Assumption~\ref{compact-action}, each $A_t\equiv\prod_{n\in N}A_{n,t_n}$ is compact. Also, all the action-distribution spaces $\Delta_{n,t_n}\equiv{\cal P}(A_{n,t_n})$, $\Delta_{-n,t_{-n}}\equiv\prod_{m\neq n}\Delta_{m,t_m}$, $\Delta_n\equiv\prod_{t_n\in T_n}\Delta_{n,t_n}$, $\Delta_{-n}\equiv\prod_{m\neq n}\Delta_m$, and $\Delta\equiv\prod_{n\in N}\Delta_n$ will be compact. By Assumption~\ref{compact-state}, each $\Omega_{n,t_n}\equiv\bigcup_{t_{-n}\in T_{-n}}\Omega_{t_n,t_{-n}}$ is compact. Consequently, all the state-distribution spaces ${\cal P}(\Omega_t)$, ${\cal P}(\Omega_{n,t_n})$, and ${\cal P}(\Omega)$ will be  compact. With Assumption~\ref{compact-payoff}, we have the compactness of the payoff-distribution space ${\cal P}(R_{n,t_n})$. Note that the payoff functions $r_{n,t}$ are defined on compact domains $A_t\times \Omega_t$. Due also to the continuity stated in Assumption~\ref{continuity}, we can obtain some much needed continuity.

\begin{proposition}\label{pi-cont}
For any $n\in N$, $t_n\in T_n$, and $t_{-n}\in T_{-n}$, the function $\pi^{\mbox a}_{n,t_n,t_{-n}}$ defined at~(\ref{mechanism}) from $A_{n,t_n}\times \Delta_{-n,t_{-n}}\times \Omega_{t_n,t_{-n}}$ to the payoff-distribution space ${\cal P}(R_{n,t_n})$ is continuous; also, the function $\pi^{\mbox d}_{n,t_n,t_{-n}}$ defined at~(\ref{oklaa}) from $\Delta_{n,t_n}\times \Delta_{-n,t_{-n}}\times \Omega_{t_n,t_{-n}}$ to ${\cal P}(R_{n,t_n})$ is continuous.
\end{proposition}

Due to the compactness of all involved spaces, we can obtain from Proposition~\ref{pi-cont} the uniform continuity of $\pi^{\mbox a}_{n,t_n,t_{-n}}$ and $\pi^{\mbox d}_{n,t_n,t_{-n}}$. So instead of $({\cal P}(R_{n,t_n}))^{\Omega_{t_n,t_{-n}}}$, we can restrict the vectors $\pi^{\mbox a}_{n,t_n,t_{-n}}(a_{n,t_n},\delta_{-n,t_{-n}})$ and $\pi^{\mbox d}_{n,t_n,t_{-n}}(\delta_{n,t_n},\delta_{-n,t_{-n}})$ defined in~(\ref{buttt}) and~(\ref{vec-a}), respectively, to the smaller ${\cal C}(\Omega_{t_n,t_{-n}},{\cal P}(R_{n,t_n}))$, the space of all uniformly continuous mappings from $\Omega_{t_n,t_{-n}}$ to ${\cal P}(R_{n,t_n})$. Since the $\Omega_t$'s are closed and disjoint, we must have
\begin{equation}\label{separatable}
d_\Omega(\Omega_t,\Omega_{t'})>0,\hspace*{.8in}\forall t,t'\in T\mbox{ with }t\neq t'.
\end{equation}
This allows us to patch up aforementioned functions through all $t_{-n}$'s to form $\pi^{\mbox a}_{n,t_n}(a_{n,t_n},\delta_{-n})$ and $\pi^{\mbox d}_{n,t_n}(\delta_{n,t_n},\delta_{-n})$
as members of $\Pi_{n,t_n}\equiv {\cal C}(\Omega_{n,t_n},{\cal P}(R_{n,t_n}))$. The payoff-distribution space ${\cal P}(R_{n,t_n})$ is bounded since the Prokhorov metric $\psi_{R_{n,t_n}}$ is always below 1. Now for the product space $({\cal P}(R_{n,t_n}))^{\Omega_{n,t_n}}$, we can define the uniform metric, namely, the supremum of all component-wise Prokhorov metrics on ${\cal P}(R_{n,t_n})$. 
Note that $\Pi_{n,t_n}$ is a closed subset of $({\cal P}(R_{n,t_n}))^{\Omega_{n,t_n}}$ according to Theorem 43.6 of Munkres \cite{MK00}. 

Moreover, uniform continuities stated in the above will lead to the following.

\begin{proposition}\label{piphi-cont}
For any player $n\in N$ and any of his types $t_n\in T_n$, the vector-valued functions $\pi^{\mbox a}_{n,t_n}$ from $A_{n,t_n}\times \Delta_{-n}$ to $\Pi_{n,t_n}$ defined through~(\ref{vec-p}) and $\pi^{\mbox d}_{n,t_n}$ from $\Delta_{n,t_n}\times \Delta_{-n}$ to $\Pi_{n,t_n}$ defined through~(\ref{vec-a}) are both continuous.
\end{proposition}

Proposition~\ref{piphi-cont} propounds two points. First, not every map $\pi^{\mbox a}_{n,t_n}(a_{n,t_n},\delta_{-n})$ or $\pi^{\mbox d}_{n,t_n}(\delta_{n,t_n},\delta_{-n})$ from states to payoff distributions will arise during any play of the preference game $\Gamma$; rather, only those continuous ones will do. Second, the continuous mappings themselves will react continuously to changes, in own actions $a_{n,t_n}$ and opponent behavioral-strategy profiles $\delta_{-n}$ in the action-based case and in own action distributions $\delta_{n,t_n}$ and opponent behavioral-strategy profiles $\delta_{-n}$ in the distribution-based case.

\subsection{Confined Definition of Preferences}

Rather than defined for the entire space $({\cal P}(R_{n,t_n}))^{\Omega_{n,t_n}}$ of payoff-distribution vectors, we can confine each preference relation $\succ_{n,t_n}$ to the smaller space $\Pi_{n,t_n}\equiv {\cal C}(\Omega_{n,t_n},{\cal P}(R_{n,t_n}))$ of continuous payoff-distribution vectors. For convenience, define the set of pairs
\begin{equation}\label{fff-def}
\varphi_{n,t_n}=\left\{(\pi,\pi')\in \Pi_{n,t_n}\times \Pi_{n,t_n}|\pi\not \succ_{n,t_n}\pi'\right\},
\end{equation}
where the left members are not less preferential than the right ones. With this definition, an alternative way to express~(\ref{maximal0}) is
\begin{equation}\label{maximal}
\hat A^{\mbox a}_{n,t_n}(\delta_{-n})=\left\{a_{n,t_n}\in A_{n,t_n}|\tilde\Pi^{\mbox a}_{n,t_n}(\delta_{-n})\times\{\pi^{\mbox a}_{n,t_n}(a_{n,t_n},\delta_{-n})\}\subseteq \varphi_{n,t_n}\right\},
\end{equation}
where
\begin{equation}\label{pi-p}
\tilde\Pi^{\mbox a}_{n,t_n}(\delta_{-n})=\left[\pi^{\mbox a}_{n,t_n}(\cdot,\delta_{-n})\right](A_{n,t_n})\equiv\left\{\pi^{\mbox a}_{n,t_n}(a'_{n,t_n},\delta_{-n})|a'_{n,t_n}\in A_{n,t_n}\right\},
\end{equation}
is the set of potential payoff-distribution vectors in $\Pi_{n,t_n}$ to be experienced by the $(n,t_n)$-player when he tries all possible pure actions in $A_{n,t_n}$ while his opponents are fixated at the behavioral-strategy profile $\delta_{-n}$. Meanwhile, an alternative way to express~(\ref{maximal-k0}) is
\begin{equation}\label{maximal-k}
\hat B^{\mbox d}_{n,t_n}(\delta_{-n})=\left\{\delta_{n,t_n}\in \Delta_{n,t_n}|\tilde\Pi^{\mbox d}_{n,t_n}(\delta_{-n})\times\{\pi^{\mbox d}_{n,t_n}(\delta_{n,t_n},\delta_{-n})\}\subseteq \varphi_{n,t_n}\right\},
\end{equation}
where
\begin{equation}\label{pi-a}
\tilde\Pi^{\mbox d}_{n,t_n}(\delta_{-n})=\left[\pi^{\mbox d}_{n,t_n}(\cdot,\delta_{-n})\right](\Delta_{n,t_n})\equiv\left\{\pi^{\mbox d}_{n,t_n}(\delta'_{n,t_n},\delta_{-n})|\delta'_{n,t_n}\in \Delta_{n,t_n}\right\},
\end{equation}
is the set of potential payoff-distribution vectors in $\Pi_{n,t_n}$ to be experienced by the $(n,t_n)$-player when he tries all possible action distributions in $\Delta_{n,t_n}$ while his opponents are fixated at the behavioral-strategy profile $\delta_{-n}$.

Due to $\pi^{\mbox a}_{n,t_n}(\cdot,\delta_{-n})$'s continuity as suggested by Proposition~\ref{piphi-cont}, the compactness of $A_{n,t_n}$ will translate into that of $\tilde\Pi^{\mbox a}_{n,t_n}(\delta_{-n})$ defined at~(\ref{pi-p}). Similarly, $\pi^{\mbox d}_{n,t_n}(\cdot,\delta_{-n})$'s continuity and $\Delta_{n,t_n}$'s compactness will together lead to the compactness of $\tilde\Pi^{\mbox d}_{n,t_n}(\delta_{-n})$ defined at~(\ref{pi-a}).
We now make an assumption on the $\succ_{n,t_n}$'s as they are defined on the spaces $\Pi_{n,t_n}$.

\begin{p-assumption}\label{ca-l}
For any player $n\in N$ and any of his types $t_n\in T_n$, the relation $\succ_{n,t_n}$ is continuous; namely, $\varphi_{n,t_n}$ defined in~(\ref{fff-def}) is a closed subset of $\Pi_{n,t_n}\times \Pi_{n,t_n}$.
\end{p-assumption}

Preference Assumption~\ref{ca-l} is routinely treated as part of the definition of a preference. We single it out just to emphasize its importance. The following is an important consequence.

\begin{proposition}\label{ca-u}
For any player $n\in N$ and any of his types $t_n\in T_n$, a compact $\Pi'\subseteq \Pi_{n,t_n}$ can always reach $\succ_{n,t_n}$-maximal; that is, there exists some $\pi\in \Pi'$ so that $\pi'\not \succ_{n,t_n} \pi$ for any $\pi'\in \Pi'$, or in other words, $\Pi'\times\{\pi\} \subseteq \varphi_{n,t_n}$.
\end{proposition}

This result is well known; see e.g., Lemma 2 of Schmeidler \cite{S69} and Theorem 5.1 of Khan and Sun \cite{KS90}. We reproduce it here for the sake of completeness.

\subsection{Existence Derivations}

In view of the compactness of $\tilde\Pi^{\mbox a}_{n,t_n}(\delta_{-n})$, Proposition~\ref{ca-u} will lead to the nonemptiness of $\hat A^{\mbox a}_{n,t_n}(\delta_{-n})$ as defined in~(\ref{maximal}). Indeed,
\begin{equation}
\left[\pi^{\mbox a}_{n,t_n}(\cdot,\delta_{-n})\right]\left(\hat A^{\mbox a}_{n,t_n}(\delta_{-n})\right)\bigcap \tilde\Pi^{\mbox a}_{n,t_n}(\delta_{-n})\neq\emptyset.
\end{equation}
Note that the nonempty $\hat A^{\mbox a}_{n,t_n}(\delta_{-n})$ is originally defined in~(\ref{maximal0}). So we will have the nonemptiness of $\hat B^{\mbox a}_{n,t_n}(\delta_{-n})$ as well, because by~(\ref{b-def}), the latter contains the Dirac measure $1_{a_{n,t_n}}$ for any $a_{n,t_n}\in \hat A^{\mbox a}_{n,t_n}(\delta_{-n})$. Meanwhile, Preference Assumption~\ref{ca-l} will together with the continuity of $\pi^{\mbox a}_{n,t_n}$ lead to the closedness of $\hat A^{\mbox a}_{n,t_n}$ as a correspondence.

\begin{proposition}\label{a-condition}
Each $\hat A^{\mbox a}_{n,t_n}(\cdot)$ defined by~(\ref{maximal0}) is closed as a correspondence.
\end{proposition}

It turns out that Proposition~\ref{a-condition} will lead to the closedness of each correspondence $\hat B^{\mbox a}_{n,t_n}$ as defined in~(\ref{b-def}), from the space $\Delta_{-n}$ of opponents' behavioral strategies to the space $\Delta_{n,t_n}$ of the $(n,t_n)$-player's behavioral strategies.

\begin{proposition}\label{b-continuity}
Each $\hat B^{\mbox a}_{n,t_n}$ defined by~(\ref{b-def}) is closed as a correspondence; also, each $\hat B^{\mbox d}_{n,t_n}$ defined by~(\ref{maximal-k0}) is closed as a correspondence.
\end{proposition}

For the distribution-based case, the definition of $\hat B^{\mbox d}_{n,t_n}(\cdot)$ in~(\ref{maximal-k0}) is almost the same as that of $\hat A^{\mbox a}_{n,t_n}(\cdot)$ in~(\ref{maximal0}), except with the earlier $a_{n,t_n},a'_{n,t_n}\in A_{n,t_n}$ replaced by $\delta_{n,t_n},\delta'_{n,t_n}\in \Delta_{n,t_n}$. So starting from $\pi^{\mbox d}_{n,t_n}$'s continuity, we can follow almost the same steps to deduce the nonemptiness of each $\hat B^{\mbox d}_{n,t_n}(\delta_{-n})$.

\begin{proposition}\label{kwaza}
The sets $\hat B^{\mbox a}(\delta)$ defined through~(\ref{zia-p}) and $\hat B^{\mbox d}(\delta)$ defined through~(\ref{zia-a}) are both nonempty at any behavioral-strategy profile $\delta\in\Delta$; in addition, both $\hat B^{\mbox a}$ and $\hat B^{\mbox d}$ are closed as correspondences from $\Delta$ to itself.
\end{proposition}

Coming back to the action-based case, the convexity of each $\hat B^{\mbox a}_{n,t_n}(\delta_{-n})$ is obvious from its definition at~(\ref{b-def}). So via~(\ref{zia-p}) each $\hat B^{\mbox a}(\delta)$ is also convex. On the other hand, each $\Delta_{n,t_n}$ is a compact and convex subset of the set of signed measures on $A_{n,t_n}$, which is itself a locally convex Hausdorff topological vector space. Thus, $\Delta\equiv\prod_{n\in N}\Delta_n\equiv\prod_{n\in N}\prod_{t_n\in T_n}\Delta_{n,t_n}$ is also a compact and convex subset of a locally convex Hausdorff topological vector space. This makes the closedness of $\hat B^{\mbox a}$ in Proposition~\ref{kwaza} equivalent to upper hemi-continuity. Therefore, we can use the Fan-Glicksberg theorem to verify the existence of a fixed point for $\hat B^{\mbox a}$.

Aside from convexity, the distribution-based case has almost all the properties enjoyed by the action-based case as shown above. To move further, we consider $\succ_{n,t_n}$ convex when
\begin{equation}\label{general-c}
\mbox{ both }\pi\not\succ_{n,t_n}\pi^0\mbox{ and }\pi\not\succ_{n,t_n}\pi^1\mbox{ will ensure }\pi\not\succ_{n,t_n} (1-\alpha)\cdot\pi^0+\alpha\cdot\pi^1\mbox{ for any }\alpha\in [0,1].
\end{equation}
When $\succ_{n,t_n}$ is complete, this concept is just ambiguity aversion seen in literature; see Schmeidler \cite{S89}. Now by~(\ref{oklab}) to~(\ref{vec-a}), we have the linearity of $\pi^{\mbox d}_{n,t_n}(\cdot,\delta_{-n})$, that
\begin{equation}
\pi^{\mbox d}_{n,t_n}[(1-\alpha)\cdot\delta^0_{n,t_n}+\alpha\cdot\delta^1_{n,t_n},\delta_{-n}]=(1-\alpha)\cdot \pi^{\mbox d}_{n,t_n}(\delta^0_{n,t_n},\delta_{-n})+\alpha\cdot \pi^{\mbox d}_{n,t_n}(\delta^1_{n,t_n},\delta_{-n}).
\end{equation}
In view of~(\ref{maximal-k0}), $\hat B^{\mbox d}_{n,t_n}(\delta_{-n})$ will be a convex subset of $\Delta_{n,t_n}$ when $\succ_{n,t_n}$ is convex. Taking similar steps to those for the action-based case, we can reach the existence of fixed points for $\hat B^{\mbox d}$ as defined at~(\ref{zia-a}). Summing up all of these, we can reach the following conclusion. 

\begin{theorem}\label{t-existence}
The game $\Gamma$ has action-based equilibria; that is, ${\cal E}^{\mbox a}\neq \emptyset$. When $\succ_{n,t_n}$ for every player $n\in N$ and any of his types $t_n\in T_n$ is a convex preference relation on $\Pi_{n,t_n}$, the game will have distribution-based equilibria, so that ${\cal E}^{\mbox d}\neq \emptyset$.
\end{theorem}

The extra condition needed by the second half of  Theorem~\ref{t-existence} somehow reflects on the ``rarity'' of distribution-based equilibria in comparison to their action-based brethren.

\subsection{Continuity of Equilibria}\label{alone-dep}


Let $\Phi_{n,t_n}$ be the space containing all closed sets of $\Pi_{n,t_n}\times\Pi_{n,t_n}$ that represent $(n,t_n)$-preferences. In view of (I) and (II) of each preference $\succ_{n,t_n}$'s definition and~(\ref{fff-def}) on each $\varphi_{n,t_n}$'s definition, $\Phi_{n,t_n}$ is the collection of all closed sets $\varphi_{n,t_n}$ of $\Pi_{n,t_n}\times\Pi_{n,t_n}$ that satisfy:\\
\indent\M (i) $(\pi,\pi)\in \varphi_{n,t_n}$ for any $\pi\in \Pi_{n,t_n}$;\\
\indent\M (ii) $(\pi,\pi'')\notin \varphi_{n,t_n}$ whenever $(\pi,\pi')\notin \varphi_{n,t_n}$ and $(\pi',\pi'')\notin \varphi_{n,t_n}$ for any $\pi,\pi',\pi''\in \Pi_{n,t_n}$. \\
\noindent We can now treat $\succ_{n,t_n}$ and $\varphi_{n,t_n}$ interchangeably. Recall that the metric $d_{\Pi_{n,t_n}}$ for the space $\Pi_{n,t_n}\equiv{\cal C}(\Omega_{n,t_n},{\cal P}(R_{n,t_n}))$ of continuous payoff-distribution vectors is such that
\begin{equation}
d_{\Pi_{n,t_n}}(\pi_1,\pi_2)=\sup_{\omega\in\Omega_{n,t_n}}\psi_{R_{n,t_n}}(\pi_1(\omega),\pi_2(\omega)),
\end{equation}
where $\psi_{R_{n,t_n}}$ is the Prokhorov metric on the payoff-distribution space ${\cal P}(R_{n,t_n})$. We can then define a metric $d_{\Pi_{n,t_n}\times\Pi_{n,t_n}}$ for the vector-pair space $\Pi_{n,t_n}\times \Pi_{n,t_n}$ by
\begin{equation}\label{federal}
d_{\Pi_{n,t_n}\times\Pi_{n,t_n}}((\pi_1,\pi'_1),(\pi_2,\pi'_2))=d_{\Pi_{n,t_n}}(\pi_1,\pi_2)\vee d_{\Pi_{n,t_n}}(\pi'_1,\pi'_2).
\end{equation}

Let ${\cal F}_{n,t_n}$ be the collection of nonempty closed subsets of $\Pi_{n,t_n}\times\Pi_{n,t_n}$. Because of (i), $\Phi_{n,t_n}$ is a subset of ${\cal F}_{n,t_n}$. A metric $d_{{\cal F}_{n,t_n}}$ for ${\cal F}_{n,t_n}$ can be defined using the Hausdorff distance; see, e.g., Hildenbrand \cite{H74} (Section B.II). For members $F_1$ and $F_2$ of ${\cal F}_{n,t_n}$,
\begin{equation}\label{def1}
d_{{\cal F}_{n,t_n}}(F_1,F_2)=\inf\left(\epsilon>0|F_1\subseteq (F_2)^\epsilon\mbox{ and }F_2\subseteq (F_1)^\epsilon\right),
\end{equation}
where the $\epsilon$-cover $F^\epsilon$ of any $F\in {\cal F}_{n,t_n}$ is given by
\begin{equation}\label{fed2}
\left\{(\pi,\pi')\in\Pi_{n,t_n}\times \Pi_{n,t_n}|d_{\Pi_{n,t_n}\times \Pi_{n,t_n}}((\pi,\pi'),(\pi_0,\pi'_0))\leq \epsilon\mbox{ for some }(\pi_0,\pi'_0)\in F\right\}.
\end{equation}

\begin{proposition}\label{kappu}
The space $\Phi_{n,t_n}$ of $(n,t_n)$-preferences is a closed subset of ${\cal F}_{n,t_n}$. 
\end{proposition}

Let $\Phi$ be $\prod_{n\in N}\prod_{t_n\in T_n}\Phi_{n,t_n}$ and ${\cal F}$ be $\prod_{n\in N}\prod_{t_n\in T_n}{\cal F}_{n,t_n}$. On ${\cal F}$, let us define the metric $d_{{\cal F}}$ so that for any members $F\equiv (F_{n,t_n})_{n\in N,t_n\in T_n}$ and $F'\equiv (F'_{n,t_n})_{n\in N,t_n\in T_n}$ of ${\cal F}$,
\begin{equation}
d_{{\cal F}}(F,F')=\max_{n\in N}\max_{t_n\in T_n}d_{{\cal F}_{n,t_n}}(F_{n,t_n},F'_{n,t_n}).
\end{equation}
By Proposition~\ref{kappu}, the space $\Phi$ of players' ambiguity attitude profiles is a closed subset of ${\cal F}$. 

Our game is parameterized by ambiguity attitude profiles $\varphi\equiv(\varphi_{n,t_n})_{n\in N,t_n\in T_n}$ in $\Phi$. To make this dependence explicit, we revise definitions~(\ref{maximal0}) through~(\ref{zia-a}). First,~(\ref{maximal0}) becomes
\begin{equation}\label{maximal0-dep}
\hat A^{\mbox a}_{n,t_n}(\delta_{-n}|\varphi_{n,t_n})=\left\{a_{n,t_n}\in A_{n,t_n}|\tilde\Pi^{\mbox a}_{n,t_n}(\delta_{-n})\times\{\pi^{\mbox a}_{n,t_n}(a_{n,t_n},\delta_{-n})\}\subseteq \varphi_{n,t_n}\right\},
\end{equation}
with $\tilde\Pi^{\mbox a}_{n,t_n}(\delta_{-n})$ defined at~(\ref{pi-p}), which is much like~(\ref{maximal}). Then,~(\ref{b-def}) becomes
\begin{equation}\label{b-def-dep}
\hat B^{\mbox a}_{n,t_n}(\delta_{-n}|\varphi_{n,t_n})=\left\{\delta_{n,t_n}\in \Delta_{n,t_n}|\delta_{n,t_n}(\hat A^{\mbox a}_{n,t_n}(\delta_{-n}|\varphi_{n,t_n}))=1\right\}.
\end{equation}
Next, following~(\ref{zia-p}), we have
\begin{equation}\label{zia-p-dep}
{\cal E}^{\mbox a}(\varphi)=\left\{\delta\equiv (\delta_{n,t_n})_{n\in N,t_n\in T_n}\in\Delta|\delta_{n,t_n}\in \hat B^{\mbox a}_{n,t_n}(\delta_{-n}|\varphi_{n,t_n})\;\;\forall n\in N,t_n\in T_n\right\}.
\end{equation}
Also,~(\ref{maximal-k0}) becomes
\begin{equation}\label{maximal-k0-dep}
\hat B^{\mbox d}_{n,t_n}(\delta_{-n}|\varphi_{n,t_n})=\left\{\delta_{n,t_n}\in \Delta_{n,t_n}|\tilde\Pi^{\mbox d}_{n,t_n}(\delta_{-n})\times\{\pi^{\mbox d}_{n,t_n}(\delta_{n,t_n},\delta_{-n})\}\subseteq \varphi_{n,t_n}\right\},
\end{equation}
with $\tilde\Pi^{\mbox d}_{n,t_n}(\delta_{-n})$ defined at~(\ref{pi-a}), which is much like~(\ref{maximal-k}). Finally, following~(\ref{zia-a}), we have
\begin{equation}\label{zia-a-dep}
{\cal E}^{\mbox d}(\varphi)=\left\{\delta\equiv (\delta_{n,t_n})_{n\in N,t_n\in T_n}\in\Delta|\delta_{n,t_n}\in \hat B^{\mbox d}_{n,t_n}(\delta_{-n}|\varphi_{n,t_n})\;\;\forall n\in N,t_n\in T_n\right\}.
\end{equation}

We have upper hemi-continuity results that are stronger than Propositions~\ref{a-condition} and~\ref{b-continuity}.

\begin{proposition}\label{p-aa}
	At each $n\in N$ and $t_n\in T_n$, the correspondence $\hat A^{\mbox a}_{n,t_n}(\cdot|\cdot)$ defined in~(\ref{maximal0-dep}) from $\Delta_{-n}\times \Phi_{n,t_n}$ to $A_{n,t_n}$ is upper hemi-continuous.
\end{proposition}

\begin{proposition}\label{p-ba}
	At each $n\in N$ and $t_n\in T_n$, the correspondence $\hat B^{\mbox a}_{n,t_n}(\cdot|\cdot)$ defined in~(\ref{b-def-dep}) from $\Delta_{-n}\times \Phi_{n,t_n}$ to $\Delta_{n,t_n}$ is upper hemi-continuous; also, the correspondence $\hat B^{\mbox d}_{n,t_n}(\cdot|\cdot)$ defined in~(\ref{maximal-k0-dep}) from $\Delta_{-n}\times \Phi_{n,t_n}$ to $\Delta_{n,t_n}$ is upper hemi-continuous.
\end{proposition}

These results would lead to the upper hemi-continuity of the equilbrium sets ${\cal E}^{\mbox a}$ and ${\cal E}^{\mbox d}$ with respect to changing ambiguity attitude profiles.

\begin{theorem}\label{t-dep}
The correspondence ${\cal E}^{\mbox a}$ defined in~(\ref{zia-p-dep}) from $\Phi$ to $\Delta$ is upper hemi-continuous; also, the correspondence ${\cal E}^{\mbox d}$ defined in~(\ref{zia-a-dep}) from $\Phi$ to $\Delta$ is upper hemi-continuous.
\end{theorem}

Theorem~\ref{t-dep} dictates that equilibria of both types at a collection of players' ambiguity attitudes would somehow limit the choices that corresponding equilibria at neighboring collections of ambiguity attitudes can have. In particular, both equilibrium sets are closed.

\section{Special Cases}\label{cases}

\subsection{The Satisfaction Version}\label{para}

When the preference relations $\succ_{n,t_n}$ are negatively transitive so that $\pi\not \succ_{n,t_n} \pi'$ and $\pi'\not \succ_{n,t_n}\pi''$ always lead to $\pi\not \succ_{n,t_n}\pi''$, the relations $\not\succ_{n,t_n}$ will become complete pre-orderings. That is, they will be reflexive, transitive, and complete, the latter of which in the sense that either $\pi\not\succ_{n,t_n}\pi'$ or $\pi'\not\succ_{n,t_n}\pi$ for any $\pi,\pi'\in \Pi_{n,t_n}$. Then, it will not take much for order-preserving utility functions over payoff distributions to emerge; see, e.g., Debreu \cite{D64}.

We suppose such is the case, so that each preference relation $\succ_{n,t_n}$ is facilitated by a function $s_{n,t_n}$ through~(\ref{stfn}) for any continuous payoff-distribution vectors $\pi,\pi'\in \Pi_{n,t_n}$. We call these $s_{n,t_n}$'s satisfaction functions; also, let us call the game $\Gamma$, with the $s_{n,t_n}$'s substantiating the $\succ_{n,t_n}$'s, a satisfaction game. Define
\begin{equation}\label{useful-pp}
s^{\mbox a}_{n,t_n}(a_{n,t_n},\delta_{-n})\equiv s_{n,t_n}\left(\pi^{\mbox a}_{n,t_n}(a_{n,t_n},\delta_{-n})\right),
\end{equation}
and
\begin{equation}\label{useful-aa}
s^{\mbox d}_{n,t_n}(\delta_{n,t_n},\delta_{-n})\equiv s_{n,t_n}\left(\pi^{\mbox d}_{n,t_n}(\delta_{n,t_n},\delta_{-n})\right).
\end{equation}
According to~(\ref{stfn}), as well as~(\ref{maximal0}) to~(\ref{zia-p}), $\delta\in\Delta$ will be considered an action-based equilibrium of the satisfaction game if and only if for any player $n\in N$ and type $t_n\in T_n$,
\begin{equation}\label{criterion-p}
\delta_{n,t_n}\left(\{a_{n,t_n}\in A_{n,t_n}|s^{\mbox a}_{n,t_n}(a_{n,t_n},\delta_{-n})\geq s^{\mbox a}_{n,t_n}(a'_{n,t_n},\delta_{-n})\;\;\forall a'_{n,t_n}\in A_{n,t_n}\}\right)=1.
\end{equation}
By~(\ref{stfn}),~(\ref{maximal-k0}), and~(\ref{zia-a}), it will be considered a distribution-based equilibrium of the game if and only if for any player $n\in N$ and type $t_n\in T_n$,
\begin{equation}\label{criterion-a}
s^{\mbox d}_{n,t_n}(\delta_{n,t_n},\delta_{-n})\geq s^{\mbox d}_{n,t_n}(\delta'_{n,t_n},\delta_{-n}),\hspace*{.8in}\forall \delta'_{n,t_n}\in \Delta_{n,t_n}.
\end{equation}

According to~(\ref{criterion-p}), an action-based equilibrium $\delta^{\mbox a}\in{\cal E}^{\mbox a}\subseteq \Delta$ for the satisfaction game $\Gamma$ is one that, at any $n\in N$ and $t_n\in T_n$, the action distribution $\delta^{\mbox a}_{n,t_n}$ has its support built on those actions $a^{\mbox a}_{n,t_n}$ that achieve $\max_{a_{n,t_n}\in A_{n,t_n}}s^{\mbox a}_{n,t_n}(a_{n,t_n},\delta^{\mbox a}_{-n})$. Meanwhile, according to~(\ref{criterion-a}), a distribution-based equilibrium $\delta^{\mbox d}\in{\cal E}^{\mbox d}\subseteq \Delta$ for the game is one that, at any $n\in N$ and $t_n\in T_n$, the action distribution $\delta^{\mbox d}_{n,t_n}$ achieves $\max_{\delta_{n,t_n}\in \Delta_{n,t_n}}s^{\mbox d}_{n,t_n}(\delta_{n,t_n},\delta^{\mbox d}_{-n})$. For equilibrium existence, let us make a continuity-related assumption.

\begin{gr-assumption}\label{cc}
For any player $n\in N$ and any of his types $t_n\in T_n$, the satisfaction function $s_{n,t_n}$ is continuous from $\Pi_{n,t_n}$ to the real line $\Re$. 	
\end{gr-assumption}

This assumption will imply Preference Assumption~\ref{ca-l}.
So for a satisfaction game, existence of action-based equilibria is guaranteed by Theorem~\ref{t-existence}. Also, with~(\ref{stfn}), the $\succ_{n,t_n}$-convexity requirement~(\ref{general-c}) will be equivalent to
\begin{equation}\label{satisfaction-c}\begin{array}{l}
\mbox{ when both }s_{n,t_n}(\pi)\leq s_{n,t_n}(\pi^0)\mbox{ and }s_{n,t_n}(\pi)\leq s_{n,t_n}(\pi^1),\\
\;\;\;\;\;\;\;\;\;\;\;\;\mbox{ it will follow that }s_{n,t_n}(\pi)\leq s_{n,t_n}((1-\alpha)\cdot\pi^0+\alpha\cdot\pi^1)\mbox{ for any }\alpha\in [0,1].
\end{array}\end{equation}
But this exactly expresses the quasi-concavity of $s_{n,t_n}$. Thus, the existence of distribution-based will be guaranteed by Theorem~\ref{t-existence} as well under quasi-concave $s_{n,t_n}$'s.

\begin{corollary}\label{e-exist-satisfaction}
The satisfaction game $\Gamma$ has action-based equilibria. When the satisfaction functions $s_{n,t_n}$ are quasi-concave, the game will have distribution-based equilibria as well. 	
\end{corollary}

The quasi-concavity of the satisfaction function $s_{n,t_n}$ translates into the added benefit bestowed to the $(n,t_n)$-player when two payoff-distribution vectors are mixed up. Thus, Corollary~\ref{e-exist-satisfaction} basically stipulates the nonemptiness of ${\cal E}^{\mbox a}$ for a general satisfaction game and that of ${\cal E}^{\mbox d}$ under ambiguity aversion. The latter message can be used to explain Azrieli and Teper's \cite{AT11} conclusions for their setting. Since $\int u_{n,t_n}\cdot d\pi\equiv (\int u_{n,t_n}\cdot d[\pi(\omega)]|\omega\in\Omega_{n,t_n})$ is linear in $\pi$, the quasi-concavity of $j_{n,t_n}$ used in~(\ref{atat}) will lead to that of its corresponding $s_{n,t_n}$. 

\subsection{The Alarmists' and Enterprising Versions}\label{alarm}

Suppose at each player $n\in N$ and type $t_n\in T_n$, there exist a real-valued utility function $u_{n,t_n}\in {\cal C}(R_{n,t_n},\Re)$ over payoffs and a closed prior set $P_{n,t_n}\subseteq {\cal P}(\Omega_{n,t_n})$ of potential state distributions, so that for any payoff-distribution vector $\pi\in \Pi_{n,t_n}$,~(\ref{s-form}) applies with
\begin{equation}\label{s0-form}
s^0_{n,t_n}(\pi,\rho)=\int_{\Omega_{n,t_n}}\{\int_{R_{n,t_n}}u_{n,t_n}(r)\cdot [\pi(\omega)](dr)\}\cdot \rho(d\omega).
\end{equation}
The above integration can be understood as an expectation of the utility $u_{n,t_n}(R)$ over the random payoff $R$, where the latter is distributed according to $\int_{\Omega_{n,t_n}}\pi(\omega)\cdot\rho(d\omega)$, essentially components $\pi(\omega)$ of the vector $\pi$ mixed over with weights assigned by the state distribution $\rho$. We will call this even more special $\Gamma$ an alarmists' game, because players are effectively on the highest alert to guard against the worst scenario. Under the convexity of $P_{n,t_n}$ and other technical restrictions,~(\ref{s-form}) and~(\ref{s0-form}) were demonstrated by Gilboa and Schmeidler \cite{GS89} to reflect features like certainty independence, continuity, monotonicity, ambiguity aversion, and non-degeneracy to be possessed by the underlying preference relation $\succ_{n,t_n}$. 

The utility function $u_{n,t_n}\in {\cal C}(R_{n,t_n},\Re)$ is not only continuous but also bounded due to the compactness of $R_{n,t_n}$. This along with the continuity of $\pi(\cdot)\in \Pi_{n,t_n}\equiv {\cal C}(\Omega_{n,t_n},{\cal P}(R_{n,t_n}))$ and the nature of the Prokhorov metric will ensure the continuity of  $\int_{R_{n,t_n}}u_{n,t_n}(r)\cdot [\pi(\cdot)](dr)$ as a function from $\Omega_{n,t_n}$ to $\Re$:
\begin{equation}\label{pro-cont}
\lim_{\omega'\rightarrow \omega}\int_{R_{n,t_n}}u_{n,t_n}(r)\cdot [\pi(\omega')](dr)=\int_{R_{n,t_n}}u_{n,t_n}(r)\cdot [\pi(\omega)](dr),
\end{equation}
just because $\lim_{\omega'\rightarrow \omega}\pi(\omega')=\pi(\omega)$. So the outer-layer integration in~(\ref{s0-form}) is well defined.

Similarly, $\int_{R_{n,t_n}}u_{n,t_n}(r)\cdot [\pi(\omega)](dr)$ is continuous in $\pi(\omega)$ at every $\omega\in\Omega_{n,t_n}$. It is also bounded across all these $\omega$'s. Moreover, the uniform metric adopted for $\Pi_{n,t_n}$ means that $\lim_{k\rightarrow +\infty}\pi^k=\pi$ always entails $\lim_{k\rightarrow +\infty}\pi^k(\omega)=\pi(\omega)$ at every $\omega\in\Omega_{n,t_n}$. With bounded convergence, we will then have $s^0_{n,t_n}(\cdot,\rho)$'s continuity as a real-valued function on $\Pi_{n,t_n}$. Since the feasible region $P_{n,t_n}$ in the optimization problem involved in~(\ref{s-form}) is independent of $\pi$, we can immediately deduce the continuity of $s_{n,t_n}$. In addition, $s^0_{n,t_n}(\cdot,\rho)$ is linear in the sense of being both concave and convex. Hence, after taking infimum, $s_{n,t_n}$ will be concave. 
The following then comes immediately from Corollary~\ref{e-exist-satisfaction}.

\begin{corollary}\label{e-exist-prior}
The alarmists' game $\Gamma$ has both action- and distribution-based equilibria. 	
\end{corollary}

Kajii and Ui \cite{KU05} can be understood as studying the alarmists' game with finite action and state spaces. They also listed different ways in which prior sets $P_{n,t_n}\subseteq {\cal P}(\Omega_{n,t_n})$ used in~(\ref{s-form}) could be generated from prior sets $P_n\subseteq {\cal P}(\Omega)$ defined for the entire state space; e.g., through the fashion of Dempster \cite{D67} or the fashion of Fagin and Halpern \cite{FH90}. 

Oppositely, we can consider what we shall call the enterprsing game, where~(\ref{radical-form}) applies. If the alarmists' game reflects aversion to ambiguity on the actual distribution of states of the world $\omega$ within the $\Omega_{n,t_n}$'s, the enterprising game reflects  players' staunch beliefs in the tendency for ambiguities to be resolved in a manner most favorable to them. In other words, the players are really ``enterprising''. The current $s_{n,t_n}$, being linked to $s^0_{n,t_n}$ through~(\ref{radical-form}), is continuous. The following thus applies.

\begin{corollary}\label{e-exist-enter}
The enterprising game $\Gamma$ has action-based equilibria. 	
\end{corollary}

In Appendix~\ref{app-dc}, we will have more to say about this game's pure equilibria, in both action- and distribution-based senses.

\subsection{The Traditional Expected-utility Version}\label{harsanyi}

A special game, which is simultaneously alarmists' and enterprising, is when the prior sets $P_{n,t_n}$ happen to be singletons $\{\rho_{n,t_n}\}$.
Then $\Gamma$'s satisfaction functions will obey
\begin{equation}\label{form1}
s_{n,t_n}(\pi)=s^0_{n,t_n}(\pi,\rho_{n,t_n}),\hspace*{.8in}\forall \pi\in\Pi_{n,t_n},
\end{equation}
with $s^0_{n,t_n}$ given in~(\ref{s0-form}). Note this agrees exactly with~(\ref{ref-tradition}). 
This special case turns out just to be the incomplete-information game as understood in the traditional sense.

Let us suppose that
\begin{equation}\label{bazaj}
p_{n,t_n|t_{-n}}\equiv\int_{\Omega_{t_n,t_{-n}}}\rho_{n,t_n}(d\omega)>0,\hspace*{.8in}\forall n\in N,\;t\equiv(t_n,t_{-n})\in T.
\end{equation}
Since $\Omega_{n,t_n}=\bigcup_{t_{-n}\in T_{-n}}\Omega_{t_n,t_{-n}}$,
\begin{equation}
\sum_{t_{-n}\in T_{-n}}p_{n,t_n|t_{-n}}=\sum_{t_{-n}\in T_{-n}}\int_{\Omega_{t_n,t_{-n}}}\rho_{n,t_n}(d\omega)=\int_{\Omega_{n,t_n}}\rho_{n,t_n}(d\omega)=1,
\end{equation}
with the last equality attributable to $\rho_{n,t_n}\in {\cal P}(\Omega_{n,t_n})$. So $p_{n,t_n}\equiv (p_{n,t_n|t_{-n}})_{t_{-n}\in T_{-n}}$ describes a distribution on opponents' type profiles, wherein every component $p_{n,t_n|t_{-n}}$ is interpretable as player $n$'s estimate on the chance for $t_{-n}$ to occur at his own type $t_n$. Also, let
\begin{equation}\label{vv1}
v_{n,t_n,t_{-n}}(a_{n,t_n},a_{-n,t_{-n}})\equiv \int_{\Omega_{t_n,t_{-n}}}u_{n,t_n}\left(r_{n,t_n,t_{-n}}(a_{n,t_n},a_{-n,t_{-n}},\omega)\right)\cdot\nu_{n,t_n,t_{-n}}(d\omega),
\end{equation}
where
\begin{equation}\label{vv2}
\nu_{n,t_n,t_{-n}}\equiv\frac{1}{p_{n,t_n|t_{-n}}}\cdot\rho_{n,t_n}|_{\Omega_{t_n,t_{-n}}}\in {\cal P}(\Omega_{t_n,t_{-n}}).
\end{equation}
Here, $\rho_{n,t_n}|_{\Omega_{t_n,t_{-n}}}$ just means the measure $\rho_{n,t_n}$ on $\Omega_{n,t_n}$ being confined to the subset $\Omega_{t_n,t_{-n}}$. 
We can treat each term $v_{n,t_n,t_{-n}}(a_{n,t_n},a_{-n,t_{-n}})$ as player $n$'s von Neumann-Morgenstern utility when his own type is $t_n$, his opponents' type profile is $t_{-n}$, he takes action $a_{n,t_n}$, and his opponents collectively adopt action profile $a_{-n,t_{-n}}$.

By plugging~(\ref{mechanism}) to~(\ref{vec-p}) into~(\ref{s0-form}), we see that $s^0_{n,t_n}(\pi^{\mbox a}_{n,t_n}(a_{n,t_n},\delta_{-n}),\rho_{n,t_n})$ equals
\begin{equation}\label{useful-ppp}\begin{array}{l}
\sum_{t_{-n}\in T_{-n}}\int_{\Omega_{t_n,t_{-n}}}\{\int_{R_{n,t_n}} u_{n,t_n}(r)\cdot[\pi^{\mbox a}_{n,t_n,t_{-n}}(a_{n,t_n},\delta_{-n,t_{-n}},\omega)](dr)\}
\cdot\rho_{n,t_n}(d\omega)\\
\;\;\;=\sum_{t_{-n}\in T_{-n}}\int_{\Omega_{t_n,t_{-n}}}\rho_{n,t_n}(d\omega)\times\\
\;\;\;\;\;\;\;\;\;\;\;\;\;\;\;\;\;\;\;\;\;\;\;\;\times\{\int_{R_{n,t_n}} u_{n,t_n}(r)\cdot[(\prod_{m\neq n}\delta_{m,t_m})\cdot (r_{n,t_n,t_{-n}}(a_{n,t_n},\cdot,\omega))^{-1}](dr)\},
\end{array}\end{equation}
which, after a change of variables, an exchange of integration orders, and the use of entities defined in~(\ref{bazaj}) through~(\ref{vv2}), would become
\begin{equation}\label{useful-p}
\sum_{t_{-n}\in T_{-n}}p_{n,t_n|t_{-n}}\cdot\int_{A_{-n,t_{-n}}}v_{n,t_n,t_{-n}}(a_{n,t_n},a_{-n,t_{-n}})\cdot \prod_{m\neq n}\delta_{m,t_m}(da_{m,t_m}).
\end{equation}
Similarly, by plugging~(\ref{oklaa}) to~(\ref{vec-a}) into~(\ref{s0-form}) and~(\ref{useful-p}), we can get
\begin{equation}\label{useful-a}
s^0_{n,t_n}\left(\pi^{\mbox d}_{n,t_n}(\delta_{n,t_n},\delta_{-n}),\rho_{n,t_n}\right)=\int_{A_{n,t_n}}s^0_{n,t_n}\left(\pi^{\mbox a}_{n,t_n}(a_{n,t_n},\delta_{-n}),\rho_{n,t_n}\right)
\cdot\delta_{n,t_n}(da_{n,t_n}).
\end{equation}

From~(\ref{useful-ppp}) to~(\ref{useful-a}), we see that among $\Gamma$'s primitives listed at the end of Section~\ref{no-need}, $\Omega$, $(\Omega_{n,t_n})_{n\in N,t_n\in T_n}$, $(R_{n,t_n})_{n\in N,t_n\in T_n}$, $(r_{n,t})_{n\in N,t\in T}$, and $(\succ_{n,t_n})_{n\in N,t_n\in T_n}$ are all not needed. In their stead, we can add the probabilities $p_{n,t_n|t_{-n}}$ and the real-valued payoffs $v_{n,t_n,t_{-n}}(a_{n,t_n}$, $a_{-n,t_{-n}})$'s. These descriptions fit the definition of a traditional game with incomplete information, albeit one without necessarily involving a common prior. Since the current one is a special alarmists' game, we know from Corollary~\ref{e-exist-prior} that both action- and distribution-based equilibria are in existence. In Section~\ref{relation}, we shall see that the linearity of $s^0_{n,t_n}(\cdot,\rho_{n,t_n})$ will lead the two types of equilibria to exactly overlap. 


\section{Relations between Equilibrium Versions}\label{relation}

\subsection{Useful Definitions}\label{some-def}

We now come to relationships between ${\cal E}^{\mbox a}$ and ${\cal E}^{\mbox d}$. Some of the conditions concerning the general preference game might be difficult to check. However, they lead to a clear message about the satisfaction game and ultimately, to the identity of the two equilibrium sets for the traditional game. Given metric spaces $X$, $Y$, and $Z$, let ${\cal K}(X,Y,Z)$ be the space of continuous kernels, so that every $\kappa\equiv (\kappa(x)|x\in X)\equiv (\kappa(x,y)|x\in X,y\in Y)\in {\cal K}(X,Y,Z)$ is a mapping from $X\times Y$ to ${\cal P}(Z)$ that also satisfies the following:\\
\indent\M (a) $[\kappa(\cdot,y)](Z')$ is Borel-measurable from $X$ to $[0,1]$ at every $y\in Y$ and $Z'\in {\cal B}(Z)$;\\
\indent\M (b) $\kappa$ is continuous from $X\times Y$ to ${\cal P}(Z)$.\\
Given $\delta\in {\cal P}(X)$, we define the integration
$\iota=\int_X \kappa(x)\cdot \delta(dx)$ in the component-wise fashion, so that $\iota\equiv (\iota(y)|y\in Y)$ and $\iota(y)=\int_X \kappa(x,y)\cdot \delta(dx)$ at every $y\in Y$; each of the latter integrations, in turn, is facilitated by
\begin{equation}\label{fitts}
[\iota(y)](Z')=\int_X [\kappa(x,y)](Z')\cdot \delta(dx),\hspace*{.8in}\forall Z'\in {\cal B}(Z).
\end{equation}
Due to (a), the above integration can be carried out. Using (a) and part of (b), we can establish $\iota$'s membership in ${\cal C}(Y,{\cal P}(Z))$.

\begin{lemma}\label{meaningful}
Given $\kappa\in {\cal K}(X,Y,Z)$ and $\delta\in {\cal P}(X)$, their integration $\iota$, whose definition ultimately relies on~(\ref{fitts}), is a member of ${\cal C}(Y,{\cal P}(Z))$.
\end{lemma}

At each $(n,t_n)$-pair, it will help to cast the action space $A_{n,t_n}$ as $X$, state space $\Omega_{n,t_n}$ as $Y$, and payoff space $R_{n,t_n}$ as $Z$. We now show that payoffs attained in the action-based sense are linked to continuous kernels. Given $\delta_{-n}\in\Delta_{-n}$, with each $\pi^{\mbox a}_{n,t_n}(a_{n,t_n},\delta_{-n})$ being patched up in the manner of~(\ref{vec-p}), we can understand the vector $\pi^{\mbox a}_{n,t_n}(\delta_{-n})\equiv (\pi^{\mbox a}_{n,t_n}(a_{n,t_n},\delta_{-n})|a_{n,t_n}\in A_{n,t_n})\equiv (\pi^{\mbox a}_{n,t_n,t_{-n}}(a_{n,t_n},\delta_{-n,t_{-n}},\omega)|a_{n,t_n}\in A_{n,t_n},t_{-n}\in T_{-n},\omega\in \Omega_{t_n,t_{-n}})$ as a mapping from $A_{n,t_n}\times \Omega_{n,t_n}$ to ${\cal P}(R_{n,t_n})$. It turns out to be a continuous kernel.

\begin{proposition}\label{beling}
At any player $n\in N$, any of his types $t_n\in T_n$, and any of his opponents' behavioral-strategy profiles $\delta_{-n}\in \Delta_{-n}$, we have $\pi^{\mbox a}_{n,t_n}(\delta_{-n})\in {\cal K}(A_{n,t_n},\Omega_{n,t_n},R_{n,t_n})$.
\end{proposition}

Comparing~(\ref{oklac}) with~(\ref{fitts}), we can have another understanding of the vector $\pi^{\mbox d}_{n,t_n}(\delta_{n,t_n},\delta_{-n})$ defined earlier at~(\ref{vec-a}). Using the current integration of continuous kernels, 
\begin{equation}\label{ahaha}
\pi^{\mbox d}_{n,t_n}(\delta_{n,t_n},\delta_{-n})=\int_{A_{n,t_n}}\pi^{\mbox a}_{n,t_n}(a_{n,t_n},\delta_{-n})\cdot\delta_{n,t_n}(da_{n,t_n}).
\end{equation}
Lemma~\ref{meaningful} would also predict the vector's membership in $\Pi_{n,t_n}\equiv{\cal C}(\Omega_{n,t_n},R_{n,t_n})$.

For any preference relation $\varphi_{n,t_n}\in\Phi_{n,t_n}$, let ${\cal W}^{\mbox a}_{n,t_n}(\varphi_{n,t_n})$ be the set that contains all the $(\kappa_{n,t_n},\delta_{n,t_n})$-pairs so that $\kappa_{n,t_n}\equiv(\kappa_{n,t_n}(a_{n,t_n})|a_{n,t_n}\in A_{n,t_n})$ is a continuous kernel in ${\cal K}(A_{n,t_n},\Omega_{n,t_n},R_{n,t_n})$, $\delta_{n,t_n}$ is an action distribution in $\Delta_{n,t_n}$, and the two satisfy
\begin{equation}\label{muck1}
\delta_{n,t_n}\left(\{a_{n,t_n}\in A_{n,t_n}|(\kappa_{n,t_n}(a'_{n,t_n}),\kappa_{n,t_n}(a_{n,t_n}))\in\varphi_{n,t_n}\;\;\forall a'_{n,t_n}\in A_{n,t_n}\}\right)=1,
\end{equation}
where the set being assessed by $\delta_{n,t_n}$ is necessarily a member of ${\cal B}(A_{n,t_n})$. Basically, any $(\kappa_{n,t_n},\delta_{n,t_n})\in {\cal W}^{\mbox a}_{n,t_n}(\varphi_{n,t_n})$ is such that the payoff-distribution vector $\kappa_{n,t_n}(a_{n,t_n})$ achieves $\varphi_{n,t_n}$-maximality among all $\kappa_{n,t_n}(a'_{n,t_n})$'s for $\delta_{n,t_n}$-almost every $a_{n,t_n}$.

Also, let ${\cal W}^{\mbox d}_{n,t_n}(\varphi_{n,t_n})$ be the set that contains all the $(\kappa_{n,t_n},\delta_{n,t_n})$-pairs so that $\kappa_{n,t_n}$ is a continuous kernel in ${\cal K}(A_{n,t_n},\Omega_{n,t_n},R_{n,t_n})$, $\delta_{n,t_n}$ is an action distribution in $\Delta_{n,t_n}$, and
\begin{equation}\label{muck3}
\left(\int_{A_{n,t_n}}\kappa_{n,t_n}(a_{n,t_n})\cdot\delta'_{n,t_n}(da_{n,t_n}),\int_{A_{n,t_n}}\kappa_{n,t_n}(a_{n,t_n})\cdot\delta_{n,t_n}(da_{n,t_n})\right)\in\varphi_{n,t_n},
\end{equation}
for any $\delta'_{n,t_n}\in \Delta_{n,t_n}$. By Lemma~\ref{meaningful}, the integrals are members of $\Pi_{n,t_n}$.  Basically, any $(\kappa_{n,t_n},\delta_{n,t_n})\in {\cal W}^{\mbox d}_{n,t_n}(\varphi_{n,t_n})$ is such that the integrated payoff-distribution vector $\int_{A_{n,t_n}}\kappa_{n,t_n}(a_{n,t_n})\cdot\delta_{n,t_n}(da_{n,t_n})$ achieves $\varphi_{n,t_n}$-maximality among all the  $\int_{A_{n,t_n}}\kappa_{n,t_n}(a_{n,t_n})\cdot \delta'_{n,t_n}(da_{n,t_n})$'s when $\delta'_{n,t_n}$ traverses through the entire action-distribution space $\Delta_{n,t_n}$.

Comparing~(\ref{maximal0-dep}) and~(\ref{b-def-dep}) with~(\ref{muck1}), we can understand $\delta_{n,t_n}\in \hat B^{\mbox a}_{n,t_n}(\delta_{-n}|\varphi_{n,t_n})$ as
\begin{equation}\label{eq-p}
\left(\left(\pi^{\mbox a}_{n,t_n}(a_{n,t_n},\delta_{-n})|a_{n,t_n}\in A_{n,t_n}\right),\delta_{n,t_n}\right)\in {\cal W}^{\mbox a}_{n,t_n}(\varphi_{n,t_n}).
\end{equation}
Comparing~(\ref{maximal-k0-dep}) with~(\ref{ahaha}) and~(\ref{muck3}), we can understand $\delta_{n,t_n}\in \hat B^{\mbox d}_{n,t_n}(\delta_{-n}|\varphi_{n,t_n})$ as
\begin{equation}\label{eq-a}
\left(\left(\pi^{\mbox a}_{n,t_n}(a_{n,t_n},\delta_{-n})|a_{n,t_n}\in A_{n,t_n}\right),\delta_{n,t_n}\right)\in {\cal W}^{\mbox d}_{n,t_n}(\varphi_{n,t_n}).
\end{equation}

\subsection{Behavioral Equilibria in General}\label{more-stuff}

When ${\cal W}^{\mbox d}_{n,t_n}(\varphi_{n,t_n})\subseteq {\cal W}^{\mbox a}_{n,t_n}(\varphi_{n,t_n})$, we say that preference relation $\varphi_{n,t_n}\in\Phi_{n,t_n}$ is individually prominent with respect to $A_{n,t_n}$. As can be seen from~(\ref{muck1}) and~(\ref{muck3}), this property entails that $\kappa_{n,t_n}(a_{n,t_n})$ being not $\varphi_{n,t_n}$-maximal among all the $\kappa_{n,t_n}(a'_{n,t_n})$ for a $\delta_{n,t_n}$-positive set of $a_{n,t_n}$'s would lead to some action distribution $\delta'_{n,t_n}$ for $\int_{A_{n,t_n}}\kappa_{n,t_n}(a)\cdot\delta'_{n,t_n}(da)$ to be strictly more preferable than  $\int_{A_{n,t_n}}\kappa_{n,t_n}(a)\cdot\delta_{n,t_n}(da)$. 
It will result in the following.

\begin{proposition}\label{a-in-p}
For the preference game $\Gamma$, we will have ${\cal E}^{\mbox d}\subseteq {\cal E}^{\mbox a}$ when every preference relation $\varphi_{n,t_n}\in \Phi_{n,t_n}$ is individually prominent with respect to $A_{n,t_n}$.
\end{proposition}

Oppositely, when ${\cal W}^{\mbox a}_{n,t_n}(\varphi_{n,t_n})\subseteq {\cal W}^{\mbox d}_{n,t_n}(\varphi_{n,t_n})$, we say that preference relation $\varphi_{n,t_n}\in \Phi_{n,t_n}$ is mixture-preserving with respect to $A_{n,t_n}$. By~(\ref{muck1}) and~(\ref{muck3}), this is when $\kappa_{n,t_n}(a_{n,t_n})$ being $\varphi_{n,t_n}$-maximal among all the $\kappa_{n,t_n}(a'_{n,t_n})$'s for $\delta_{n,t_n}$-almost every $a_{n,t_n}$ would lead to   $\int_{A_{n,t_n}}\kappa_{n,t_n}(a)\cdot\delta_{n,t_n}(da)$ being $\varphi_{n,t_n}$-maximal among all the $\int_{A_{n,t_n}}\kappa_{n,t_n}(a)\cdot\delta'_{n,t_n}(da)$'s. The property plays a decisive role in the other direction of equilibrium-set inclusion.

\begin{proposition}\label{p-in-a}
For the preference game $\Gamma$, we will have ${\cal E}^{\mbox a}\subseteq {\cal E}^{\mbox d}$ when every preference relation $\varphi_{n,t_n}\in\Phi_{n,t_n}$ is mixture-preserving with respect to $A_{n,t_n}$.
\end{proposition}

When the relations $\varphi_{n,t_n}$ are facilitated by satisfaction functions $s_{n,t_n}$, we will show that individual prominence is linked to ambiguity seeking. Meanwhile, mixture preservation seems more stringent as so far its guarantors involve both ambiguity aversion and seeking. In this backdrop, Propositions~\ref{a-in-p} and~\ref{p-in-a} may be found to be consistent with the notion that distribution-based equilibria are ``rarer'' than their action-based counterparts.

Due to (b), any continuous kernel $\kappa_{n,t_n}\in {\cal K}(A_{n,t_n},\Omega_{n,t_n},R_{n,t_n})$ can be viewed as a continuous mapping from $A_{n,t_n}$ to $\Pi_{n,t_n}\equiv {\cal C}(\Omega_{n,t_n},{\cal P}(R_{n,t_n}))$. Now consider satisfaction function $s_{n,t_n}$ defined on $\Pi_{n,t_n}$ that meets Satisfaction Assumption~\ref{cc}, i.e., the continuity of $s_{n,t_n}$ as a function from $\Pi_{n,t_n}$ to $\Re$. Then, $s_{n,t_n}(\kappa_{n,t_n}(\cdot))$ is a continuous and hence measurable mapping from $A_{n,t_n}$ to $\Re$. We say such $s_{n,t_n}$ strongly concave with respect to $A_{n,t_n}$ when for any continuous kernel $\kappa_{n,t_n}\in {\cal K}(A_{n,t_n},\Omega_{n,t_n},R_{n,t_n})$ and action  distribution $\delta_{n,t_n}\in \Delta_{n,t_n}$,
\begin{equation}\label{s-concave}
s_{n,t_n}\left(\int_{A_{n,t_n}}\kappa_{n,t_n}(a_{n,t_n})\cdot\delta_{n,t_n}(da_{n,t_n})\right)\geq \int_{A_{n,t_n}}s_{n,t_n}(\kappa_{n,t_n}(a_{n,t_n}))\cdot \delta_{n,t_n}(da_{n,t_n}).
\end{equation}
We say $s_{n,t_n}$ strongly convex with respect to $A_{n,t_n}$ when the inequality opposite to~(\ref{s-concave}) is always true. We say $s_{n,t_n}$ strongly linear with respect to $A_{n,t_n}$ when it is both strongly concave and convex with respect to $A_{n,t_n}$. Here come the links between $s_{n,t_n}$'s strong properties and earlier notions about the $s_{n,t_n}$-based preference relation $\varphi_{n,t_n}$.

\begin{proposition}\label{mip-ko}
With respect to the same $A_{n,t_n}$, any preference relation $\varphi_{n,t_n}\in\Phi_{n,t_n}$ that is based on a strongly convex satisfaction function $s_{n,t_n}$ will be individually prominent. 
\end{proposition}

\begin{proposition}\label{mip-ok}
With respect to the same $A_{n,t_n}$, any preference relation $\varphi_{n,t_n}\in\Phi_{n,t_n}$ that is based on a strongly linear satisfaction function $s_{n,t_n}$ will be both individually prominent and mixture-preserving.
\end{proposition}

So far, we have not found any intermediate result which guarantees mixture preservation without ensuring individual prominence. On the other hand, strong concavity/convexity of $s_{n,t_n}$ with respect to $A_{n,t_n}$ is certainly stronger than $s_{n,t_n}$'s ordinary concavity/convexity except when $A_{n,t_n}$ is a singleton, at which time the strong properties reduce to truisms. The converse is actually true when the payoff space $R_{n,t_n}$ is finite.

\begin{proposition}\label{heroic}
For any player $n\in N$ and any of his types $t_n\in T_n$, suppose the payoff space $R_{n,t_n}$ is finite; also, suppose $s_{n,t_n}$ is a satisfaction function over $\Pi_{n,t_n}$ that meets Satisfaction Assumption~\ref{cc}. Then, its ordinary concavity/convexity will lead to its strong concavity/convexity.
\end{proposition}

Confine to the case where the payoff spaces $R_{n,t_n}$ are finite. Combining Propositions~\ref{a-in-p},~\ref{mip-ko}, and~\ref{heroic}, we see that ${\cal E}^{\mbox d}\subseteq{\cal E}^{\mbox a}$ will happen when the satisfaction functions are convex; whereas, combining everything from Propositions~\ref{a-in-p},~\ref{p-in-a},~\ref{mip-ok}, and~\ref{heroic}, we see that both the previous and ${\cal E}^{\mbox a}\subseteq{\cal E}^{\mbox d}$ will happen when satisfaction functions are linear. We can now reach the following for the satisfaction game introduced in Section~\ref{para}.

\begin{theorem}\label{ep-satisfaction}
For the satisfaction game $\Gamma$ with finite payoff spaces $R_{n,t_n}$, we have ${\cal E}^{\mbox d}\subseteq {\cal E}^{\mbox a}\neq\emptyset$ when the satisfaction functions $s_{n,t_n}$ are convex; furthermore, we have ${\cal E}^{\mbox d}={\cal E}^{\mbox a}\neq\emptyset$ when the functions are linear.
\end{theorem}

Note that ${\cal E}^{\mbox a}\neq\emptyset$ is attributable to Corollary~\ref{e-exist-satisfaction}. Now the message is clear for the satisfaction game. Any distribution-based equilibrium will be an action-based one when all players are ambiguity-seeking. When viewing Theorem~\ref{ep-satisfaction} in conjunction with Corollary~\ref{e-exist-satisfaction}, we may speculate that distribution-based equilibria are ``easier'' to come by when players are ``more'' ambiguity-averse. The theorem also leads to more understanding on the enterprising game studied in Section~\ref{alarm}.

\begin{corollary}\label{ep-enterprising}
For the enterprising game $\Gamma$ with finite payoff spaces, ${\cal E}^{\mbox d}\subseteq {\cal E}^{\mbox a}\neq\emptyset$. 	 
\end{corollary}

Although the nonemptiness of ${\cal E}^{\mbox a}$ is guaranteed, 
so far that for the enterprising game's ${\cal E}^{\mbox d}$ remains unclear. On the other hand, as shall be clear in Appendix~\ref{app-dc}, ambiguity seeking will not prevent distribution-based equilibria from emerging. For the traditional game, we can obtain a general result on equilibrium equivalence by appealing to Propositions~\ref{a-in-p} to~\ref{mip-ok}. 


\begin{theorem}\label{ep-ea}
In the traditional game $\Gamma$, all the satisfaction functions $s_{n,t_n}$ are strongly linear with respect to the corresponding action spaces $A_{n,t_n}$. Consequently, ${\cal E}^{\mbox d}={\cal E}^{\mbox a}\neq\emptyset$. 	
\end{theorem}

Theorem~\ref{ep-ea} offers the justification on why traditionally, one does not have to worry much about how behavioral equilibria are interpreted and enforced.

\subsection{Pure Equilibria in Particular}\label{e-for-p}

When focusing on pure equilibria where players do not use chance outcomes in their own strategies, we can reach simple conclusions without resorting to finite payoff spaces. When action distribution $\delta_{n,t_n}\in \Delta_{n,t_n}$ happens to be the Dirac measure $1_{a_{n,t_n}}$ concentrating on one pure action $a_{n,t_n}\in A_{n,t_n}$, we have from~(\ref{oklab}) and~(\ref{oklac}) that
\begin{equation}
\pi^{\mbox d}_{n,t_n,t_{-n}}(1_{a_{n,t_n}},\delta_{-n,t_{-n}},\omega)=\pi^{\mbox a}_{n,t_n,t_{-n}}(a_{n,t_n},\delta_{-n,t_{-n}},\omega).
\end{equation}
So in view of~(\ref{vec-p}) and~(\ref{vec-a}),
\begin{equation}\label{jit}
\pi^{\mbox d}_{n,t_n}(1_{a_{n,t_n}},\delta_{-n})=\pi^{\mbox a}_{n,t_n}(a_{n,t_n},\delta_{-n}).
\end{equation}
This simple observation would render it necessary that a pure distribution-based equilibrium is also a pure action-based equilibrium. Let $1_{A_{n,t_n}}\equiv\{1_{a_{n,t_n}}|a_{n,t_n}\in A_{n,t_n}\}$, $1_{A_n}\equiv \prod_{t_n\in T_n}1_{A_{n,t_n}}$, and $1_A\equiv\prod_{n\in N}1_{A_n}$. The last is the space of all pure strategy profiles. 

\begin{theorem}\label{pure-pa}
For the preference game $\Gamma$, we have $1_A\cap {\cal E}^{\mbox d}\subseteq 1_A \cap {\cal E}^{\mbox a}$. 	
\end{theorem}




Let us revisit the enterprising game defined in Section~\ref{alarm}. Recall the definition of $s^{\mbox a}_{n,t_n}$ at~(\ref{useful-pp}) and that of $s^{\mbox d}_{n,t_n}$ at~(\ref{useful-aa}).

\begin{proposition}\label{much-ep}
It is true that $s^{\mbox a}_{n,t_n}(a_{n,t_n},\delta_{-n})=s^{\mbox d}_{n,t_n}(1_{a_{n,t_n}},\delta_{-n})$.
\end{proposition}

When the action spaces $A_{n,t_n}$ are finite, the behavioral-strategy spaces $\Delta_{n,t_n}$ are simplices embedded in $\Re^{A_{n,t_n}}$. Then, pure strategies in $1_{A_{n,t_n}}$ constitute all extreme points of $\Delta_{n,t_n}$. As the supremums of convex functions over convex sets come from extreme points,
\begin{equation}\label{convex-con}
\sup_{\delta'_{n,t_n}\in 1_{A_{n,t_n}}}s^{\mbox d}_{n,t_n}(\delta'_{n,t_n},\delta_{-n})=\sup_{\delta'_{n,t_n}\in \Delta_{n,t_n}}s^{\mbox d}_{n,t_n}(\delta'_{n,t_n},\delta_{-n}).
\end{equation}
For a more general case, we resort to the following result.

\begin{lemma}\label{l-okla}
Suppose $X$ is a compact subset of a finite-dimensional real Euclidean space and $f$ is a continuous and convex real-valued function defined on ${\cal P}(X)$. Then, $\sup_{\xi\in {\cal P}(X)}f(\xi)$ can be achieved at the Dirac measure $1_x$ for some $x\in X$.
\end{lemma}	

In the enterprising game $\Gamma$, we now suppose that all action spaces $A_{n,t_n}$ are compact subsets of finite-dimensional real Euclidean spaces. Note that $s^{\mbox d}_{n,t_n}(\cdot,\delta_{-n})$ is not only convex but also continuous. For the latter, just follow the continuity of $\pi^{\mbox d}_{n,t_n}(\cdot,\delta_{-n})$ as stated in Proposition~\ref{piphi-cont}, $s^{\mbox d}_{n,t_n}(\cdot,\delta_{-n})$'s definition at~(\ref{useful-aa}), and the continuity of the $s_{n,t_n}$ defined at~(\ref{radical-form}) that was covered right before Corollary~\ref{e-exist-enter}. By identifying the function $s^{\mbox d}_{n,t_n}(\cdot,\delta_{-n})$ with $f$ in Lemma~\ref{l-okla} and the set $A_{n,t_n}$ with $X$ in the lemma, we can again reach~(\ref{convex-con}). This and the earlier Proposition~\ref{much-ep} turn out to be pivotal for the opposite of either Corollary~\ref{ep-enterprising} or Theorem~\ref{pure-pa}. We can obtain the following when all these are combined.

\begin{theorem}\label{chedan}
For the enterprising game $\Gamma$, we have $1_A\cap {\cal E}^{\mbox d}=1_A \cap {\cal E}^{\mbox a}$. 	
\end{theorem}

According to Theorem~\ref{chedan}, there will be a unified set $1_{\cal E}\equiv 1_A\cap {\cal E}^{\mbox d}=1_A\cap {\cal E}^{\mbox a}$ of pure equilibria for the enterprising game when action spaces are mildly regulated.
For a special case which involves strategic complementarities to be detailed in Appendix~\ref{app-dc}, not only is $1_{\cal E}$ nonempty, but we can also identify from it those well-behaving ones.

\section{Auctions with Ambiguities on Item's Worth}\label{application}

\subsection{A Framework for Auctions}\label{g-auction}

Let us treat a single-item auction. Here, players are bidders and they may only receive crude signals about the actual worths to them and others of the item being auctioned. For each player $n\in N\equiv \{1,...,\bar n\}$, let type space $T_n\equiv\{1,...,\bar t_n\}$ denote the set of signals that player $n$ can receive. A compact metric space $\Omega$, decomposable into $(\Omega_{n,t_n})_{t_n\in T_n}$ for any $n\in N$ and into $(\Omega_t)_{t\in T}$ with each $\Omega_t$ equal to $\bigcap_{n\in N}\Omega_{n,t_n}$, holds states of the world. To an $(n,t_n)$-bidder, $\omega$ can come from anywhere in $\Omega_{n,t_n}$ prior to the auction. Players' ambiguity attitudes are expressible by preference relations $\succ_{n,t_n}$, which may or may not be substantiated by satisfaction functions $s_{n,t_n}$, which may or may not be substantiated by utility functions $u_{n,t_n}$ and prior sets $P_{n,t_n}$ in ``$\inf$''- or ``$\sup$''-fashions, and so on and so forth.

In our framework, what set an auction apart from any other incomplete-information game are its action spaces $A_{n,t_n}$ and payoff functions $r_{n,t_n,t_{-n}}$. For each $(n,t_n)$-bidder, let real compact interval $A_{n,t_n}\equiv [\underline a_{n,t_n},\overline a_{n,t_n}]$ hold prices that the bidder can offer. For any type profile $t\equiv (t_n)_{n\in N}\in T\equiv \prod_{n\in N}T_n$, action profile $a_t\equiv (a_{n,t_n})_{n\in N}\in \prod_{n\in N}A_{n,t_n}$, and potential state of the world  $\omega\in\Omega_t$, let $\iota_{n,t_n,t_{-n}}(a_{n,t_n},a_{-n,t_{-n}},\omega)$ be the 0-1 indicator on whether or not player $n$ is the one who wins the item. Since at most one bidder can win out of any $(t,a_t,\omega)$-combination, we require that
\begin{equation}\label{shemo}
\sum_{n\in N}\iota_{n,t_n,t_{-n}}(a_{n,t_n},a_{-n,t_{-n}},\omega)=0\mbox{ or }1.
\end{equation}
We also use  $\tau_{n,t_n,t_{-n}}(a_{n,t_n},a_{-n,t_{-n}},\omega)$ for the $(n,t_n)$-bidder's payment to the auctioneer. When losers have to pay as well, it can be nonzero even when $\iota_{n,t_n,t_{-n}}(a_{n,t_n},a_{-n,t_{-n}},\omega)=0$.
In addition, let $\upsilon_{n,t_n,t_{-n}}(\omega)$ be the actual worth of the item to the bidder when the external factor turns out to be $\omega$. 
The payoff $r_{n,t_n,t_{-n}}(a_{n,t_n},a_{-n,t_{-n}},\omega)$ equals
\begin{equation}\label{ex-u}
\iota_{n,t_n,t_{-n}}(a_{n,t_n},a_{-n,t_{-n}},\omega)\cdot \upsilon_{n,t_n,t_{-n}}(\omega)-\tau_{n,t_n,t_{-n}}(a_{n,t_n},a_{-n,t_{-n}},\omega).
\end{equation}
This reflects that a bidder will earn the difference between the item's actual worth and his payment when he wins the bid, and otherwise he will still make a payment.

Auction models of both Lo \cite{L98} and Bose, Ozdenoren, and Pape \cite{BOP06} incorporated ambiguity. However, bidders there are fully aware of the item's true worths to themselves, which also happen to be independent of other bidders' value assessments. The current framework allows for more flexibilities. First, bidders' ambiguity attitudes can be expressed by multiple priors along with~(\ref{s-form}) or~(\ref{radical-form}), or more general satisfaction functions, or even mere preference relations. Second, types may convey just what bidders can receive prior to bidding, while the state space $\Omega$ can contain more information on factors that influence the item's eventual worth to the winner. Third, the functions $\iota_{n,t_n,t_{-n}}$, $\nu_{n,t_n,t_{-n}}$, and $\tau_{n,t_n,t_{-n}}$ can take various forms to suit particular modeling needs.

For the most general case, we can use Theorem~\ref{t-existence} to predict that action-based equilibria, likely of the mixed nature, would exist. If bidders are alarmists, we can also predict the existence of distribution-based equilibria through Corollary~\ref{e-exist-satisfaction}. Under more auction-specific assumptions, however, we may hope to obtain results regarding the existence of pure equilibria of either type, their behaviors with respect to changing bidder types, and comparative statics properties of theirs when model primitives alter. Bidders in an auction are likely to be enterprising. If such is the case, then according to Theorem~\ref{chedan}, we would not even have to distinguish between the two types of equilibria when they are pure. 

\subsection{A Concrete Example}

For a concrete example, consider a case where all type spaces $T_n$ are the same $\{1,...,\bar t\}$. Also, regardless of the bidder identity $n$ and type $t_n$, the action space is the same real interval $A_{n,t_n}\equiv [0,\bar w]$. Here, $\bar w$ serves as the highest possible worth of the item. The state space $\Omega$ is $\Psi_{\bar n}\times \{1,...,\bar t\}^{\bar n}\times [0,\bar w]^{\bar n}$, where $\Psi_{\bar n}$ contains all the $\bar n!$ permutations $\psi\equiv (\psi_n)_{n=1,...,\bar n}$ of the numbers $1,...,\bar n$. Every element $\omega\in\Omega$ can be understood as the tuple $(\psi,t,w)\equiv ((\psi_n)_{n=1,...,\bar n},(t_n)_{n=1,...,\bar n},(w_n)_{n=1,...,\bar n})$, where $\psi\equiv (\psi_n)_{n=1,...,\bar n}$ is potentially useful during the auction for tie-breaking purposes and to each bidder $n$, $t_n$ is the signal he receives before the auction and $w_n$ is the true worth of the item to him. For each type profile $t\equiv (t_n)_{n=1,...,\bar n}\in \{1,...,\bar t\}^{\bar n}$, the $\Omega_t$ is merely $\Psi_{\bar n}\times\{t\}\times [0,\bar w]^{\bar n}$. This is the space of all states of the world $\omega$ when bidders' private types are known to form $t$. When bidder $n$ only knows his own type $t_n$, the state $\omega$ would come from $\Omega_{n,t_n}=\bigcup_{t_{-n}\in \{1,...,\bar t\}^{\bar n-1}}\Omega_{t_n,t_{-n}}=\{t_n\}\times\left(\Psi_{\bar n}\times \{1,...,\bar t\}^{\bar n-1}\times[0,\bar w]^{\bar n}\right)$. So bidder $n$ could have ambiguity on the way in which ties are broken, other bidders' types, and true valuations of all bidders including his own.

We let $\iota_{n,t_n,t_{-n}}(a_{n,t_n},a_{-n,t_{-n}},\omega)=1$ if $a_{n,t_n}>a_{m,t_m}$ for any $m\neq n$ or if $a_{n,t_n}\geq a_{m,t_m}$ for any $m\neq n$ while $\omega=(\psi,t,w)$ is such that $\psi_n\leq \psi_m-1$ whenever $a_{m,t_m}=a_{n,t_n}$. This satisfies~(\ref{shemo}) and reflects that the item goes to the highest bidder with ties broken equitably. Since $w_n$ reflects the item's true worth, we have $\upsilon_{n,t_n,t_{-n}}(\psi,t,w)=w_n$. If the auction is first-price, we will have  $\tau_{n,t_n,t_{-n}}(a_{n,t_n},a_{-n,t_{-n}},\psi,t,w)=\iota_{n,t_n,t_{-n}}(a_{n,t_n},a_{-n,t_{-n}},\psi,t,w)\cdot a_{n,t_n}$. In view of~(\ref{ex-u}), we will further have
\begin{equation}\label{revenue}\begin{array}{l}
r_{n,t_n,t_{-n}}(a_{n,t_n},a_{-n,t_{-n}},\psi,t,w)=(w_n-a_{n,t_n})\times[{\bf 1}(a_{n,t_n}>\max_{m\neq n}a_{m,t_m})\\
\;\;\;\;\;\;\;\;\;\;\;\;+{\bf 1}(a_{n,t_n}=\max_{m\neq n}a_{m,t_m})\cdot{\bf 1}(\psi_n\leq \psi_m-1\mbox{ whenever }a_{m,t_m}=a_{n,t_n})].
\end{array}\end{equation}
We will have to replace $w_n-a_{n,t_n}$ with $w_n-\max_{m\neq n}a_{m,t_m}$ if the auction is second-price. Regardless, we can let payoff spaces be $R_{n,t_n}=[-\bar w,\bar w]$.

Bidders can be alarmists, so that~(\ref{s-form}) and~(\ref{s0-form}) apply. Or, they can be enterprising with~(\ref{radical-form}) and~(\ref{s0-form}) being true. In either case, let the utility functions $u_{n,t_n}$ be identity maps and let there be a nonempty closed subset $Q_{n,t_n}$ of ${\cal P}(\{1,...,\bar t\}^{\bar n-1}\times [0,\bar w]^{\bar n})$, so that the prior set used in~(\ref{s-form}) or~(\ref{radical-form}), as a nonempty subset of ${\cal P}(\Omega_{n,t_n})\equiv {\cal P}(\Psi_{\bar n}\times \{1,...,\bar t\}^{\bar n-1}\times [0,\bar w]^{\bar n})$, satisfies
\begin{equation}\label{ozaji-alarm}
P_{n,t_n}=\{\upsilon\}\times Q_{n,t_n}.
\end{equation}
In~(\ref{ozaji-alarm}), $\upsilon$ stands for the uniform distribution on $\Psi_{\bar n}$ wherein every member $\psi$ receives a $1/n!$ chance. Every $q\in {\cal P}(\{1,...,\bar t\}^{\bar n-1}\times [0,\bar w]^{\bar n})$ can be understood as being made up of two components $p\equiv (p_{t_{-n}})_{t_{-n}\in\{1,...,\bar t\}^{\bar n-1}}$ and $\nu\equiv (\nu_{t_{-n}})_{t_{-n}\in \{1,...,\bar t\}^{\bar n-1}}$, where $p$ is a probability mass function on opponents' type profiles and given each opponent-type profile $t_{-n}\in \{1,...,\bar t\}^{\bar n-1}$, the element $\nu_{t_{-n}}\in {\cal P}([0,\bar w]^{\bar n})$ is a distribution on the worth profile. The end effect for $q\equiv(p,\nu)$ is such that, for any $T'\subseteq \{1,...,\bar t\}^{\bar n-1}$ and $W'\in {\cal B}([0,\bar w]^{\bar n})$,
\begin{equation}\label{peipei}
q(T'\times W')=\sum_{t_{-n}\in T'}p_{t_{-n}}\cdot \nu_{t_{-n}}(W').
\end{equation}

When interested in pure action-based equilibria, for instance, consider $\tilde s_{n,t_n}(a_{n,t_n},a_{-n})=s^{\mbox a}_{n,t_n}(a_{n,t_n},1_{a_{-n}})$, where the latter function is defined at~(\ref{useful-pp}). By~(\ref{criterion-p}),  we will have pure strategy $1_a\in 1_A\cap {\cal E}^{\mbox a}$ if and only if $a_{n,t_n}\in\tilde B_{n,t_n}(a_{-n})$ for every $n$ and $t_n$, where
\begin{equation}\label{best-def}
\tilde B_{n,t_n}(a_{-n})=\left\{a_{n,t_n}\in A_{n,t_n}|\tilde s_{n,t_n}(a_{n,t_n},a_{-n})\geq \tilde s_{n,t_n}(a'_{n,t_n},a_{-n})\;\;\forall a'_{n,t_n}\in A_{n,t_n}\right\}.
\end{equation}
Thus, $1_a\in 1_A$ will be a pure equilibrium if and only if $a\equiv (a_{n,t_n})_{n=1,...,\bar n,t_n=1,...,\bar t}\in A\equiv \prod_{n=1}^{\bar n}\prod_{t=1}^{\bar t}A_{n,t_n}$ is that for a corresponding agent-based normal-form game where payoffs are the $\tilde s_{n,t_n}$'s. For the first-price and alarmists' case, by~(\ref{s-form}),~(\ref{mechanism}) to~(\ref{vec-p}),~(\ref{useful-pp}),~(\ref{s0-form}), and~(\ref{revenue}) to~(\ref{peipei}), as well as other model primitives, we can obtain
\begin{equation}\label{1s-def-new-alarm}
\tilde s_{n,t_n}(a_{n,t_n},a_{-n})=\inf_{(p,\nu)\in Q_{n,t_n}}\tilde  u_{n,t_n}(a_{n,t_n},a_{-n},p,\nu),
\end{equation}
where $a_{n,t_n}\in [-\bar w,\bar w]$ and  $a_{-n}\equiv (a_{m,t_m})_{m\neq n,t_m\in T_m}\in [-\bar w,\bar w]^{(\bar n-1)\cdot\bar t}$; also,
\begin{equation}\label{1w-def-new-alarm}
\tilde  u_{n,t_n}(a_{n,t_n},a_{-n},p,\nu)=\sum_{t_{-n}\in \{1,...,\bar t\}^{\bar n-1}}p_{t_{-n}}\cdot\tilde  v_{n,t_{-n}}(a_{n,t_n},a_{-n,t_{-n}},\nu_{t_{-n}}),
\end{equation}
with
\begin{equation}\label{1v-def-new-alarm}\begin{array}{l}
\tilde  v_{n,t_{-n}}(a_{n,t_n},a_{-n,t_{-n}},\nu_{t_{-n}})=\int_0^{\bar w}(w_n-a_{n,t_n})\cdot\nu_{t_{-n}}(dw)\times[{\bf 1}(a_{n,t_n}>\max_{m\neq n}a_{m,t_m})\\
\;\;\;\;\;\;\;\;\;\;\;\;\;\;\;\;\;\;\;\;\;\;\;\;\;\;\;\;\;\;\;\;\;\;\;\;\;\;\;\;\;\;\;\;\;\;\;\;+{\bf 1}(a_{n,t_n}=\max_{m\neq n}a_{m,t_m})/\#(\mbox{argmax}\;a_{m,t_m})],
\end{array}\end{equation}
where $\#(\mbox{argmax}\;a_{m,t_m})$ counts the number of bidders that tie in the highest bid.

The model will be symmetric when $Q_{n,t_n}$ is independent of $n$, any of $Q_{n,t_n}$'s member $(p,\nu)\equiv ((p_{t_{-n}})_{t_{-n}\in\{1,...,\bar t\}^{\bar n-1}},(\nu_{t_{-n}})_{t_{-n}\in\{1,...,\bar t\}^{\bar n-1}})$ satisfies that $p_{t_{-n}}$ and $\nu_{t_{-n}}$ are both functions of the empirical type distribution resulting from $t_{-n}$, and any transformation that turns any $(w_n,w_{-n})$ into $(w_n,w'_{-n})$ with $w'_{-n}$ having the same empirical worth distribution as $w_{-n}$ is measure-preserving for $\nu_{t_{-n}}$. Whether pure action-based equilibria exist remains so far unresolved, although according to earlier discussion, ordinary action-based equilibria are guaranteed to exist and when bidders are alarmists, so do ordinary distribution-based ones.

\subsection{Existing Models}

With the exception of requiring more general type spaces, several existing auction models can also be cast into the framework set up in Section~\ref{g-auction}.

Lo \cite{L98} let every bidder be sure about the true worth to himself of the item being auctioned, while making the bidder uncertain as to which distribution of other bidders' valuations to assume out of an own-valuation-independent set. When bidders are alarmists, equilibrium own-valuation-dependent bidding prices could be found by examining an associated traditional auction in which every bidder allows opponents the most optimistic outlooks ever supportable by his given set of other-valuation distributions.

With a zero reservation price, Lo's \cite{L98} model can be understood as having continuous type spaces $T_n=[0,\bar w]$, with each state $t_n\in T_n$ being the true worth to bidder $n$ of the item being auctioned. Its focus is on $\bar n=2$ but according to the author, can be extended to the case with a general $\bar n$ number of bidders. 
The state space $\Omega$ can be understood as $\Psi_{\bar n}\times \prod_{n\in N}T_n=\Psi_{\bar n}\times [0,\bar w]^{\bar n}$, with $\Psi_{\bar n}$ again serving tie-breaking purposes. Consequently, each $\Omega_{n,t_n}=\Psi_{\bar n}\times \{t_n\}\times \prod_{m\neq n}T_m=\Psi_{\bar n}\times \{t_n\}\times [0,\bar w]^{\bar n-1}$ and each $\Omega_t=\Psi_{\bar n}\times \{t\}=\Psi_{\bar n}\times\{(t_n)_{n=1,...,\bar n}\}$. Again, we can let action spaces be $A_{n,t_n}=[0,\bar w]$. 

Note that every $\omega\in\Omega$ is made up of two components, a permutation $\psi$ and a worth profile $t\equiv (t_n)_{n=1,...,\bar n}$. For the first-price case, we now have
\begin{equation}\label{revenue-lo}\begin{array}{l}
r_{n,t_n,t_{-n}}(a_{n,t_n},a_{-n,t_{-n}},\psi,t)=(t_n-a_{n,t_n})\times[{\bf 1}(a_{n,t_n}>\max_{m\neq n}a_{m,t_m})\\
\;\;\;\;\;\;\;\;\;\;\;\;+{\bf 1}(a_{n,t_n}=\max_{m\neq n}a_{m,t_m})\cdot{\bf 1}(\psi_n\leq \psi_m-1\mbox{ whenever }a_{m,t_m}=a_{n,t_n})].
\end{array}\end{equation}
Lo \cite{L98} let bidders be alarmists that use identity utility functions $u_{n,t_n}$ in~(\ref{s0-form}). For the prior set $P_{n,t_n}$ involved in~(\ref{s-form}), he let there be a nonempty closed subset $Q_n$ of ${\cal P}([0,\bar w]^{\bar n-1})$ that is independent of $t_n$, so that $P_{n,t_n}=\{\upsilon\}\times Q_{n}$, where $\upsilon$ is again the uniform distribution on $\Psi_{\bar n}$. In pursuing pure action-based equilibria, Lo \cite{L98} used the same criterion~(\ref{best-def}), with
\begin{equation}\label{1s-def-new-alarm-lo}
\tilde s_{n,t_n}(a_{n,t_n},a_{-n})=\inf_{q\in Q_n}\tilde  u_{n,t_n}(a_{n,t_n},a_{-n},q),
\end{equation}
\begin{equation}\label{1w-def-new-alarm-lo}
\tilde  u_{n,t_n}(a_{n,t_n},a_{-n},q)=\int_{[0,\bar w]^{\bar n-1}}\tilde  v_{n,t_n,t_{-n}}(a_{n,t_n},a_{-n,t_{-n}})\cdot q(dt_{-n}),
\end{equation}
\begin{equation}\label{1v-def-new-alarm-lo}\begin{array}{l}
\tilde  v_{n,t_n,t_{-n}}(a_{n,t_n},a_{-n,t_{-n}})=(t_n-a_{n,t_n})\times[{\bf 1}(a_{n,t_n}>\max_{m\neq n}a_{m,t_m})\\
\;\;\;\;\;\;\;\;\;\;\;\;\;\;\;\;\;\;\;\;\;\;\;\;\;\;\;\;\;\;\;\;\;\;\;\;\;\;\;\;\;\;\;\;\;\;\;\;+{\bf 1}(a_{n,t_n}=\max_{m\neq n}a_{m,t_m})/\#(\mbox{argmax}\;a_{m,t_m})].
\end{array}\end{equation}
The model will be symmetric when $Q_n$ is independent of $n$ and for any of $Q_n$'s member $q$, it would render a transformation measure-preserving whenever the latter turns any $t_{-n}$ into any $t'_{-n}$ with the same empirical type distribution. Concerning Bose, Ozdenoren, and Pape's \cite{BOP06} study of optimal auctions, their setup regarding bidders can be similarly casted; the only exception is that~(\ref{1v-def-new-alarm-lo}) needs to adapt with the particular auction design.

The traditional auction can be considered as having the same type, action, and state spaces, as well as the payoff structure~(\ref{revenue-lo}).  Identity utility functions $u_{n,t_n}$ and single priors $\rho_{n,t_n}$ would together give rise to satisfaction functions through~(\ref{s0-form}) and~(\ref{form1}). Moreover, $\rho_{n,t_n}=\upsilon\times q_{n,t_n}$ where $\upsilon$ is the uniform distribution on $\Psi_{\bar n}$ and $q_{n,t_n}\in {\cal P}([0,\bar w]^{\bar n-1})$. Often, $q_{n,t_n}$ is independent of $t_n$ and is of a product form. These reflect that every bidder views every competitor's valuation as being independently drawn from a given distribution.

Cases considered in the above would all fit within our framework, except for the fact that more general type spaces need to be tolerated. So a generalization in this direction poses as our most immediate task for future considerations.

\section{Concluding Remarks}\label{conclusion}

We have allowed ambiguities on external factors to be treated in games involving incomplete information. As players can be ambiguous about opponents' types, while all players can have type-sensitive ambiguity attitudes and behaviors, our setting has indirectly enabled ambiguities on opponents' ambiguity attitudes and behaviors as well. For the two proposed behavioral-equilibrium concepts, we arrived to various results concerning their existence, continuity, and mutual relationships. The enterprising game in which players are optimistic about the resolutions of their ambiguities delivered more concrete results. Not only are pure equilibria unified in such a game, but also their existence and monotone features are guaranteed when strategic complementarities are present.

More than providing normative answers to how participants should behave in situations involving both incomplete information and diverse ambiguity attitudes, some of our results might lead to explanations for phenomena already observed in real life. For instance, for auctions of works of art, offshore oilfields, electromagnetic spectra, etc., we speculate that uncertainties about the worths of items being auctioned and players' opportunistic attitudes toward the eventual resolutions of ambiguities might give extra impetus to upward movements of bidding prices. Hence, the winner's curse could be made even worse. 

The model's confinement to finite type spaces could certainly hamper its applicabilities in some occasions. To deal with more general type spaces whilst still countenancing general ambiguity attitudes, it seems that information structures different from the current one revolving around the $\Omega_{n,t_n}$ sets are warranted. 
Issues concerning topologies on preference spaces, like those covered in Hildenbrand \cite{H74} and Klein and Thompson \cite{KT84}, will likely arise.


\vspace*{.6in}

\noindent{\bf\Large Appendices}\appendix
\numberwithin{equation}{section}

\section{Proofs for Section~\ref{analysis}}\label{app-a}

\noindent{\bf Proof of Proposition~\ref{pi-cont}: }By Assumptions~\ref{compact-action},~\ref{compact-state}, and~\ref{continuity}, we know that $r_{n,t_n,t_{-n}}$ is uniformly continuous. 
By a well known convergence result (Hildenbrand \cite{H74}, D.I.(38)), this will lead to the continuity of each $\pi^{\mbox a}_{n,t_n,t_{-n}}$, defined at~(\ref{mechanism}) as a function from $A_{n,t_n}\times\Delta_{-n,t_{-n}}\times\Omega_{t_n,t_{-n}}$ to ${\cal P}(R_{n,t_n})$, in $(a_{n,t_n},\omega)$ and also $\delta_{-n,t_{-n}}$. Indeed, suppose $\lim_{k\rightarrow +\infty}(a^k_{n,t_n},\omega^k)=(a_{n,t_n},\omega)$ and $\lim_{k\rightarrow +\infty}\delta^k_{-n,t_{-n}}=\delta_{-n,t_{-n}}$. Then, by~(\ref{mechanism}) and $r_{n,t_n,t_{-n}}$'s continuity in $(a_{n,t_n},\omega)$ at an  $a_{-n,t_{-n}}$-independent rate, $\lim_{k\rightarrow +\infty}\pi^{\mbox a}_{n,t_n,t_{-n}}(a^k_{n,t_n},\delta_{-n,t_{-n}},\omega^k)$ equals
\begin{equation}\begin{array}{l}
\lim_{k\rightarrow +\infty}(\prod_{m\neq n}\delta_{m,t_m})\cdot (r_{n,t_n,t_{-n}}(a^k_{n,t_n},\cdot,\omega^k))^{-1}\\
\;\;\;\;\;\;\;\;\;\;\;\;\;\;\;\;\;\;\;\;\;\;\;\;=(\prod_{m\neq n}\delta_{m,t_m})\cdot (r_{n,t_n,t_{-n}}(a_{n,t_n},\cdot,\omega))^{-1},
\end{array}\end{equation}
which in turn equals $\pi^{\mbox a}_{n,t_n,t_{-n}}(a_{n,t_n},\delta_{-n,t_{-n}},\omega)$. Due to Assumption~\ref{compact-action}, the spaces $A_{m,t_m}$ are all compact and hence separable. Then, $\lim_{k\rightarrow +\infty}\delta^k_{-n,t_{-n}}=\delta_{-n,t_{-n}}$ would result with
$\lim_{k\rightarrow +\infty}\prod_{m\neq n}\delta^k_{m,t_m}=\prod_{m\neq n}\delta_{m,t_m}$. So similarly, by~(\ref{mechanism}) and $r_{n,t_n,t_{_n}}$'s continuity in $a_{-n,t_{-n}}$, we will have $\lim_{k\rightarrow +\infty}\pi^{\mbox a}_{n,t_n,t_{-n}}(a_{n,t_n},\delta^k_{-n,t_{-n}},\omega)$ equal to
\begin{equation}
\lim_{k\rightarrow +\infty}(\prod_{m\neq n}\delta^k_{m,t_m})\cdot (r_{n,t_n,t_{-n}}(a_{n,t_n},\cdot,\omega))^{-1}=(\prod_{m\neq n}\delta_{m,t_m})\cdot (r_{n,t_n,t_{-n}}(a_{n,t_n},\cdot,\omega))^{-1}.
\end{equation}
The following shows that $\pi^{\mbox a}_{n,t_n,t_{-n}}$'s continuity in $(a_{n,t_n},\omega)$ can be at a rate independent of the $\delta_{-n,t_{-n}}$ present, just also because $r_{n,t_n,t_{-n}}$'s continuity in $(a_{n,t_n},\omega)$ is independent of $a_{-n,t_{-n}}$.\vspace*{.02in}

\noindent{\em Lemma}\;\;Let $X$ and $Y$ be separable metric spaces, and $u$ and $v$ be measurable functions from $X$ to $Y$. Then, any $\rho\in {\cal P}(X)$ satisfies
\begin{equation}
\psi_Y(\rho\cdot u^{-1},\rho\cdot v^{-1})\leq \sup_{x\in X}d_Y(u(x),v(x)),
\end{equation}
where the right-hand side is independent of the $\rho$ involved. 	\\
\noindent {\em Proof}: Let $\epsilon=\sup_{x\in X}d_Y(u(x),v(x))$. There is nothing to prove if $\epsilon=0$ because then $u=v$. So suppose $\epsilon>0$. For any $Y'\in {\cal B}(Y)$, we observe that
\begin{equation}
u^{-1}(Y')\subseteq v^{-1}((Y')^\epsilon).
\end{equation}
Thus,
\begin{equation}
(\rho\cdot u^{-1})(Y')\leq (\rho\cdot v^{-1})((Y')^\epsilon)<(\rho\cdot v^{-1})((Y')^\epsilon)+\epsilon.
\end{equation}
So by the definition of the Prokhorov metric $\psi_Y$, we have the desired inequality.

Combine the continuity in $(a_{n,t_n},\omega)$ at a $\delta_{-n,t_{-n}}$-independent rate and continuity in $\delta_{-n,t_{-n}}$, and we get $\pi^{\mbox a}_{n,t_n,t_{-n}}$'s continuity in $(a_{n,t_n},\delta_{-n,t_{-n}},\omega)$. We can similarly tackle $\pi^{\mbox d}_{n,t_n,t_{-n}}$'s continuity in $(\delta_{n,t_n},\delta_{-n,t_{-n}},\omega)$. \qed

\noindent{\bf Proof of Proposition~\ref{ca-u}: }We prove by contradiction. Suppose such a $\succ_{n,t_n}$-maximal $\pi$ does not exist in $\Pi'$. Now for any $\pi'\in \Pi'$, define
\begin{equation}\label{gff-def}
L(\pi')=\{\pi\in\Pi'|\pi'\succ_{n,t_n}\pi\}.
\end{equation}
By the earlier hypothesis, every $\pi\in \Pi'$ has a corresponding $\pi'$ so that $\pi\in L(\pi')$. Thus,
\begin{equation}\label{ozaji}
\Pi'=\bigcup_{\pi'\in\Pi'} L(\pi').
\end{equation}
By~(\ref{fff-def}) and~(\ref{gff-def}), each $L(\pi')$ is a projection of $G_{n,t_n}\equiv (\Pi_{n,t_n}\times\Pi_{n,t_n})\setminus \varphi_{n,t_n}$ to the second $\Pi_{n,t_n}$. Due to Preference Assumption~\ref{ca-l}, the set $G_{n,t_n}$ is open. So must be every $L(\pi')$.

Since $\Pi'$ is compact, from~(\ref{ozaji}) we can infer that $\Pi'$ has a finite subcover from among the $L(\pi')$'s. Pick a subcover with the smallest number of elements say $k$, involving open sets say $L(\pi_1),...,L(\pi_k)$. Suppose $k\geq 2$. By $\succ_{n,t_n}$'s irreflexibility and~(\ref{gff-def}), we know
\begin{equation}\label{k-imp}
\pi_k\notin L(\pi_k).
\end{equation}
It must be the case that $\pi_k\in L(\pi_l)$ for some $l\leq k-1$. Consider any $\pi\in L(\pi_k)$. Note that $\pi_l\succ_{n,t_n}\pi_k$ and $\pi_k\succ_{n,t_n}\pi$ by~(\ref{gff-def}). By $\succ_{n,t_n}$'s transitivity, we must then have $\pi_l\succ_{n,t_n}\pi$; that is, $\pi\in L(\pi_l)$. But this just means that $L(\pi_k)\subseteq L(\pi_l)$, a contradiction to the minimality of $k$. The only choice is $k=1$. But this forces $\pi_1\in L(\pi_1)$, an impossibility in view of~(\ref{k-imp}). \qed

\noindent{\bf Proof of Proposition~\ref{a-condition}: }Suppose $(a^k_{n,t_n},\delta^k_{-n})\in A_{n,t_n}\times \Delta_{-n}$ for $k=1,2,...$ and $(a_{n,t_n},\delta_{-n})\in A_{n,t_n}\times \Delta_{-n}$ are such that $\lim_{k\rightarrow +\infty}a^k_{n,t_n}=a_{n,t_n}$, $\lim_{k\rightarrow +\infty}\delta^k_{-n}=\delta_{-n}$, and $a^k_{n,t_n}\in\hat A^{\mbox a}_{n,t_n}(\delta^k_{-n})$ for each $k=1,2,...$. We are to show that $a_{n,t_n}\in \hat A^{\mbox a}_{n,t_n}(\delta_{-n})$ as well.

Let $a'_{n,t_n}$ be arbitrarily chosen from $A_{n,t_n}$. By the membership of the $a^k_{n,t_n}$'s in the corresponding spaces $\hat A^{\mbox a}_{n,t_n}(\delta^k_{-n})$ and~(\ref{maximal0}),
\begin{equation}\label{comb0}
\pi^{\mbox a}_{n,t_n}(a'_{n,t_n},\delta^k_{-n})\not \succ_{n,t_n} \pi^{\mbox a}_{n,t_n}(a^k_{n,t_n},\delta^k_{-n}),\hspace*{.8in}\forall k=1,2,....
\end{equation}
Proposition~\ref{piphi-cont} and the first two conditions above will together lead to
\begin{equation}\label{comb1}
\lim_{k\rightarrow +\infty}\pi^{\mbox a}_{n,t_n}(a'_{n,t_n},\delta^k_{-n})=\pi^{\mbox a}_{n,t_n}(a'_{n,t_n},\delta_{-n}),
\end{equation}
\begin{equation}\label{comb2}
\pi^{\mbox a}_{n,t_n}(a_{n,t_n},\delta_{-n})=\lim_{k\rightarrow +\infty}\pi^{\mbox a}_{n,t_n}(a_{n,t_n},\delta^k_{-n})=\lim_{k\rightarrow +\infty}\pi^{\mbox a}_{n,t_n}(a^k_{n,t_n},\delta^k_{-n}).
\end{equation}
Combining~(\ref{comb0}) to~(\ref{comb2}), as well as Preference Assumption~\ref{ca-l}, we get
\begin{equation}
\pi^{\mbox a}_{n,t_n}(a'_{n,t_n},\delta_{-n})\not \succ_{n,t_n} \pi^{\mbox a}_{n,t_n}(a_{n,t_n},\delta_{-n}).
\end{equation}
Due to~(\ref{maximal0}) again and the arbitrariness of $a'_{n,t_n}\in A_{n,t_n}$, it follows that $a_{n,t_n}\in \hat A^{\mbox a}_{n,t_n}(\delta_{-n})$. \qed

\noindent{\bf Proof of Proposition~\ref{b-continuity}: }Suppose $\delta^k\equiv (\delta^k_m)_{m\in N}\equiv (\delta^k_{m,\tau_m})_{m\in N,\tau_m\in T_m}\in \Delta\equiv \prod_{m\in N}\Delta_m\equiv \prod_{m\in N}\prod_{\tau_m\in T_m}\Delta_{m,\tau_m}$ for $k=1,2,...$ and $\delta\equiv (\delta_m)_{m\in N}\equiv (\delta_{m,\tau_m})_{m\in N,\tau_m\in T_m}\in\Delta$ are such that $\lim_{k\rightarrow +\infty}\delta^k_{n,t_n}=\delta_{n,t_n}$, $\lim_{k\rightarrow +\infty}\delta^k_{-n}=\delta_{-n}$, and $\delta^k_{n,t_n}\in\hat B^{\mbox a}_{n,t_n}(\delta^k_{-n})$ for each $k=1,2,...$. We are to show that $\delta_{n,t_n}\in \hat B^{\mbox a}_{n,t_n}(\delta_{-n})$ as well. For this purpose, the closedness of $\hat A^{\mbox a}_{n,t_n}(\cdot)$ as shown in Proposition~\ref{a-condition} and the convergence of $\delta^k_{-n}$ to $\delta_{-n}$ will now lead to
\begin{equation}\label{firsty}
\mbox{Ls}\left((\hat A^{\mbox a}_{n,t_n}(\delta^k_{-n}))_{k=1,2,...}\right)\subseteq\hat A^{\mbox a}_{n,t_n}(\delta_{-n}),
\end{equation}
where Ls$(\cdot)$ stands for a set sequence's topological limes superior; see Hildenbrand \cite{H74} (Section B.II). Let $\epsilon>0$ be given. Since $\delta^k_{n,t_n}$ converges to $\delta_{n,t_n}$, as long as $l$ is large enough,
\begin{equation}\label{secondy}
\delta^l_{n,t_n}\left[\left(\mbox{Ls}((\hat A^{\mbox a}_{n,t_n}(\delta^k_{-n}))_{k=1,2,...})\right)^\epsilon\right]\leq \delta_{n,t_n}\left[\left(\mbox{Ls}((\hat A^{\mbox a}_{n,t_n}(\delta^k_{-n}))_{k=1,2,...})\right)^{2\epsilon}\right]+\epsilon.
\end{equation}
By the definition of Ls$(\cdot)$ and $A_{n,t_n}$'s compactness, we have, when $l$ is further large enough,
\begin{equation}\label{thirdy}
\hat A^{\mbox a}_{n,t_n}(\delta^l_{-n})\subseteq \left(\mbox{Ls}\left((\hat A^{\mbox a}_{n,t_n}(\delta^k_{-n}))_{k=1,2,...}\right)\right)^\epsilon.
\end{equation}
Combining the above, we obtain
\begin{equation}\begin{array}{l}
\delta_{n,t_n}\left[(\hat A^{\mbox a}_{n,t_n}(\delta_{-n}))^{2\epsilon}\right]\geq \delta_{n,t_n}\left[\left(\mbox{Ls}((\hat A^{\mbox a}_{n,t_n}(\delta^k_{-n}))_{k=1,2,...})\right)^{2\epsilon}\right]\\
\;\;\;\;\;\;\;\;\;\;\;\;\geq \delta^l_{n,t_n}\left[\left(\mbox{Ls}((\hat A^{\mbox a}_{n,t_n}(\delta^k_{-n}))_{k=1,2,...})\right)^\epsilon\right]-\epsilon\geq \delta^l_{n,t_n}\left[\hat A^{\mbox a}_{n,t_n}(\delta^l_{-n})\right]-\epsilon=1-\epsilon,
\end{array}\end{equation}
where the first inequality is due to~(\ref{firsty}), the second inequality is due to~(\ref{secondy}), the third inequality is due to~(\ref{thirdy}), and the last equality comes from~(\ref{b-def}) and the membership of $\delta^l_{n,t_n}$ in $\hat B^{\mbox a}_{n,t_n}(\delta^l_{-n})$. For any $k=1,2,...$, this means that
\begin{equation}\label{qiuqiu}
\delta_{n,t_n}\left[\bigcap_{l=k}^{+\infty}\left(\hat A^{\mbox a}_{n,t_n}(\delta_{-n})\right)^{1/l}\right]\geq 1-\frac{1}{2k}.
\end{equation}
According to Proposition~\ref{a-condition}, $\hat A^{\mbox a}_{n,t_n}(\delta_{-n})$ is closed and hence is equal to $\bigcap_{l=k}^{+\infty}(\hat A^{\mbox a}_{n,t_n}(\delta_{-n}))^{1/l}$ for any $k=1,2,...$. So the above~(\ref{qiuqiu}) will result with $\delta_{n,t_n}(\hat A^{\mbox a}_{n,t_n}(\delta_{-n}))=1$, translating into $\delta_{n,t_n}$'s membership in $\hat B^{\mbox a}_{n,t_n}(\delta_{-n})$ by~(\ref{b-def}).

We can follow almost the same steps used in the proof of Proposition~\ref{a-condition} to deduce that as a correspondence from $\Delta_{-n}$ to $\Delta_{n,t_n}$, each $\hat B^{\mbox d}_{n,t_n}$ is closed. \qed


\noindent{\bf Proof of Proposition~\ref{kappu}: }Let $\varphi^k_{n,t_n}$ be members of $\Phi_{n,t_n}$ for $k=1,2,...$ and let $F$ be a member of ${\cal F}_{n,t_n}$. Suppose $\lim_{k\rightarrow +\infty}d_{{\cal F}_{n,t_n}}(\varphi^k_{n,t_n},F)=0$. We show that $F\in \Phi_{n,t_n}$ as well.

According to Theorem B.II.1 of Hildenbrand \cite{H74},
\begin{equation}\label{budeliao}
\mbox{Li}\left((\varphi^k_{n,t_n})_{k=1,2,...}\right)=F=\mbox{Ls}\left((\varphi^k_{n,t_n})_{k=1,2,...}\right).
\end{equation}
In the above, $\mbox{Li}((\varphi^k_{n,t_n})_{k=1,2,...})$ is the topological limes inferior of the sequence $(\varphi^k_{n,t_n})_{k=1,2,...}$: any $(\pi,\pi')\in \Pi_{n,t_n}\times\Pi_{n,t_n}$ will belong to the set if and only if there is a sequence $((\pi_k,\pi'_k))_{k=1,2,...}$ so that $(\pi_k,\pi'_k)\in \varphi^k_{n,t_n}$ and $\lim_{k\rightarrow +\infty}d_{\Pi_{n,t_n}\times \Pi_{n,t_n}}((\pi_k,\pi'_k),(\pi,\pi'))=0$; also, $\mbox{Ls}((\varphi^k_{n,t_n})_{k=1,2,...})$ is the topological limes superior of the sequence $(\varphi^k_{n,t_n})_{k=1,2,...}$: any $(\pi,\pi')\in \Pi_{n,t_n}\times\Pi_{n,t_n}$ will belong to the set if and only if there is a subsequence $((\pi_{q^k},\pi'_{q^k}))_{k=1,2,...}$ so that $(\pi_{q^k},\pi'_{q^k})\in \varphi^{q^k}_{n,t_n}$ and $\lim_{k\rightarrow +\infty}d_{\Pi_{n,t_n}\times \Pi_{n,t_n}}((\pi_{q^k},\pi'_{q^k}),(\pi,\pi'))=0$.

We first show that $F$ satisfies (i). For any $\pi\in \Pi_{n,t_n}$, we know from the membership of each $\varphi^k_{n,t_n}$ in $\Phi_{n,t_n}$ that $(\pi,\pi)\in \varphi^k_{n,t_n}$. Then, it must be true that $(\pi,\pi)\in \mbox{Li}((\varphi^k_{n,t_n})_{k=1,2,...})$. But by~(\ref{budeliao}), this means that $(\pi,\pi)\in F$.

We next show that $F$ satisfies (ii). Suppose members $\pi$, $\pi'$, and $\pi''$ of $\Pi_{n,t_n}$ are such that $(\pi,\pi')\notin F$ and $(\pi',\pi'')\notin F$. Since $F$ is closed, there must exist $\epsilon>0$, so that for any $\pi_1$ in the closed $\epsilon$-neighborhood of $\pi$, any $\pi'_1$ in the closed $\epsilon$-neighborhood of $\pi'$, and any $\pi''_1$ in the closed $\epsilon$-neighborhood of $\pi''$, neither  $(\pi_1,\pi'_1)$ nor $(\pi'_1,\pi''_1)$ is in $F$.

If $\varphi^k_{n,t_n}\cap (\{(\pi,\pi')\})^{\epsilon}\neq\emptyset$ occurred for an infinite number of $k$'s, then due to the compactness of $\Pi_{n,t_n}\times \Pi_{n,t_n}$, we would have
\begin{equation} \mbox{Ls}\left((\varphi^k_{n,t_n})_{k=1,2,...}\right)\cap\left(\{(\pi,\pi')\}\right)^{\epsilon}\neq\emptyset.
\end{equation}
But by~(\ref{budeliao}), this would lead to $F\cap(\{(\pi,\pi')\})^{\epsilon}\neq\emptyset$, a contradiction. So there must exist $k'$, so that $\varphi^k_{n,t_n}\cap(\{(\pi,\pi')\})^{\epsilon}=\emptyset$ for $k=k',k'+1,...$. Similarly, there must exist $k''$, so that $\varphi^k_{n,t_n}\cap(\{(\pi',\pi'')\})^{\epsilon}=\emptyset$ for $k=k'',k''+1,...$. Then, for $k=k'\vee k'',k'\vee k''+1,...$, the set $\varphi^k_{n,t_n}$ will intersect neither $(\{\pi,\pi'\})^{\epsilon}$ nor $(\{\pi',\pi'')\})^{\epsilon}$.

That is, for any $\pi_1$ in the closed $\epsilon$-neighborhood of $\pi$, $\pi'_1$ in the closed $\epsilon$-neighborhood of $\pi'$, and $\pi''_1$ in the closed $\epsilon$-neighborhood of $\pi''$, we have $(\pi_1,\pi'_1)\notin \varphi^k_{n,t_n}$ and $(\pi'_1,\pi''_1)\notin \varphi^k_{n,t_n}$ for any $k=k'\vee k'',k'\vee k''+1,...$. But since each such $\varphi^k_{n,t_n}$ is a member of $\Phi_{n,t_n}$ and hence satisfies (ii), $\varphi^k_{n,t_n}$ must not contain $(\pi_1,\pi''_1)$ either. Therefore,
\begin{equation}
\varphi^k_{n,t_n}\cap (\{(\pi,\pi'')\})^\epsilon=\emptyset,\hspace*{.8in}\forall k=k'\vee k'',k'\vee k''+1,....
\end{equation}
This will lead to $(\pi,\pi'')\notin \mbox{Ls}((\varphi^k_{n,t_n})_{k=1,2,...})$ which, in view of~(\ref{budeliao}), amounts to $(\pi,\pi'')\notin F$.

Due to the satisfaction of both (i) and (ii), we know that $F\in \Phi_{n,t_n}$ as well. \qed

\noindent{\bf Proof of Proposition~\ref{p-aa}: }Suppose sequence $(\delta^k_{-n})_{k=1,2,...}$ in $\Delta_{-n}$ converges to its member $\delta_{-n}$, and sequence $(\varphi^k_{n,t_n})_{k=1,2,...}$ in $\Phi_{n,t_n}$ converges to its member $\varphi_{n,t_n}$ so that
\begin{equation}\label{budeliao1}
\mbox{Ls}\left((\varphi^k_{n,t_n})_{k=1,2,...}\right)\subseteq \varphi_{n,t_n}.
\end{equation}
Also, suppose sequence $(a^k_{n,t_n})_{k=1,2,...}$ in $A_{n,t_n}$ converges to its member $a_{n,t_n}$; in addition, $a^k_{n,t_n}$ is inside $\hat A^{\mbox a}_{n,t_n}(\delta^k_{-n}|\varphi^k_{n,t_n})$ for every $k=1,2,...$.

Then, by Proposition~\ref{piphi-cont}, the sequence $(\pi^{\mbox a}_{n,t_n}(a^k_{n,t_n},\delta^k_{-n}))_{k=1,2,...}$ in $\Pi_{n,t_n}$ would converge to its member $\pi^{\mbox a}_{n,t_n}(a_{n,t_n},\delta_{-n})$. Fix any member $\pi$ of $\tilde \Pi^{\mbox a}_{n,t_n}(\delta_{-n})$. By the same continuity result and~(\ref{pi-p}), we also know there is a sequence $(\pi^k)_{k=1,2,...}$ in $\Pi_{n,t_n}$ that converges to $\pi$ and for each $k=1,2,...$, the payoff-distribution vector $\pi^k$ is inside $\tilde\Pi^{\mbox a}_{n,t_n}(\delta^k_{-n})$.

For any $k=1,2,...$, since $a^k_{n,t_n}\in \hat A^{\mbox a}_{n,t_n}(\delta^k_{-n}|\varphi^k_{n,t_n})$ and $\pi^k\in\tilde{\Pi}^{\mbox a}_{n,t_n}(\delta^k_{-n})$, we have from~(\ref{maximal0-dep}) that $(\pi^k,\pi^{\mbox a}_{n,t_n}(a^k_{n,t_n},\delta^k_{-n}))\in \varphi^k_{n,t_n}$. Now the convergence of $(\pi^k)_{k=1,2,...}$ to $\pi$ and that of $(\pi^{\mbox a}_{n,t_n}(a^k_{n,t_n},\delta^k_{-n}))_{k=1,2,...}$ to $\pi^{\mbox a}_{n,t_n}(a_{n,t_n},\delta_{-n})$ would lead to
\begin{equation}
\left(\pi,\pi^{\mbox a}_{n,t_n}(a_{n,t_n},\delta_{-n})\right)\in \mbox{Li}\left((\varphi^k_{n,t_n})_{k=1,2,...}\right)\subseteq \mbox{Ls}\left((\varphi^k_{n,t_n})_{k=1,2,...}\right).
\end{equation}
By~(\ref{budeliao1}), we will then have $(\pi,\pi^{\mbox a}_{n,t_n}(a_{n,t_n},\delta_{-n}))\in \varphi_{n,t_n}$. Going over all such $\pi$'s, we obtain
\begin{equation}
\tilde\Pi^{\mbox a}_{n,t_n}(\delta_{-n})\times\{\pi^{\mbox a}_{n,t_n}(a_{n,t_n},\delta_{-n})\}\subseteq \varphi_{n,t_n}.
\end{equation}
By~(\ref{maximal0-dep}), this will just mean that $a_{n,t_n}\in \hat A^{\mbox a}_{n,t_n}(\delta_{-n}|\varphi_{n,t_n})$.\qed

\noindent{\bf Proof of Proposition~\ref{p-ba}: }Suppose sequence $(\delta^k_{-n})_{k=1,2,...}$ in $\Delta_{-n}$ converges to its member $\delta_{-n}$ and sequence $(\varphi^k_{n,t_n})_{k=1,2,...}$ in $\Phi_{n,t_n}$ converges to its member $\varphi_{n,t_n}$. Also, suppose sequence $(\delta^k_{n,t_n})_{k=1,2,...}$ in $\Delta_{n,t_n}$ converges to its member $\delta_{n,t_n}$; in addition, $\delta^k_{n,t_n}$ is inside $\hat B^{\mbox a}_{n,t_n}(\delta^k_{-n}|\varphi^k_{n,t_n})$ for every $k=1,2,...$. 

Then, the upper hemi-continuity of $\hat A^{\mbox a}_{n,t_n}(\cdot|\cdot)$ as shown in Proposition~\ref{p-aa}, the convergence of  $\delta^k_{-n}$ to $\delta_{-n}$, and the convergence of $\varphi^k_{n,t_n}$ to $\varphi_{n,t_n}$, will together lead to
\begin{equation}\label{firsty-dep}
\mbox{Ls}\left((\hat A^{\mbox a}_{n,t_n}(\delta^k_{-n}|\varphi^k_{n,t_n}))_{k=1,2,...}\right)\subseteq\hat A^{\mbox a}_{n,t_n}(\delta_{-n}|\varphi_{n,t_n}),
\end{equation}
which is just like~(\ref{firsty}) in the proof of Proposition~\ref{b-continuity}, except with $\delta^k_{-n}|\varphi^k_{n,t_n}$ replacing $\delta^k_{-n}$. The rest of the proof can just follow the earlier proof with the same replacement. In the end, we can show that $\delta_{n,t_n}(\hat A^{\mbox a}_{n,t_n}(\delta_{-n}|\varphi_{n,t_n}))=1$, translating into $\delta_{n,t_n}$'s membership in $\hat B^{\mbox a}_{n,t_n}(\delta_{-n}|\varphi_{n,t_n})$ by~(\ref{b-def-dep}).

We can use similar arguments in the proof of Proposition~\ref{p-aa}, involving Proposition~\ref{piphi-cont}, to show that $\hat B^{\mbox d}_{n,t_n}(\cdot|\cdot)$ defined in~(\ref{maximal-k0-dep}) is upper hemi-continuous. \qed

\noindent{\bf Proof of Theorem\ref{t-dep}: }Suppose sequence $(\varphi^k)_{k=1,2,...}$ in $\Phi$ converges to its member $\varphi$, sequence $(\delta^k)_{k=1,2,...}$ in $\Delta$ converges to its member $\delta$, and for each $k=1,2,...$, $\delta^k$ is a member of ${\cal E}^{\mbox a}(\varphi^k)$. By~(\ref{zia-p-dep}), we would have $\delta^k_{n,t_n}\in\hat B_{n,t_n}(\delta^k_{-n}|\varphi^k_{n,t_n})$ for each $n\in N$ and $t_n\in T_n$.

Fix some $n\in N$ and $t_n\in T_n$. By the convergence of the sequence $\delta^k_{n,t_n}$ to $\delta_{n,t_n}$ and each $\delta^k_{n,t_n}$'s membership in $\hat B_{n,t_n}(\delta^k_{-n}|\varphi^k_{n,t_n})$, we would have
\begin{equation}\label{ayu}
\delta_{n,t_n}\in \mbox{Li}\left((B_{n,t_n}(\delta^k_{-n}|\varphi^k_{n,t_n}))_{k=1,2,...}\right)\subseteq\mbox{Ls}\left((B_{n,t_n}(\delta^k_{-n}|\varphi^k_{n,t_n}))_{k=1,2,...}\right).
\end{equation}
By Proposition~\ref{p-ba} on the upper hemi-continuity of $\hat B^{\mbox a}_{n,t_n}(\cdot|\cdot)$, the convergence of $\delta^k_{-n}$ to $\delta_{-n}$, and the convergence of $\varphi^k_{n,t_n}$ to $\varphi_{n,t_n}$, we can conclude that
\begin{equation}\label{byu}
\mbox{Ls}\left((B_{n,t_n}(\delta^k_{-n}|\varphi^k_{n,t_n}))_{k=1,2,...}\right)\subseteq \hat B^{\mbox a}_{n,t_n}(\delta_{-n}|\varphi_{n,t_n}).
\end{equation}
Combining~(\ref{ayu}) and~(\ref{byu}), we can get $\delta_{n,t_n}\in \hat B^{\mbox a}_{n,t_n}(\delta_{-n}|\varphi_{n,t_n})$. Note that this is true for every $n\in N$ and $t_n\in T_n$. So by~(\ref{zia-p-dep}), we have $\delta\in {\cal E}^{\mbox a}(\varphi)$.

We can use similar arguments, involving Proposition~\ref{p-ba}, to show that ${\cal E}^{\mbox d}$ defined in~(\ref{zia-a-dep}) is upper hemi-continuous. \qed

\section{Proofs for Section~\ref{relation}}\label{app-b}

\noindent{\bf Proof of Lemma~\ref{meaningful}: }We first prove that $\iota(y)\in {\cal P}(Z)$ at every $y\in Y$. Just because $\kappa(x,y)\in {\cal P}(Z)$ at every $x\in X$, we can easily see from~(\ref{fitts}) that $[\iota(y)](\emptyset)=0$ and $[\iota(y)](Z')+[\iota(y)](Z\setminus Z')=1$ at every $Z'\in {\cal B}(Z)$. Given non-overlapping subsets $Z^1,Z^2,...$ in ${\cal B}(Z)$, bounded convergence applied to~(\ref{fitts}) will also lead to
\begin{equation}
[\iota(y)]\left(\bigcup_{k=1}^{+\infty}Z^k\right)=\sum_{k=1}^{+\infty}[\iota(y)](Z^k).
\end{equation}
Thus, $\iota(y)$ is a probability measure on the measurable space $(Z,{\cal B}(Z))$.

We next show that $\iota$ is continuous from $Y$ to ${\cal P}(Z)$. For sequence $y^1,y^2,...$ that converges to $y$ in $Y$, we know from (b) that $\lim_{k\rightarrow +\infty}\kappa(x,y^k)=\kappa(x,y)$ at every $x\in X$. By the nature of the Prokhorov metric, this amounts to that, for every open subset $Z'$ of $Z$,
\begin{equation}\label{weakc}
[\kappa(x,y)](Z')\leq \liminf_{k\rightarrow +\infty}[\kappa(x,y^k)](Z').
\end{equation}
Now we can obtain
\begin{equation}\label{eve}\begin{array}{l}
[\iota(y)](Z')=\int_X [\kappa(x,y)](Z')\cdot \delta(dx)\leq \int_X \{\liminf_{k\rightarrow +\infty}[\kappa(x,y^k)](Z')\}\cdot\delta(dx)\\
\;\;\;\;\;\;\;\;\;\;\;\;\leq \liminf_{k\rightarrow +\infty}\int_X [\kappa(x,y^k)](Z')\cdot \delta(dx)=\liminf_{k\rightarrow +\infty}[\iota(y^k)](Z'),
\end{array}\end{equation}
where the first equality is due to~(\ref{fitts}), the first inequality uses~(\ref{weakc}), the second inequality comes from Fatou's lemma, and the last equality is again due to~(\ref{fitts}). Since~(\ref{eve}) applies to every open subset $Z'$ of $Z$, it amounts to $\lim_{k\rightarrow +\infty}\iota(y^k)=\iota(y)$. Hence, $\iota\in {\cal C}(Y,{\cal P}(Z))$. \qed

\noindent{\bf Proof of Proposition~\ref{beling}: }We show that $\pi^{\mbox a}_{n,t_n}(\delta_{-n})$ satisfies both (a) and (b) with $X=A_{n,t_n}$, $Y=\Omega_{n,t_n}\equiv\bigcup_{t_{-n}\in T_{-n}}\Omega_{t_n,t_{-n}}$, and $Z=R_{n,t_n}$.

By Assumption~\ref{continuity}, the payoff function $r_{n,t_n,t_{-n}}(\cdot,\cdot,\omega)$ at every $\omega\in\Omega_{t_n,t_{-n}}$ is continuous and hence measurable. So for any $R'_{n,t_n}\in {\cal B}(R_{n,t_n})$, the set $(r_{n,t_n,t_{-n}}(\cdot,\cdot,\omega))^{-1}(R'_{n,t_n})$
is a member of ${\cal B}(A_{n,t_n}\times A_{-n,t_{-n}})$. Meanwhile, we can obtain from~(\ref{mechanism}) and~(\ref{aha}) that $[\pi^{\mbox a}_{n,t_n,t_{-n}}(a_{n,t_n},\delta_{-n,t_{-n}},\omega)](R'_{n,t_n})$ is the integration of the indicator function of the measurable set  $(r_{n,t_n,t_{-n}}(\cdot,\cdot,\omega))^{-1}(R'_{n,t_n})$ over $a_{-n,t_{-n}}\in A_{-n,t_{-n}}$ under the measure $\prod_{m\neq n}\delta_{m,t_m}$. So by Fubini's theorem, $[\pi^{\mbox a}_{n,t_n,t_{-n}}(\cdot,\delta_{-n,t_{-n}},\omega)](R'_{n,t_n})$ is a Borel-measurable mapping from $A_{n,t_n}$ to $[0,1]$. This means that (a) is satisfied by the vector $\pi^{\mbox a}_{n,t_n}(\delta_{-n})$. Also, Proposition~\ref{pi-cont} provides the continuity of $\pi^{\mbox a}_{n,t_n,t_{-n}}(\cdot,\delta_{-n,t_{-n}},\cdot)$ as a mapping from $A_{n,t_n}\times \Omega_{t_n,t_{-n}}$ to ${\cal P}(R_{n,t_n})$ at every $t_{-n}\in T_{-n}$. In view of the separation condition~(\ref{separatable}), we can obtain (b) as well. \qed


\noindent{\bf Proof of Proposition~\ref{a-in-p}: }It will suffice to prove $\hat B^{\mbox d}_{n,t_n}(\delta_{-n}|\varphi_{n,t_n})\subseteq \hat B^{\mbox a}_{n,t_n}(\delta_{-n}|\varphi_{n,t_n})$ for every $(n,t_n)$-pair and $\delta_{-n}$. But by~(\ref{eq-p}) and~(\ref{eq-a}), the conclusion is immediate in view of the relation ${\cal W}^{\mbox d}_{n,t_n}(\varphi_{n,t_n})\subseteq {\cal W}^{\mbox a}_{n,t_n}(\varphi_{n,t_n})$. \qed

\noindent{\bf Proof of Proposition~\ref{p-in-a}: }It will suffice to prove $\hat B^{\mbox a}_{n,t_n}(\delta_{-n}|\varphi_{n,t_n})\subseteq \hat B^{\mbox d}_{n,t_n}(\delta_{-n}|\varphi_{n,t_n})$ for every $(n,t_n)$-pair and $\delta_{-n}$. But by~(\ref{eq-p}) and~(\ref{eq-a}), the conclusion is immediate in view of the relation ${\cal W}^{\mbox a}_{n,t_n}(\varphi_{n,t_n})\subseteq {\cal W}^{\mbox d}_{n,t_n}(\varphi_{n,t_n})$. \qed

\noindent{\bf Proof of Proposition~\ref{mip-ko}: }Suppose for continuous kernel $\kappa_{n,t_n}\in {\cal K}(A_{n,t_n},\Omega_{n,t_n},R_{n,t_n})$ and action distribution $\delta_{n,t_n}\in \Delta_{n,t_n}$, the value $s_{n,t_n}(\kappa_{n,t_n}(a_{n,t_n}))$ is strictly below $s_{n,t_n}(\kappa_{n,t_n}(a'_{n,t_n}))$ for some $a'_{n,t_n}\in A_{n,t_n}$ at a $\delta_{n,t_n}$-positive set of $a_{n,t_n}$'s. Let
\begin{equation}
\overline s_{n,t_n}=\sup_{a_{n,t_n}\in A_{n,t_n}}s_{n,t_n}(\kappa_{n,t_n}(a_{n,t_n})),
\end{equation}
which is finite as $s_{n,t_n}$ is a continuous map on the compact space $\Pi_{n,t_n}$. Our hypothesis indicates that $\delta_{n,t_n}(A'_{n,t_n})>0$ for $A'_{n,t_n}=\{a_{n,t_n}\in A_{n,t_n}|s_{n,t_n}(\kappa_{n,t_n}(a_{n,t_n}))<\overline s_{n,t_n}\}$. Note that $A'_{n,t_n}=\bigcup_{l=1}^{+\infty}A^l_{n,t_n}$, where
\begin{equation}
A^l_{n,t_n}=\left\{a_{n,t_n}\in A_{n,t_n}|s_{n,t_n}(\kappa_{n,t_n}(a_{n,t_n}))<\overline s_{n,t_n}-\frac{1}{l}\right\},\hspace*{.5in}\forall l=1,2,....
\end{equation}
So for some $l$, we have $\delta_{n,t_n}(A^l_{n,t_n})>1/l>0$. Identify for this $l$ an $a^l_{n,t_n}\in A_{n,t_n}$ so that \begin{equation}
s_{n,t_n}(\kappa_{n,t_n}(a^l_{n,t_n}))\geq\overline s_{n,t_n}-\frac{1}{2l^2}\geq \overline s_{n,t_n}-\frac{1}{2l}.
\end{equation}

Now let $\delta'_{n,t_n}\in {\cal P}(A_{n,t_n})$ be the Dirac measure on the point $a^l_{n,t_n}$. By this construction,
\begin{equation}
\begin{array}{l}
s_{n,t_n}\left(\int_{A_{n,t_n}}\kappa_{n,t_n}(a'_{n,t_n})\cdot \delta'_{n,t_n}(da'_{n,t_n})\right)=s_{n,t_n}(\kappa_{n,t_n}(a^l_{n,t_n}))\\
\;\;\;\;\;\;\geq\int_{A_{n,t_n}}s_{n,t_n}(\kappa_{n,t_n}(a_{n,t_n}))\cdot \delta_{n,t_n}(da_{n,t_n})+\delta_{n,t_n}(a^l_{n,t_n})/(2l)-1/(2l^2)\cdot (1-1/l)\\
\;\;\;\;\;\;>\int_{A_{n,t_n}}s_{n,t_n}(\kappa_{n,t_n}(a_{n,t_n}))\cdot \delta_{n,t_n}(da_{n,t_n}),
\end{array}\end{equation}
which, by $s_{n,t_n}$'s strong convexity with respect to $A_{n,t_n}$, is greater than
\begin{equation}
s_{n,t_n}\left(\int_{A_{n,t_n}}\kappa_{n,t_n}(a_{n,t_n})\cdot \delta_{n,t_n}(da_{n,t_n})\right).
\end{equation}
Therefore, the $s_{n,t_n}$-based $\varphi_{n,t_n}$ is individually prominent with respect to $A_{n,t_n}$. \qed

\noindent{\bf Proof of Proposition~\ref{mip-ok}: }Since $s_{n,t_n}$'s strong linearity implies its strong convexity, we know that $\varphi_{n,t_n}$ is individually prominent by Proposition~\ref{mip-ko}.

For mixture preservation, suppose given continuous kernel $\kappa_{n,t_n}\in {\cal K}(A_{n,t_n},\Omega_{n,t_n},R_{n,t_n})$ and action distribution $\delta_{n,t_n}\in \Delta_{n,t_n}$, the value $s_{n,t_n}(\kappa_{n,t_n}(a_{n,t_n}))$ is above $s_{n,t_n}(\kappa_{n,t_n}(a'_{n,t_n}))$ for every $a'_{n,t_n}\in A_{n,t_n}$ at $\delta_{n,t_n}$-almost every $a_{n,t_n}$. Then by the strong linearity of $s_{n,t_n}$,
\begin{equation}\begin{array}{l}
s_{n,t_n}\left(\int_{A_{n,t_n}}\kappa_{n,t_n}(a'_{n,t_n})\cdot\delta'_{n,t_n}(da'_{n,t_n})\right)=\int_{A_{n,t_n}}s_{n,t_n}(\kappa_{n,t_n}(a'_{n,t_n}))\cdot\delta'_{n,t_n}(da'_{n,t_n})\\
\;\leq \int_{A_{n,t_n}}s_{n,t_n}(\kappa_{n,t_n}(a_{n,t_n}))\cdot\delta_{n,t_n}(da_{n,t_n})=s_{n,t_n}\left(\int_{A_{n,t_n}}\kappa_{n,t_n}(a_{n,t_n})\cdot\delta_{n,t_n}(da_{n,t_n})\right),
\end{array}\end{equation}
for any $\delta'_{n,t_n}\in \Delta_{n,t_n}$.
Thus, the $s_{n,t_n}$-based $\varphi_{n,t_n}$ is mixture-preserving with respect to $A_{n,t_n}$.\qed

\noindent{\bf Proof of Proposition~\ref{heroic}: }Under Assumption~\ref{compact-state}, the state space $\Omega_{n,t_n}$ is compact and hence separable. Since $R_{n,t_n}$ is finite, ${\cal P}(R_{n,t_n})$ is homeomorphic to a compact subset of a finite-dimensional  Euclidean space. Therefore, $\Pi_{n,t_n}\equiv {\cal C}(\Omega_{n,t_n},{\cal P}(R_{n,t_n}))$ equipped with the uniform metric based on the Prokhorov metric for $R_{n,t_n}$ is homeomorphic to a subset of the infinite-dimensional Euclidean space $\Re^\infty$. The latter, being equipped with the metric induced from the $l^\infty$-norm, is a real topological vector space. Now, since $\kappa_{n,t_n}\in {\cal K}(A_{n,t_n},\Omega_{n,t_n},R_{n,t_n})$ is continuous from $A_{n,t_n}$ to $\Pi_{n,t_n}$, it can be treated as a continuous and hence measurable mapping from $A_{n,t_n}$ to $\Re^\infty$. That is, $\kappa_{n,t_n}$ is equivalent to a random variable say $K_{n,t_n}$ with domain in the probability space $(A_{n,t_n},{\cal B}(A_{n,t_n}),\delta_{n,t_n})$ and range in the measurable space $(\Re^\infty,{\cal B}(\Re^\infty))$. Incidentally,~(\ref{s-concave}) can be written as
\begin{equation}
s_{n,t_n}(\mathbb{E}[K_{n,t_n}])\geq \mathbb{E}[s_{n,t_n}(K_{n,t_n})].
\end{equation}
Then, using the general Jensen's inequality, we can deduce that ordinary concavity/convexity will lead to the so-called strong concavity/convexity. \qed

\noindent{\bf Proof of Theorem~\ref{ep-ea}: }Let continuous kernel $\kappa_{n,t_n}\in {\cal K}(A_{n,t_n},\Omega_{n,t_n},R_{n,t_n})$ be given. By~(\ref{s0-form}) and~(\ref{form1}), we know with a utility function $u_{n,t_n}\in {\cal C}(R_{n,t_n},\Re)$ and a single prior $\rho_{n,t_n}\in {\cal P}(R_{n,t_n})$, the traditional game's satisfaction function $s_{n,t_n}$ will produce, at each $a_{n,t_n}\in A_{n,t_n}$,
\begin{equation}
s_{n,t_n}(\kappa_{n,t_n}(a_{n,t_n}))=\int_{\Omega_{n,t_n}}\{\int_{R_{n,t_n}}u_{n,t_n}(r)\cdot [\kappa_{n,t_n}(a_{n,t_n},\omega)](dr)\}\cdot\rho_{n,t_n}(d\omega).
\end{equation}
The above is Borel-measurable in $a_{n,t_n}$ because $u_{n,t_n}$ is the limit of a sequence of simple real-valued functions on $R_{n,t_n}$.
For any distribution $\delta_{n,t_n}\in \Delta_{n,t_n}$, by understanding $u_{n,t_n}$ as a sequence of simple real-valued functions on $R_{n,t_n}$, we can obtain
\begin{equation}\begin{array}{l}
\int_{A_{n,t_n}}s_{n,t_n}(\kappa_{n,t_n}(a_{n,t_n}))\cdot\delta_{n,t_n}(da_{n,t_n})\\
\;\;\;\;\;\;=\int_{A_{n,t_n}}\int_{\Omega_{n,t_n}}\{\int_{R_{n,t_n}}u_{n,t_n}(r)\cdot [\kappa_{n,t_n}(a_{n,t_n},\omega)](dr)\}\cdot\rho_{n,t_n}(d\omega)\cdot\delta_{n,t_n}(da_{n,t_n})\\
\;\;\;\;\;\;=\int_{R_{n,t_n}}u_{n,t_n}(r)\cdot[\int_{A_{n,t_n}}\int_{\Omega_{n,t_n}}\kappa_{n,t_n}(a_{n,t_n},\omega)\cdot\rho_{n,t_n}(d\omega)\cdot\delta_{n,t_n}(da_{n,t_n})](dr)\\
\;\;\;\;\;\;=\int_{\Omega_{n,t_n}}\{\int_{R_{n,t_n}}u_{n,t_n}(r)\cdot[\int_{A_{n,t_n}}\kappa_{n,t_n}(a_{n,t_n},\omega)\cdot\delta_{n,t_n}(da_{n,t_n})](dr)\}\cdot\rho_{n,t_n}(d\omega)\\
\;\;\;\;\;\;=s_{n,t_n}\left(\int_{A_{n,t_n}}\kappa_{n,t_n}(a_{n,t_n})\cdot\delta_{n,t_n}(da_{n,t_n})\right).
\end{array}\end{equation}
So the satisfaction functions $s_{n,t_n}$ for the traditional game are strongly linear with respect to their corresponding $A_{n,t_n}$'s. 
Due to Propositions~\ref{mip-ko} and~\ref{mip-ok}, each $\varphi_{n,t_n}$ induced by $s_{n,t_n}$ will be both individually prominent and mixture-preserving with respect to $A_{n,t_n}$. By Propositions~\ref{a-in-p} and~\ref{p-in-a}, we then have ${\cal E}^{\mbox d}={\cal E}^{\mbox a}$. Meanwhile, the existence of both types of equilibria can come from Theorem~\ref{t-existence}.\qed


\noindent{\bf Proof of Theorem~\ref{pure-pa}: }Let us use 
$1_{a_{-n}}$ for $(1_{a_{m,t_m}})_{m\neq n,t_m\in T_m}$. Suppose $1_a\in {\cal E}^{\mbox d}$ for some $a\equiv(a_{n,t_n})_{n\in N,t_n\in T_n}$ in the product action space $\prod_{n\in N}\prod_{t_n\in T_n}A_{n,t_n}$. Then, by~(\ref{maximal-k0}) and~(\ref{zia-a}),
\begin{equation}
\pi^{\mbox d}_{n,t_n}(\delta'_{n,t_n},1_{a_{-n}})\not\succ_{n,t_n}\pi^{\mbox d}_{n,t_n}(1_{a_{n,t_n}},1_{a_{-n}}),\hspace*{.8in}\forall n\in N,\;t_n\in T_n,\; \delta'_{n,t_n}\in \Delta_{n,t_n}.
\end{equation}
Since any $1_{A_{n,t_n}}$ is merely a subset of its corresponding $\Delta_{n,t_n}$, this results in
\begin{equation}
\pi^{\mbox d}_{n,t_n}(1_{a'_{n,t_n}},1_{a_{-n}})\not\succ_{n,t_n}\pi^{\mbox d}_{n,t_n}(1_{a_{n,t_n}},1_{a_{-n}}),\hspace*{.8in}\forall  n\in N,\;t_n\in T_n,\;  a'_{n,t_n}\in A_{n,t_n}.
\end{equation}
Due to~(\ref{jit}), this is the same as
\begin{equation}
\pi^{\mbox a}_{n,t_n}(a'_{n,t_n},1_{a_{-n}})\not\succ_{n,t_n}\pi^{\mbox a}_{n,t_n}(a_{n,t_n},1_{a_{-n}}),\hspace*{.8in}\forall  n\in N,\;t_n\in T_n,\; a'_{n,t_n}\in A_{n,t_n}.
\end{equation}
But~(\ref{maximal0}) to~(\ref{zia-p}) will give this the meaning of $1_a\in {\cal E}^{\mbox a}$. \qed

\noindent{\bf Proof of Proposition~\ref{much-ep}: }Due to~(\ref{radical-form}) and~(\ref{useful-aa}), 
\begin{equation}\label{such-e}
s^{\mbox d}_{n,t_n}(\delta_{n,t_n},\delta_{-n})=\sup_{\rho\in P_{n,t_n}}s^0_{n,t_n}\left(\pi^{\mbox d}_{n,t_n}(\delta_{n,t_n},\delta_{-n}),\rho\right),
\end{equation}
where $s^0_{n,t_n}$ is defined at~(\ref{s0-form}). We have from~(\ref{useful-ppp}) and~(\ref{useful-a}) that
\begin{equation}\label{usefully}
s^0_{n,t_n}\left(\pi^{\mbox d}_{n,t_n}(\delta_{n,t_n},\delta_{-n}),\rho\right)=\int_{A_{n,t_n}}s^0_{n,t_n}\left(\pi^{\mbox a}_{n,t_n}(a_{n,t_n},\delta_{-n}),\rho\right)\cdot\delta_{n,t_n}(da_{n,t_n}).
\end{equation}
This means that $s^0_{n,t_n}(\pi^{\mbox d}_{n,t_n}(\cdot,\delta_{-n}),\rho)$ is linear. With~(\ref{such-e}) showing it to be the supremum of linear functions, we can thus conclude that $s^{\mbox d}_{n,t_n}(\cdot,\delta_{-n})$ is convex.

From~(\ref{usefully}), it is also clear that
\begin{equation}\label{paat}
s^0_{n,t_n}\left(\pi^{\mbox d}_{n,t_n}(1_{a_{n,t_n}},\delta_{-n}),\rho\right)=s^0_{n,t_n}\left(\pi^{\mbox a}_{n,t_n}(a_{n,t_n},\delta_{-n}),\rho\right).
\end{equation}
Hence,~(\ref{radical-form}),~(\ref{useful-pp}), and~(\ref{useful-aa}) will together lead to the desired result. \qed

\noindent{\bf Proof of Lemma~\ref{l-okla}: }The compactness of $X$ will lead to that of ${\cal P}(X)$. Because $f$ is continuous, some $\xi^0\in {\cal P}(X)$ will have achieved the supremum. The measure's support $\mbox{supp}(\xi^0)$ is a closed subset of $X$. We are done if it is already a singleton. Suppose otherwise. Since $X$ is compact in a finite-dimensional real Euclidean space say $\Re^k$, it is bounded. Some closed rectangle $Y^0$ with a finite total edge length say $e^0\equiv e^0_1+\cdots+e^0_k>0$ must have covered $\mbox{supp}(\xi^0)$. Without loss of generality, suppose $e^0_1$ is the largest among all of $Y^0$'s edge lengths. Note that $e^0_1\geq e^0/k$.

Consider the closed rectangle $Y^0_L$ which takes the left half of $Y^0$'s first edge and the rest of its edges. Note that the closure of $Y^0\setminus Y^0_L$ is $Y^0_R$, the closed rectangle which takes the right half of $Y^0$'s first edge and the rest of its edges. For either the left- or right-half closed rectangle, the total edge length $e^1$ is at most $(2k-1)/(2k)$ times $e^0$. Suppose $\xi^0(X\cap Y^0_L)=0$. Then, $\mbox{supp}(\xi^0)$ is indeed covered by the smaller rectangle $Y^0_R$. Suppose $\xi^0(X\cap (Y^0\setminus Y^0_L))=0$. Then, $\mbox{supp}(\xi^0)$ is indeed covered by the smaller rectangle $Y^0_L$.

If neither is true, then we have both $p^0_L\equiv\xi^0(X\cap Y^0_L)>0$ and $p^0_R\equiv \xi^0(X\cap (Y^0\setminus Y^0_L))=1-p^0_L>0$. Consider members of ${\cal P}(X)$,
\begin{equation}
\xi^0_L\equiv \frac{1}{p^0_L}\cdot \xi^0|_{X\cap Y^0_L},\;\;\;\;\;\;\mbox{ and }\;\;\;\;\;\;
\xi^0_R\equiv \frac{1}{p^0_R}\cdot \xi^0|_{X\cap (Y^0\setminus Y^0_L)}.
\end{equation}
Note that $\mbox{supp}(\xi^0_L)\subseteq Y^0_L$ and $\mbox{supp}(\xi^0_R)\subseteq Y^0_R$. Also,
\begin{equation}
\xi^0=p^0_L\cdot \xi^0_L+p^0_R\cdot \xi^0_R.
\end{equation}
By the convexity of $f$,
\begin{equation}
p^0_L\cdot f(\xi^0_L)+p^0_R\cdot f(\xi^0_R)\geq f(\xi^0).
\end{equation}
Since $\xi^0$ has already achieved the supremum, both $\xi^0_L$ and $\xi^0_R$ must have too.

So no matter whichever one of the above three cases is present, we will be able to identify some supremum-reaching $\xi^1\in {\cal P}(X)$, whose support $\mbox{supp}(\xi^1)$ is covered by a closed rectangle $Y^1$, inside the original support-covering rectangle $Y^0$ and with a total edge length $e^1$ that is at most $(2k-1)/(2k)$ times the original $e^0$.

We can repeat the whole procedure from $\xi^0$ to $\xi^1$ incessantly. Then, we will get a sequence $(\xi^n)_{n=0,1,...}$ of supremum-reaching distributions in ${\cal P}(X)$, whose supports are covered by increasingly nested closed rectangles $Y^n$ with total edge lengths satisfying
\begin{equation}
e^{n+1}\leq \frac{2k-1}{2k}\cdot e^n,\hspace*{.8in}\forall n=0,1,....
\end{equation}
Thus, $\lim_{n\rightarrow +\infty}e^n=0$, and hence $(X\cap Y^n)_{n=0,1,...}$ is a nested sequence of nonempty closed sets with shrinking dimensions in the compact set $X$. There must be one and only one member say $x\in X$ inside all of the rectangles $Y^n$.

For any given $\epsilon>0$, we know $e^n$ will be smaller than it when $n$ is large enough. For any closed subset $F$ of $\Re^k$, let $F^\epsilon$ be the set of all points that are within $\epsilon$-distance of $F$, where the distance between two points in $\Re^k$ is measured through the $l_1$-norm. Because $x$ is inside $Y^n$, whose total edge length $e^n$ is below $\epsilon$, we can conclude that
\begin{equation}
x\notin X\cap F^\epsilon\Longrightarrow X\cap F\cap Y^n=\emptyset,
\end{equation}
which will further lead to $(X\cap F)\cap \mbox{supp}(\xi^n)=\emptyset$ because $\mbox{supp}(\xi^n)$ is in $Y^n$. So depending on whether or not $x\in X\cap F^\epsilon$, we have either
\begin{equation}\label{either-a}
1_x(X\cap F^\epsilon)+\epsilon=1+\epsilon>1\geq \xi^n(X\cap F),
\end{equation}
or
\begin{equation}\label{or-a}
1_x(X\cap F^\epsilon)+\epsilon=\epsilon>0=\xi^n(X\cap F).
\end{equation}
But~(\ref{either-a}) and~(\ref{or-a}) together mean that
\begin{equation}
\psi_{X}(1_x,\xi^n)\leq \epsilon.
\end{equation}
That is, when measured by the Prokhorov metric $\psi_X$ adopted to the distribution space ${\cal P}(X)$, the sequence $(\xi^n)_{n=0,1,...}$ converges to $1_x$. By $f$'s continuity,
\begin{equation}
f(\xi^0)=f(\xi^1)=\cdots=f(1_x),
\end{equation}
and hence the Dirac measure $1_x$ has achieved the supremum. \qed

\noindent{\bf Proof of Theorem~\ref{chedan}: }Combining Proposition~\ref{much-ep} and~(\ref{convex-con}), we obtain
\begin{equation}\label{convex-con-1}
\sup_{a'_{n,t_n}\in A_{n,t_n}}s^{\mbox a}_{n,t_n}(a'_{n,t_n},\delta_{-n})=\sup_{\delta'_{n,t_n}\in 1_{A_{n,t_n}}}s^{\mbox d}_{n,t_n}(\delta'_{n,t_n},\delta_{-n})=\sup_{\delta'_{n,t_n}\in \Delta_{n,t_n}}s^{\mbox d}_{n,t_n}(\delta'_{n,t_n},\delta_{-n}).
\end{equation}
Now suppose $1_a\in {\cal E}^{\mbox a}$ for some $a\equiv (a_{n,t_n})_{n\in N,t_n\in T_n}$ in the space $\prod_{n\in N}\prod_{t_n\in T_n}A_{n,t_n}$. Then, due to~(\ref{criterion-p}),
\begin{equation}
s^{\mbox a}_{n,t_n}(a_{n,t_n},1_{a_{-n}})=\sup_{a'_{n,t_n}\in A_{n,t_n}}s^{\mbox a}_{n,t_n}(a'_{n,t_n},1_{a_{-n}}),\hspace*{.8in}\forall n\in N,\;t_n\in T_n.
\end{equation}
By~(\ref{convex-con-1}), this will lead to
\begin{equation}
s^{\mbox a}_{n,t_n}(a_{n,t_n},1_{a_{-n}})=\sup_{\delta'_{n,t_n}\in \Delta_{n,t_n}}s^{\mbox d}_{n,t_n}(\delta'_{n,t_n},1_{a_{-n}}),\hspace*{.8in}\forall n\in N,\;t_n\in T_n.
\end{equation}
But due to Proposition~\ref{much-ep}, we have further that
\begin{equation}
s^{\mbox d}_{n,t_n}(1_{a_{n,t_n}},1_{a_{-n}})=\sup_{\delta'_{n,t_n}\in \Delta_{n,t_n}}s^{\mbox d}_{n,t_n}(\delta'_{n,t_n},1_{a_{-n}}),\hspace*{.8in}\forall n\in N,\;t_n\in T_n.
\end{equation}
According to~(\ref{criterion-a}), this exactly means that $1_a\in {\cal E}^{\mbox d}$. The above results with $1_A\cap {\cal E}^{\mbox a}\subseteq 1_A\cap {\cal E}^{\mbox d}$. We can reach our desired conclusion by combining this with
Theorem~\ref{pure-pa}. \qed

\section{A Special Enterprising Game}\label{app-dc}

\subsection{No Ambiguity on Opponent-type Distributions}\label{okba}

Due to Theorem~\ref{chedan}'s unification of its two types of pure equilibria, we only have to deal with pure action-based equilibria for the enterprising game. 
Define
\begin{equation}\label{s-tilde}
\tilde s_{n,t_n}(a_{n,t_n},a_{-n})=s^{\mbox a}_{n,t_n}(a_{n,t_n},1_{a_{-n}}),
\end{equation}
where $s^{\mbox a}_{n,t_n}$ is given by~(\ref{useful-pp}) and $a_{-n}\in A_{-n}\equiv \prod_{m\neq n}A_m\equiv\prod_{m\neq n}\prod_{t_m\in T_m}A_{m,t_m}$ represents opponents' pure-action profile. Due to~(\ref{radical-form}),~(\ref{mechanism}) to~(\ref{vec-p}),~(\ref{useful-pp}),~(\ref{s0-form}), and~(\ref{s-tilde}),
\begin{equation}\label{amazing}
\tilde s_{n,t_n}(a_{n,t_n},a_{-n})=\sup_{\rho\in P_{n,t_n}} w_{n,t_n}(a_{n,t_n},a_{-n},\rho),
\end{equation}
where
\begin{equation}\label{ww-def}
w_{n,t_n}(a_{n,t_n},a_{-n},\rho)=\sum_{t_{-n}\in T_{-n}}\int_{\Omega_{t_n,t_{-n}}}\tilde u_{n,t_n,t_{-n}}(a_{n,t_n},a_{-n,t_{-n}},\omega)\cdot\rho|_{\Omega_{t_n,t_{-n}}}(d\omega),
\end{equation}
and $\tilde u_{n,t_n,t_{-n}}\equiv u_{n,t_n}\circ r_{n,t_n,t_{-n}}$ is the continuous real-valued composite payoff-utility function defined on the compact $A_t\times \Omega_t\equiv A_{n,t_n}\times A_{-n,t_{-n}}\times \Omega_{t_n,t_{-n}}$.

By~(\ref{criterion-p}),  we will have pure strategy $1_a\in 1_A\cap {\cal E}^{\mbox a}$ if and only if $a_{n,t_n}\in\tilde B_{n,t_n}(a_{-n})$, where
\begin{equation}\label{best-def}
\tilde B_{n,t_n}(a_{-n})=\left\{a_{n,t_n}\in A_{n,t_n}|\tilde s_{n,t_n}(a_{n,t_n},a_{-n})\geq \tilde s_{n,t_n}(a'_{n,t_n},a_{-n})\;\;\forall a'_{n,t_n}\in A_{n,t_n}\right\}.
\end{equation}
Thus, $1_a\in 1_A$ will be a pure equilibrium for $\Gamma$ if and only if $a\in A$ is that for a corresponding agent-based normal-form game where payoffs are the $\tilde s_{n,t_n}$'s given at~(\ref{amazing}) and~(\ref{ww-def}).

We find a special case to be further analyzable. In this case,\\
\indent\M (a) all the action spaces $A_{n,t_n}$'s across different $t_n$'s are the same;\\
\indent\M (b) there is a compact metric space $\tilde\Omega$, so that every $\Omega_t$ is merely $\{t\}\times\tilde{\Omega}$;\\
\indent\M (c) for each player $n\in N$ and type $t_n\in T_n$, there are distribution $p_{n,t_n}\equiv (p_{n,t_n|t_{-n}})_{t_{-n}\in T_{-n}}$ and nonempty subset ${\cal Q}_{n,t_n}$ of $({\cal P}(\tilde{\Omega}))^{T_{-n}}$, so that the prior set used in the definition~(\ref{radical-form}), as a nonempty subset of ${\cal P}(\Omega_{n,t_n})\equiv{\cal P}(T_{-n}\times\tilde{\Omega})$, satisfies
\begin{equation}\label{sp-form}
P_{n,t_n}=\left\{p_{n,t_n|t_{-n}} \times\nu_{t_{-n}}|\nu\equiv(\nu_{t_{-n}})_{t_{-n}\in T_{-n}}\in {\cal Q}_{n,t_n}\right\}.
\end{equation}
In most works on games involving incomplete information, (a) was assumed. Due to this point, we can just use $A_n$ for the action space of player $n$ regardless of his type and now use $A_{-n}$ for $\prod_{m\neq n}A_m$. By (b), every $\Omega_{n,t_n}=T_{-n}\times \tilde{\Omega}$ and $\Omega=T\times\tilde{\Omega}$, indicating that clear cuts can be made between players' types and other external factors which affect all players. For convenience, we still call each $\tilde{\omega}$ a state. The domain of every payoff-utility function $\tilde u_{n,t_n,t_{-n}}$ is $A_n\times A_{-n}\times \Omega_{t_n,t_{-n}}$. Because different $\Omega_{t_n,t_{-n}}$'s are disjoint, we can patch up all the $\tilde u_{n,t_n,t_{-n}}$ to obtain $\tilde u_n:A_n\times A_{-n}\times \Omega\rightarrow \Re$. But with (b), $\omega=(t_n,t_{-n},\tilde{\omega})$. So the just gotten $\tilde u_n(a_n,a_{-n},\omega)$ can be further rewritten as $\tilde u_{n,t_n,t_{-n}}(a_n,a_{-n},\tilde{\omega})$. This function is still continuous on a compact space. Meanwhile, (c) means the following. With probability $p_{n,t_n|t_{-n}}$ player $n$ believes unambiguously that opponents' type profile is at some $t_{-n}$; his ambiguity on other external factors, on the other hand, is reflected by the membership of the prior vector $\nu\equiv (\nu_{t_{-n}})_{t_{-n}\in T_{-n}}$ in the set ${\cal Q}_{n,t_n}$.

With~(\ref{sp-form}) in place,~(\ref{amazing}) and~(\ref{ww-def}) can be rewritten as
\begin{equation}\label{s-def-new}
\tilde s_{n,t_n}(a_{n,t_n},a_{-n})=\sup_{\nu\in {\cal Q}_{n,t_n}}\tilde w_{n,t_n}(a_{n,t_n},a_{-n},\nu),
\end{equation}
where $a_{n,t_n}\in A_n$ and $a_{-n}\equiv (a_{m,t_m})_{m\neq n,t_m\in T_m}\in \prod_{m\neq n}A_m^{\;T_m}$; also,
\begin{equation}\label{w-def-new}
\tilde w_{n,t_n}(a_n,a_{-n},\nu)=\sum_{t_{-n}\in T_{-n}}p_{n,t_n|t_{-n}}\cdot\tilde v_{n,t_n,t_{-n}}(a_n,a_{-n,t_{-n}},\nu_{t_{-n}}),
\end{equation}
where $a_n\in A_n$ and
\begin{equation}\label{v-def-new}
\tilde v_{n,t_n,t_{-n}}(a_n,a_{-n},\mu)=\int_{\tilde\Omega}\tilde u_{n,t_n,t_{-n}}(a_n,a_{-n},\tilde\omega)\cdot\mu(d\tilde\omega),
\end{equation}
where this time $a_{-n}\in A_{-n}$ and $\mu\in {\cal P}(\tilde\Omega)$.

Later, it might help to understand the ${\cal Q}_{n,t_n}$ used in~(\ref{sp-form}) as follows:
\begin{equation}\label{p-def-new}
{\cal Q}_{n,t_n}=\left(\prod_{t_{-n}\in T_{-n}}\tilde P_{n,t_n,t_{-n}}\right)\bigcap {\cal K}_{n,t_n},
\end{equation}
where $\tilde P_{n,t_n,t_{-n}}\subseteq {\cal P}(\tilde{\Omega})$ for each $t_{-n}\in T_{-n}$ and ${\cal K}_{n,t_n}\subseteq ({\cal P}(\tilde{\Omega}))^{T_{-n}}$. For instance, we can always let $\tilde P_{n,t_n,t_{-n}}={\cal P}(\tilde{\Omega})$ and ${\cal K}_{n,t_n}={\cal Q}_{n,t_n}$.
Our special enterprising game is denotable by $\Gamma\equiv (N,(T_n)_{n\in N},(A_n)_{n\in N},\tilde\Omega,(\tilde u_{n,t})_{n\in N,t\in T}$, $(p_{n,t_n})_{n\in N,t_n\in T_n},(\tilde P_{n,t})_{n\in N,t\in T},({\cal K}_{n,t_n})_{n\in N,t_n\in T_n})$. In it, there is no ambiguity on the opponent-type distribution. Rather, each $(n,t_n)$-player is uncertain about distributions of the non-type factors.


\subsection{Enter Strategic Complementarities}\label{special}

Let each action space $A_n$ be a finite set or compact interval within the real line $\Re$, and equip it with the ordinary order. For each $A_{-n}$, we adopt the component-wise partial order. For two partially ordered sets $X$ and $Y$, we use ${\cal M}(X,Y)$ to denote the subset of $Y^X$ that contains all monotone mappings from $X$ to $Y$, i.e., mappings $y:X\rightarrow Y$ so that $y(x^1)\leq y(x^2)$ whenever $x^1,x^2\in X$ satisfy $x^1\leq x^2$. We let the component-wise partial order be adopted for $\prod_{m\neq n}{\cal M}(T_m,A_m)$ as well.

We further suppose that the state space  $\tilde{\Omega}=\prod_{k=1}^{\bar k}\tilde{\Omega}_k$ where $\bar k$ is a natural number and each $\tilde{\Omega}_k$ is a finite set or compact interval within the real line $\Re$. Also, we equip $\tilde{\Omega}$ with the component-wise partial order. For the state-distribution space ${\cal P}(\tilde\Omega)$, we adopt the usual stochastic order, so that $\mu^1,\mu^2\in {\cal P}(\tilde\Omega)$ is considered to satisfy $\mu^1\leq \mu^2$ when for any monotone function $q\in {\cal M}(\tilde\Omega,\Re)$ that is integrable under both $\mu^1$ and $\mu^2$,
\begin{equation} \int_{\tilde\Omega}q(\tilde\omega)\cdot\mu^1(d\tilde\omega)\leq \int_{\tilde\Omega}q(\tilde\omega)\cdot\mu^2(d\tilde\omega).
\end{equation}

The above is equivalent to $\mu^1(\tilde{\Omega}\cap U)\leq\mu^2(\tilde{\Omega}\cap U)$ for every of $\Re^{\bar k}$'s upper sets $U$, a set satisfying $\omega^2\in U$ whenever $\omega^1\in U$ and $\omega^1\leq \omega^2$; see, e.g., Section 6.B.1 of Shaked and Shanthikumar \cite{SS07}. Given $\mu^1,\mu^2\in {\cal P}(\tilde{\Omega})$, we can construct $\mu^1\vee\mu^2\in {\cal P}(\tilde{\Omega})$ by forcing its value at $\tilde{\Omega}\cap K$ for every upper rectangular set $K$ be $\mu^1(\tilde{\Omega}\cap K)\vee \mu^2(\tilde{\Omega}\cap K)$. Similarly, we can obtain $\mu^1\wedge\mu^2\in {\cal P}(\tilde{\Omega})$. Thus, ${\cal P}(\tilde{\Omega})$ is a lattice under the usual stochastic order. For a partial order between sublattices of ${\cal P}(\tilde{\Omega})$, we can adopt
the induced set order; see Theorem 2.4.1 of Topkis \cite{T98}. For sublattices $P^1$ and $P^2$ of ${\cal P}(\tilde{\Omega})$, we consider $P^1\leq P^2$ in the induced set order sense when $\mu^1\in P^1$ and $\mu^2\in P^2$ will always lead to
\begin{equation}
\mu^1\wedge \mu^2\in P^1,\hspace*{.8in}\mu^1\vee\mu^2\in P^2.
\end{equation}

We can adopt the component-wise partial order for each lattice $({\cal P}(\tilde{\Omega}))^{T_{-n}}$. This way, $\nu^1\equiv (\nu^1_{t_{-n}})_{t_{-n}\in T_{-n}},\nu^2\equiv(\nu^2_{t_{-n}})_{t_{-n}\in T_{-n}}\in ({\cal P}(\tilde{\Omega}))^{T_{-n}}$ are considered to satisfy $\nu^1\leq \nu^2$ when
\begin{equation}
\nu^1_{t_{-n}}\leq\nu^2_{t_{-n}},\hspace*{.8in}\forall t_{-n}\in T_{-n}.
\end{equation}
This partial order certainly applies to the smaller lattice ${\cal M}(T_{-n},{\cal P}(\tilde{\Omega}))$ as well. For the latter's sublattices, we can similarly adopt the induced set order.

We now make further assumptions on the game's model primitives.

\begin{e-assumption}\label{as-bongaji}
	For any $n\in N$, $t_n\in T_n$, $t_{-n}\in T_{-n}$, $a_n\in A_n$, and $a_{-n}\in A_{-n}$, the payoff-utility $\tilde u_{n,t_n,t_{-n}}(a_n,a_{-n},\tilde{\omega})$ is increasing in $\tilde{\omega}\in\tilde{\Omega}$.
\end{e-assumption}	

\begin{e-assumption}\label{as-utility}
	For any $n\in N$, the payoff-utility function $\tilde u_{n,t_n,t_{-n}}(a_n,a_{-n},\tilde{\omega})$ has increasing differences between $a_n\in A_n$ and $(t_n,t_{-n},a_{-n},\tilde{\omega})\in T_n\times T_{-n}\times A_{-n}\times\tilde{\Omega}$, as well as between $(t_n,t_{-n},a_{-n})\in T_n\times T_{-n}\times A_{-n}$ and $\tilde\omega\in\tilde\Omega$. 
\end{e-assumption}

\begin{e-assumption}\label{as-mono-probability}
	For any $n\in N$, the distribution $p_{n,t_n}$ is monotone in $t_n\in T_n$ in the usual stochastic order, so that for any $t^1_n,t^2_n\in T_n$ with $t^1_n\leq t^2_n$ and any $f\in {\cal M}(T_{-n},\Re)$,
	\[ \sum_{t_{-n}\in T_{-n}}p_{n,t^1_n|t_{-n}}\cdot f_{t_{-n}}\leq  \sum_{t_{-n}\in T_{-n}}p_{n,t^2_n|t_{-n}}\cdot f_{t_{-n}}.\]
\end{e-assumption}

\begin{e-assumption}\label{as-prior}
	For any $n\in N$, $t_n\in T_n$, and $t_{-n}\in T_{-n}$, the prior set $\tilde P_{n,t_n,t_{-n}}$ is a sublattice of the lattice ${\cal P}(\tilde{\Omega})$.
\end{e-assumption}

\begin{e-assumption}\label{as-mono-prior}
	For any $n\in N$ and $t_{-n}\in T_{-n}$, the prior set $\tilde P_{n,t_n,t_{-n}}$ is increasing in $t_n\in T_n$.
\end{e-assumption}

Monotonic Assumption~\ref{as-bongaji} essentially associates higher $\tilde\omega$ values with better payoffs. In Monotonic Assumption~\ref{as-utility}, the payoff-utility function's increasing differences between player $n$'s own action $a_n$ and the type-action profile $(t_n,t_{-n},a_{-n})$ is quite anticipated for a game involving strategic complementarites. They indicate the increasing efficiency of a player under ever more friendly environments. These properties are also required in the traditional expected-utility version as well; see, e.g., van Zandt and Vives \cite{VZV07}. The full plate of increasing differences involving the newly added factor $\tilde{\omega}$, which resemble those for the action $a_n$, suggest that the latter should bear the interpretation of not only an efficiency booster but also somehow a surrogate action.

Note that the action space $A_n$ is a subset of the single-dimensional real line; also, $\tilde u_{n,t_n,t_{-n}}$ is already assumed to be continuous. So we have no need for additional supermodularity and continuity requirements on $\tilde u_{n,t_n,t_{-n}}(\cdot,a_{-n},\tilde{\omega})$. For any $\tilde u^0_{n,t_n,t_{-n}}(a_n,a_{-n})$ already suitable as a payoff function for the traditional game, $\tilde u_{n,t_n,t_{-n}}(a_n,a_{-n},\tilde{\omega})$ defined in terms of
\begin{equation}
\tilde u_{n,t_n,t_{-n}}(a_n,a_{-n},\tilde{\omega})=\tilde u^0_{n,t_n,t_{-n}}(a_n,a_{-n})+(\alpha_n t_n+\beta_n t_{-n}+\gamma_n a_n+\upsilon_n a_{-n}+\zeta_n)\cdot\tilde\omega,
\end{equation}
where $\alpha_n$, $\beta_n$, $\gamma_n$, and $\upsilon_n$ are positive constants, and $\zeta_n$ is a constant guaranteeing the positivity of the entire multiplier in front of $\tilde{\omega}$,  will satisfy Monotonic Assumption~\ref{as-utility}.

Meanwhile, Monotonic Assumption~\ref{as-mono-probability} suggests that a player's own type is positively correlated with his opponents' types. It has been assumed for the traditional game as well; see van Zandt and Vives \cite{VZV07}. Finally, Monotonic Assumptions~\ref{as-prior} and~\ref{as-mono-prior} collectively indicate that a player's own type is positively correlated with the external factors. Taken together, the latter two points both highlight the informational value of a player's own type.

\subsection{Monotone Pure Equilibria}\label{pureq}

Now we can obtain an intermediate result of the order-theoretic nature.

\begin{proposition}\label{common}
	For $\tilde v_{n,t_n,t_{-n}}(a_n,a_{-n},\mu)$ defined at~(\ref{v-def-new}), it is both supermodular and submodular in $\mu\in {\cal P}(\tilde{\Omega})$. Also, it has increasing differences between $a_n\in A_n$ and $(t_n,t_{-n},a_{-n},\mu)$ $\in T_n\times T_{-n}\times A_{-n}\times {\cal P}(\tilde{\Omega})$, as well as between $(t_n,t_{-n},a_{-n})\in T_n\times T_{-n}\times A_{-n}$ and $\mu\in {\cal P}(\tilde{\Omega})$.
\end{proposition}

To go any further, however, we find it necessary to consider separately two special scenarios. In scenario A, the distributions $p_{n,t_n}$ are independent of $t_n$. We can thus use $p^A_{n|t_{-n}}$ to stand for each probability $p_{n,t_n|t_{-n}}$. Also, the prior sets $\tilde P_{n,t_n,t_{-n}}$ can be some general $\tilde P^A_{n,t_n,t_{-n}}$'s. However, each ${\cal K}_{n,t_n}$ is equal to the set of monotone mappings ${\cal M}(T_{-n},{\cal P}(\tilde{\Omega}))$. In view of~(\ref{p-def-new}), the latter two facts together lead to
\begin{equation}\label{p-def-new-a}
{\cal Q}^A_{n,t_n}\equiv\left(\prod_{t_{-n}\in T_{-n}}\tilde P^A_{n,t_n,t_{-n}}\right)\bigcap {\cal M}(T_{-n},{\cal P}(\tilde{\Omega})).
\end{equation}
Here, player $n$ should not expect to gain from the identity of his own type $t_n$ any information about opponents' types $t_{-n}$; yet, he should anticipate the latter types to be positively correlated with the external factor $\tilde{\omega}$.

In scenario B, the probabilities $p_{n,t_n|t_{-n}}$'s can be some general $p^B_{n,t_n|t_{-n}}$'s. However, the prior set $\tilde P_{n,t_n,t_{-n}}$ is independent of $t_{-n}$, and hence is representable by $\tilde P^B_{n,t_n}$. In addition, each ${\cal K}_{n,t_n}$ is equal to $1(T_{-n},{\cal P}(\tilde\Omega))$, the set of constant mappings from $T_{-n}$ to ${\cal P}(\tilde\Omega)$. Note that $1(T_{-n},{\cal P}(\tilde\Omega))\subseteq {\cal M}(T_{-n},{\cal P}(\tilde{\Omega}))$. In view of~(\ref{p-def-new}), the above would  lead to
\begin{equation}\label{p-def-new-b}
{\cal Q}^B_{n,t_n}\equiv \left((\tilde P^B_{n,t_n})^{T_{-n}}\right)\bigcap 1(T_{-n},{\cal P}(\tilde\Omega)).
\end{equation}
Here, player $n$ can learn from his own type $t_n$ something about opponents' types $t_{-n}$; yet, these latter types will play no role in shaping his understanding of the external factor $\tilde{\Omega}$.

\begin{proposition}\label{magazj}
	For $\tilde w_{n,t_n}(a_n,a_{-n},\nu)$ defined at~(\ref{w-def-new}), it is both supermodular and submodular in $\nu\in {\cal M}(T_{-n},{\cal P}(\tilde{\Omega}))$, and has increasing differences between $a_n\in A_n$ and $(t_n,a_{-n},\nu)\in T_n\times \prod_{m\neq n}{\cal M}(T_m,A_m)\times {\cal M}(T_{-n},{\cal P}(\tilde{\Omega}))$, as well as between $a_{-n}\in \prod_{m\neq n}{\cal M}(T_m,$ $A_m)$ and $\nu\in {\cal M}(T_{-n},{\cal P}(\tilde{\Omega}))$.
	In addition, the function has increasing differences between $t_n\in T_n$ and $\nu\in {\cal M}(T_{-n},{\cal P}(\tilde \Omega))$ in scenario A and between $t_n\in T_n$ and $\nu\in 1(T_{-n},{\cal P}(\tilde \Omega))$ in scenario B.
\end{proposition}

\begin{proposition}\label{boring}
	For ${\cal Q}_{n,t_n}$ defined at~(\ref{p-def-new}), regardless of the scenario that prevails, at each fixed $t_n$ it is a nonempty sublattice of $({\cal P}(\tilde\Omega))^{T_{-n}}$ and hence a nonempty lattice in its own right. Also, it is increasing in $t_n$.
\end{proposition}

It is noteworthy that in Proposition~\ref{magazj}, we have restricted $a_{-n}\in\prod_{m\neq n}A_m^{\;T_m}$ to monotone opponent strategies in $\prod_{m\neq n}{\cal M}(T_m,A_m)$ and $\nu\in ({\cal P}(\tilde{\Omega}))^{T_{-n}}$ to monotone opponent-type-to-state-distribution maps in ${\cal M}(T_{-n},{\cal P}(\tilde{\Omega}))$. Moreover, the need there to prove the increasing differences that $\tilde w_{n,t_n}(a_n,a_{-n},\nu)$ has between $t_n$ and $\nu$ has prevented us from considering the more general case, where the conditional probabilities $p_{n,t_n|t_{-n}}$ take the more general B-version and the prior sets ${\cal Q}_{n,t_n}$ take the more general A-version.

The following lemma is about the preservation of increasing differences after maximization in the nature of~(\ref{s-def-new}). It is likely to be useful in other circumstances.

\begin{lemma}\label{l-import}
	Given partially ordered sets $X$ and $Y$, as well as lattice $Z$, let $f$ be a real-valued function defined on $X\times Y\times Z$, and $\tilde Z(y)$ be a subset of $Z$ at each $y\in Y$. Suppose that (I) $f$ has increasing differences between $x\in X$ and $(y,z)\in Y\times Z$, that (II) $f$ is supermodular in $z\in Z$, that (III) each $\tilde Z(y)$ is a sublattice, and that (IV) $\tilde Z(\cdot)$ is increasing in $y$. Also, suppose that (V) $f$ has increasing differences between $y\in Y$ and $z\in Z$. Then, for
	\[ g(x,y)=\sup_{z\in \tilde Z(y)}f(x,y,z),\]
	it will follow that $g$ has increasing differences between $x\in X$ and $y\in Y$.
\end{lemma}

We have singled out hypothesis (V) in Lemma~\ref{l-import} regarding $f$'s increasing differences between $y$ and $z$, because it seems the most demanding to us. This is the reason why Proposition~\ref{magazj} is concerned even with increasing differences between $a_{-n}$ and $\nu$, which ripple back to similar requirements in Proposition~\ref{common} and to Monotonic Assumption~\ref{as-utility}. Note that Theorem 2.7.6 of Topkis \cite{T98} goes from the supermodularity of $f$ in $(x,y)$ and lattice nature of $Y$ to the supermodularity of $g$ as defined in $ g(x)=\sup_{y\in Y}f(x,y)$. Our result is of a similar nature. With it, we can obtain a result key to equilibrium analysis.

\begin{proposition}\label{challenging}
	For $\tilde s_{n,t_n}(a_{n,t_n},a_{-n})$ defined at~(\ref{s-def-new}), it has increasing differences between $a_n\in A_n$ and $(t_n,a_{-n})\in T_n\times \prod_{m\neq n}{\cal M}(T_m,A_m)$.
\end{proposition}

Recall that $\tilde u_{n,t_n,t_{-n}}$ is continuous on a compact space. 
Through~(\ref{s-def-new}) to~(\ref{v-def-new}), this will lead to the continuity of $\tilde s_{n,t_n}(\cdot,a_{-n})$. Since each $A_n$ is a finite set or closed interval within the real line, not only is each $A_n$ a complete lattice but $\tilde s_{n,t_n}(\cdot,a_{-n})$ is also supermodular. Now the complete-lattice nature of the $A_{n}$'s, as well as the continuity, supermodularity, and Proposition~\ref{challenging}'s increasing differences will have provided essential elements of a game possessing strategic complementarities. Our ensuing analysis can lean on existing works such as Milgrom and Roberts \cite{MR90}, Milgrom and Shannon \cite{MS94}, and Yang and Qi \cite{YQ13}.

Combining Theorems 1 and 2 of Milgrom and Roberts \cite{MR90} (also summarized as Fact 2 of Yang and Qi \cite{YQ13}), we can conclude that each best-response action set $\tilde B_{n,t_n}(a_{-n})$ defined at~(\ref{best-def}) is a nonempty complete sublattice of $A_n$. By Milgrom and Shannon \cite{MS94} (also summarized as Fact 3 of Yang and Qi \cite{YQ13}), we further know that $\tilde B_{n,t_n}(a_{-n})$ is increasing in $(t_n,a_{-n})\in T_n\times\prod_{m\neq n}{\cal M}(T_m,A_m)$. The remainder of the development closely follows Yang and Qi \cite{YQ13}. As noted by it, each ${\cal M}(T_n,A_n)$, the space of monotone type-to-action mappings of player $n$, is a complete lattice. Now for any $n\in N$, define correspondence $\tilde{\cal B}_n:\prod_{m\neq n}{\cal M}(T_m,A_m)\rightrightarrows {\cal M}(T_n,A_n)$ from the space of monotone type-to-action mappings of other players to the space of the current player's monotone type-to-action mappings:
\begin{equation}\label{m-def}
\tilde{\cal B}_n(a_{-n})=\left\{a_n\equiv (a_{n,t_n})_{t_n\in T_n}\in {\cal M}(T_n,a_n)|a_{n,t_n}\in \tilde B_{n,t_n}(a_{-n})\;\forall t_n\in T_n\right\},
\end{equation}
for any $a_{-n}\in \prod_{m\neq n}{\cal M}(T_m,A_m)$. The following is a useful characterization.

\begin{proposition}\label{significant0}
	For the correspondence $\tilde{\cal B}_n$ defined at~(\ref{m-def}), it is a nonempty complete sublattice of ${\cal M}(T_n,A_n)$ at each $a_{-n}\in \prod_{m\neq n}{\cal M}(T_m,A_m)$. Also, it is increasing in $a_{-n}$.
\end{proposition}

Define a correspondence $\tilde{\cal B}$ from the complete lattice $\prod_{n\in N}{\cal M}(T_n,A_n)$ to itself so that
\begin{equation}\label{zia-ep}
a'\in \tilde{\cal B}(a)\;\;\mbox{ if and only if }\;\;a'_n\in \tilde{\cal B}_n(a_{-n})\mbox{ for any }n\in N.
\end{equation}
Now Proposition~\ref{significant0} will make $\tilde{\cal B}(a)$ a nonempty complete sublattice of $\prod_{n\in N}{\cal M}(T_n,A_n)$ at every $a\in \prod_{n\in N}{\cal M}(T_n,A_n)$ that is increasing in $a$. According to the discussion around~(\ref{best-def}), fixed points of $\tilde{\cal B}$ will form pure and type-monotone equilibria of the special enterprising game $\Gamma$ sense. Following the fixed point theorem of Zhou \cite{Z94}, which is a generalization of the classical result of Tarski \cite{T55}, we have the following existence result.

\begin{theorem}\label{big-t}
	The set of $\tilde{\cal B}$'s fixed points, $\tilde {\cal E}\equiv \{a\in\prod_{n\in N}{\cal M}(T_n,A_n)|a\in \tilde{\cal B}(a)\}$, is a nonempty complete lattice. Thus, $\Gamma$ has pure and monotone equilibria. 	
\end{theorem}

In languages used earlier, Theorem~\ref{big-t} will result with
\begin{equation}
1_A\cap{\cal E}^{\mbox a}=1_A\cap{\cal E}^{\mbox d}\supseteq 1_{\tilde{\cal E}}\equiv\{1_a\in 1_A|a\in \tilde{\cal E}\}\neq\emptyset.
\end{equation}
As $\tilde{\cal E}$ is a nonempty complete lattice, it has both the smallest and largest members. Let us denote them by $\tilde a_*$ and $\tilde a^*$, respectively.

\subsection{Monotone Comparative Statics}\label{comps}

Let $\Lambda$ be a partially ordered set, and let $(\Gamma(\lambda)|\lambda\in\Lambda)$ be a family of special enterprising games finalized in Section~\ref{special}. For $\lambda\in\Lambda$, suppose games $\Gamma(\lambda)$ share a common set of players $N$, state space $\tilde\Omega$, type-space vector $(T_n)_{n\in N}$, and action-space vector $(A_n)_{n\in N}$. However, the utility functions $(\tilde u_{n,t}(\lambda))_{n\in N,t\in T}$, distributions $(p_{n,t_n}(\lambda))_{n\in N,t_n\in T_n}$, and prior sets $(\tilde {\cal Q}_{n,t_n}(\lambda))_{n\in N,t_n\in T_n}$ are allowed to be $\lambda$-dependent.

We can define $\tilde v_{n,t_n,t_{-n}}(a_n,a_{-n},\nu|\lambda)$, $\tilde w_{n,t_n}(a_n,a_{-n},\nu|\lambda)$, and $\tilde s_{n,t_n}(a_{n,t_n},a_{-n}|\lambda)$, respectively, using almost the same albeit $\lambda$-dependent~(\ref{v-def-new}),~(\ref{w-def-new}), and~(\ref{s-def-new}). We can then define $\tilde B_{n,t_n}(a_{-n}|\lambda)$, $\tilde{\cal B}_n(a_{-n}|\lambda)$, and $\tilde{\cal B}(a|\lambda)$, respectively, using almost the same albeit $\lambda$-dependent~(\ref{best-def}),~(\ref{m-def}), and~(\ref{zia-ep}). It is possible to predict how $\tilde{\cal B}(\cdot|\lambda)$'s extremal fixed points $\tilde a_*(\lambda)$ and $\tilde a^*(\lambda)$ would evolve with $\lambda$ when the game $\Gamma(\lambda)$'s dependence on $\lambda$ follows certain conditions. Let us list the latter in the following.

\begin{e-assumption-p}\label{as-utility-p}
	For any $n\in N$, $t_n\in T_n$, $t_{-n}\in T_{-n}$, and $a_{-n}\in A_{-n}$, the payoff-utility function $\tilde u_{n,t_n,t_{-n}}(a_n,a_{-n},\tilde{\omega}|\lambda)$ has increasing differences between $(a_n,\tilde{\omega})\in A_n\times\tilde{\Omega}$ and $\lambda\in\Lambda$.
\end{e-assumption-p}

\begin{e-assumption-p}\label{as-mono-probability-p}
	For any $n\in N$ and $t_n\in T_n$, the probability $p^A_{n}\equiv(p^A_{n|t_{-n}})_{t_{-n}\in T_{-n}}$ is invariant in $\lambda\in \Lambda$; also,
	the probability $p^B_{n,t_n}(\lambda)\equiv (p^B_{n,t_n|t_{-n}}(\lambda))_{t_{-n}\in T_{-n}}$ is monotone in $\lambda\in \Lambda$ in the usual stochastic order.
\end{e-assumption-p}

\begin{e-assumption-p}\label{as-mono-prior-p}
	For any $n\in N$ and $t_n\in T_n$, the prior set $\tilde P^A_{n,t_n,t_{-n}}(\lambda)$ is increasing in $\lambda\in \Lambda$  for every $t_{-n}\in T_{-n}$ and the prior set $\tilde P^B_{n,t_n}(\lambda)$ is increasing in $\lambda\in \Lambda$.
\end{e-assumption-p}

The increasing differences between $a_n$ and $\lambda$ in Parametric Assumption~\ref{as-utility-p} and the monotonicity in $\lambda$ of probabilities $p^B_{n,t_n}(\lambda)$ in Parametric Assumption~\ref{as-mono-probability-p} are required for even the traditional game; see van Zandt and Vives \cite{VZV07}. When ambiguities on the external factors $\tilde\omega$ are further involved, it should not be surprising that increasing differences between $\tilde{\omega}$ and $\lambda$ be postulated in Parametric Assumption~\ref{as-utility-p} and the  monotonicity in $\lambda$ of prior sets be postulated in Parametric Assumption~\ref{as-mono-prior-p}.

These assumptions will lead to the following intermediate results of the order-theoretic nature. The requirement in Parametric Assumption~\ref{as-mono-probability-p} that the probabilities $p^A_{n}$ be invariant in $\lambda$ is especially needed for the proof of Proposition~\ref{magazj-p}.

\begin{proposition}\label{common-p}
	For $\tilde v_{n,t_n,t_{-n}}(a_n,a_{-n},\mu|\lambda)$ defined at the $\lambda$-dependent version of~(\ref{v-def-new}), it has increasing differences between $(a_n,\mu)\in A_n\times {\cal P}(\tilde{\Omega})$ and $\lambda\in \Lambda$.
\end{proposition}

\begin{proposition}\label{magazj-p}
	For $\tilde w_{n,t_n}(a_n,a_{-n},\nu|\lambda)$ defined at the $\lambda$-dependent version of~(\ref{w-def-new}), 
	it has increasing differences between $a_n\in A_n$ and $\lambda\in\Lambda$. In addition, the function has increasing differences between $\nu\in {\cal M}(T_{-n},{\cal P}(\tilde{\Omega}))$ and $\lambda\in \Lambda$ in scenario A and between $\nu\in 1(T_{-n},{\cal P}(\tilde{\Omega}))$ and $\lambda\in \Lambda$ in scenario B.
\end{proposition}

\begin{proposition}\label{boring-p}
	For ${\cal Q}_{n,t_n}(\lambda)$ defined at the $\lambda$-dependent version of~(\ref{p-def-new}), regardless of the scenario that prevails, it is increasing in $\lambda$ at each fixed $n\in N$ and $t_n\in T_n$.
\end{proposition}

\begin{proposition}\label{challenging-p}
	For $\tilde s_{n,t_n}(a_{n,t_n},a_{-n}|\lambda)$ defined at the $\lambda$-dependent version of~(\ref{s-def-new}), it has increasing differences between $a_{n,t_n}\in A_n$ and $\lambda\in\Lambda$ at each fixed $n\in N$, $t_n\in T_n$, and $a_{-n}\in \prod_{m\neq n}{\cal M}(T_m,A_m)$.	
\end{proposition}

The key to the rest of the derivation is the monotonicity of the correspondence defined at~(\ref{m-def}). Ideas from Yang and Qi \cite{YQ13} can be borrowed in its proof.

\begin{proposition}\label{significant0-p}
	For the correspondence $\tilde{\cal B}_n(a_{-n}|\lambda)$ defined at the $\lambda$-dependent~(\ref{m-def}), it is monotonically increasing in $\lambda\in\Lambda$ at each fixed $a_{-n}\in \prod_{m\neq n}{\cal M}(T_m,A_m)$.
\end{proposition}

From Proposition~\ref{significant0-p} and the $\lambda$-dependent version of~(\ref{zia-ep}), we can immediately have the monotonicity of $\tilde{\cal B}(a|\lambda)$ in $\lambda$ at each fixed $a\in \prod_{n\in N}{\cal M}(T_n,A_n)$. Using Lemma 4 of Yang and Qi \cite{YQ13}, a counterpart to Tarski's \cite{T55} monotone comparative statics result in Zhou's \cite{Z94} setting, we can then achieve monotonicity of the extremal equilibria as $\lambda$ varies.

\begin{theorem}\label{big-t-p}
	The enterprising game $\Gamma(\lambda)$' smallest and largest pure and monotone equilibria, $\tilde a_*(\lambda)$ and $\tilde a^*(\lambda)$, are both increasing in $\lambda\in\Lambda$.
\end{theorem}

For scenario A, i.e., the one involving special type distributions but more general ambiguity attitudes on external factors, our general message about monotone equilibria that evolve monotonically over exogenous parameters is consistent with the one derived for the traditional expected-utility case. For the latter, see, e.g., van Zandt and Vives \cite{VZV07}. For scenario B, i.e., the one involving general type distributions but special ambiguity attitudes, barring some minutiae our results can even be thought of as generalizations of existing ones.

\subsection{Competitive Pricing with Uncertain Demand}

We now consider a situation that fits the theory developed so far in Appendix~\ref{app-dc}.

Players from the set $N\equiv \{1,...,\bar n\}$ are firms engaged in price competition in a common market for their manufactured product items. Suppose it costs $\underline a_n$ for firm $n$ to manufacture a unit item. Also, let the firm's type $t_n\in T_n\equiv\{1,...,\bar t_n\}\subseteq \Re$ stand for a factor of the demand that it is to face. The firm's action space $A_n\equiv [\underline a_n,\overline a_n]\subseteq \Re$ denotes the range of prices that it can charge. 
Suppose the state space $\tilde\Omega\equiv [0,\overline\omega]$ contains positive global additive factors to demands faced by all firms.
We take the demand faced by firm $n$ to be
\begin{equation}
\phi_{n,t_n}(a_n,a_{-n},\tilde\omega)=\bar b_n-\bar c_n\cdot a_n+\sum_{m\neq n}\bar d_{nm}\cdot a_m+\bar e_n\cdot t_n+\bar f_n\cdot \tilde\omega+\bar g_n\cdot t_n\tilde\omega,
\end{equation}
where $\bar b_n$, $\bar c_n$, $(\bar d_{nm})_{m\neq n}$, $\bar e_n$, $\bar f_n$, and $\bar g_n$ are positive constants. Basically, demand to firm $n$ will decline when the firm raises its price; but it will rise when competitors raise their prices. Moreover, both $t_n$ and $\tilde\omega$ serve as demand boosters, with the former being locally confined and the latter globally felt. The last term indicates that their effects may be compounded.

The profit that firm $n$ can earn is therefore
\begin{equation}\label{uuu}\begin{array}{l}
\tilde u_{n,t_n,t_{-n}}(a_n,a_{-n},\tilde\omega)=(a_n-\underline a_n)\cdot \phi_{n,t_n}(a_n,a_{-n},\tilde\omega)\\
\;\;\;\;\;\;\;\;\;\;\;\;=(a_n-\underline a_n)\cdot (\bar b_n-\bar c_n\cdot a_n+\sum_{m\neq n}\bar d_{nm}\cdot a_m+\bar e_n\cdot t_n+\bar f_n\cdot \tilde\omega+\bar g_n\cdot t_n\tilde\omega),
\end{array}\end{equation}
which is independent of $t_{-n}$. More importantly, the function is increasing in $\tilde\omega$. So Monotonic Assumption~\ref{as-bongaji} is satisfied. Taking derivatives, we obtain
\begin{equation}\label{uuu1}
\frac{\partial\tilde u_{n,t_n,t_{-n}}}{\partial a_n}(a_n,a_{-n},\tilde\omega)=\underline a_n\bar c_n+\bar b_n-2\bar c_n\cdot a_n+\sum_{m\neq n}\bar d_{nm}\cdot a_m+\bar e_n\cdot t_n+\bar f_n\cdot \tilde\omega+\bar g_n\cdot t_n\tilde\omega,
\end{equation}
which is increasing in $(t_n,t_{-n},a_{-n},\tilde\omega)$; also,
\begin{equation}\label{uuu2}
\frac{\partial\tilde u_{n,t_n,t_{-n}}}{\partial\tilde\omega}(a_n,a_{-n},\tilde\omega)=\bar f_n\cdot a_n-\underline a_n\bar f_n+\bar g_n\cdot(a_n-\underline a_n)\cdot t_n,
\end{equation}
which is increasing in $(t_n,t_{-n},a_{-n})$. Hence, Monotonic Assumption~\ref{as-utility} is satisfied.

For local and global demand signals, suppose scenario A of Section~\ref{special} takes over. This means that players are unambiguous about their local signals but ambiguous about the global one. Also, Monotonic Assumption~\ref{as-mono-probability} is automatic. In particular, each firm $n$ believes that other firms' types are distributed according to some $p^A_n\equiv (p^A_{n|t_{-n}})_{t_{-n}\in T_{-n}}$, irrespective of its own type $t_n$; moreover, there are prior sets $\tilde P^A_{n,t_n,t_{-n}}$ so that the ${\cal Q}_{n,t_n}$ used in~(\ref{s-def-new}) of its decision making process is defined through~(\ref{p-def-new-a}).

Now, suppose the $P^A_{n,t_n,t_{-n}}$'s are sublattices of ${\cal P}(\tilde\Omega)$ that also increase with $t_n$. The latter monotonicity connotes a certain positive correlation between a firm's local demand signal and the global one. Then, Monotonic Assumptions~\ref{as-prior} and~\ref{as-mono-prior} will be satisfied. Thus, Theorem~\ref{big-t} can be used to predict that firms will be able to reach highest equilibrium pricing policies $\tilde a^*_{n,t_n}$ that are increasing in their observed local signals $t_n$.

A partially ordered set $\Lambda$ may provide parameters to the pricing game, so that the constants $\bar b_n$, $(\bar d_{nm})_{m\neq n}$, $\bar e_n$, $\bar f_n$, and $\bar g_n$ and prior sets $\tilde P^A_{n,t_n,t_{-n}}$ are all functions of $\lambda\in\Lambda$. From~(\ref{uuu}), we can explain the monotonicity of those constants with respect to $\lambda$ by an expanded demand base and demand's heightened sensitivities to other players' prices, as well as local and global signals. When this is the case, we will be able to learn from~(\ref{uuu1}) and~(\ref{uuu2}) the satisfaction of Parametric Assumption~\ref{as-utility-p}.

With the distributions $p^A_n$ invariant in $\lambda$, Parametric Assumption~\ref{as-mono-probability-p} is automatic. Suppose further that a higher $\lambda$ also reflects firms' bullish forecasts on the market, to the effect that the prior sets $\tilde P^A_{n,t_n,t_{-n}}$ increase in $\lambda$ as well. Then, Parametric Assumption~\ref{as-mono-prior-p} will be satisfied. The end result is that Theorem~\ref{big-t-p} can now be used to predict the increase of the highest monotone equilibrium $\tilde a^*$ with respect to the changing $\lambda$. This result is quite anticipated, as bigger markets, more reactive demands, and brightened outlooks will embolden firms to price more aggressively.

\section{Proofs for Appendix~\ref{app-dc}}\label{app-c}

\noindent{\bf Proof of Proposition~\ref{common}: }At any fixed $n\in N$, $t_n\in T_n$, and $t_{-n}\in T_{-n}$, define $q$ by
\begin{equation}\label{qq-dd}
q(\tilde\omega)\equiv \tilde u_{n,t_n,t_{-n}}(a_n,a_{-n},\tilde\omega).
\end{equation}
The continuity of $\tilde u_{n,t_n,t_{-n}}$ suggests that $q$, as a real-valued function defined on the compact $\tilde\Omega$, is lower bounded by some $\underline u_{n,t_n,t_{-n}}(a_n,a_{-n})$. Monotonic Assumption~\ref{as-bongaji} also states that it is increasing. The former property indicates that $\mu\cdot q^{-1}$ for any $\mu\in {\cal P}(\tilde\Omega)$ is a cumulative distribution function (cdf) say $F$ on the real interval $[\underline u_{n,t_n,t_{-n}}(a_n,a_{-n}),+\infty)$. The latter property means that $q^{-1}([\underline u_{n,t_n,t_{-n}}(a_n,a_{-n}),x])$ is a lower set in the sense that $\tilde\omega^1\in q^{-1}([\underline u_{n,t_n,t_{-n}}(a_n,a_{-n}),x])$ whenever $\tilde\omega^2\in q^{-1}([\underline u_{n,t_n,t_{-n}}(a_n,a_{-n}),x])$ and $\tilde\omega^1\leq \tilde\omega^2$. Thus, due to the partial order we have chosen for ${\cal P}(\tilde\Omega)$, for any members $\mu^1$ and $\mu^2$,
\begin{equation}\begin{array}{l}
[(\mu^1\vee\mu^2)\cdot q^{-1}]([\underline u_{n,t_n,t_{-n}}(a_n,a_{-n}),x])\\
\;\;\;\;\;\;\;\;\;\;\;\;=(\mu^1\cdot q^{-1})([\underline u_{n,t_n,t_{-n}}(a_n,a_{-n}),x])\wedge (\mu^2\cdot q^{-1})([\underline u_{n,t_n,t_{-n}}(a_n,a_{-n}),x]).
\end{array}\end{equation}
The same applies to the opposite combination of ``$\vee$'' and ``$\wedge$'' as well. Then, for cdf's $F^1\equiv \mu^1\cdot q^{-1}$ and $F^2\equiv \mu^2\cdot q^{-1}$, we have
\begin{equation}
F^1\wedge F^2=(\mu^1\vee\mu^2)\cdot q^{-1},\hspace*{.8in}F^1\vee F^2=(\mu^1\wedge\mu^2)\cdot q^{-1}.
\end{equation}
Hence,
\begin{equation}\label{superb}\begin{array}{l}
\int_{\tilde\Omega}q(\tilde\omega)\cdot (\mu^1\vee\mu^2)(d\tilde\omega)+\int_{\tilde\Omega}q(\tilde\omega)\cdot (\mu^1\wedge\mu^2)(d\tilde\omega)\\
\;\;\;\;\;\;\;\;\;\;\;\;=\int_{\underline u_{n,t_n,t_{-n}}(a_n,a_{-n})}^{+\infty}x\cdot (F^1\wedge F^2)(dx)+\int_{\underline u_{n,t_n,t_{-n}}(a_n,a_{-n})}^{+\infty}x\cdot (F^1\vee F^2)(dx)\\
\;\;\;\;\;\;\;\;\;\;\;\;=2\cdot\underline u_{n,t_n,t_{-n}}(a_n,a_{-n})+\int_{\underline u_{n,t_n,t_{-n}}(a_n,a_{-n})}^{+\infty}(1-F^1(x)\wedge F^2(x))\cdot dx\\
\;\;\;\;\;\;\;\;\;\;\;\;\;\;\;\;\;\;\;\;\;\;\;\;\;\;\;\;\;\;+\int_{\underline u_{n,t_n,t_{-n}}(a_n,a_{-n})}^{+\infty}(1-F^1(x)\vee F^2(x))\cdot dx\\
\;\;\;\;\;\;\;\;\;\;\;\;=2\cdot\underline u_{n,t_n,t_{-n}}(a_n,a_{-n})+\int_{\underline u_{n,t_n,t_{-n}}(a_n,a_{-n})}^{+\infty}(1-F^1(x))\cdot dx\\
\;\;\;\;\;\;\;\;\;\;\;\;\;\;\;\;\;\;\;\;\;\;\;\;\;\;\;\;\;\;+\int_{\underline u_{n,t_n,t_{-n}}(a_n,a_{-n})}^{+\infty}(1-F^2(x))\cdot dx\\
\;\;\;\;\;\;\;\;\;\;\;\;=\int_{\underline u_{n,t_n,t_{-n}}(a_n,a_{-n})}^{+\infty}x\cdot F^1(dx)+\int_{\underline u_{n,t_n,t_{-n}}(a_n,a_{-n})}^{+\infty}x\cdot F^2(dx)\\
\;\;\;\;\;\;\;\;\;\;\;\;=\int_{\tilde\Omega}q(\tilde\omega)\cdot \mu^1(d\tilde\omega)+\int_{\tilde\Omega}q(\tilde\omega)\cdot \mu^2(d\tilde\omega),
\end{array}\end{equation}
where the first and fifth equalities are achieved through changes of variables, the second and fourth equalities come from integrals by parts, and the third equality is due to the fact that
\begin{equation}
F^1(x)\wedge F^2(x)+F^1(x)\vee F^2(x)=F^1(x)+F^2(x).
\end{equation}
In view of~(\ref{v-def-new}) and~(\ref{qq-dd}), we have from~(\ref{superb}) that $\tilde v_{n,t_n,t_{-n}}(a_n,a_{-n},\cdot)$ is both supermodular and submodular as a real-valued function defined on ${\cal P}(\tilde{\Omega})$.

Now suppose $\mu\in {\cal P}(\tilde{\Omega})$ is fixed. Then, due to the average in~(\ref{v-def-new}) and the increasing differences stated in Monotonic Assumption~\ref{as-utility}, $\tilde v_{n,t_n,t_{-n}}(a_n,a_{-n},\mu)$ will have increasing differences between $a_n\in A_n$ and $(t_n,t_{-n},a_{-n})\in T_n\times T_{-n}\times A_{-n}$. Next, suppose $t^1_n,t^2_n\in T_n$ satisfy $t^1_n\leq t^2_n$, $t^1_{-n},t^2_{-n}\in T_{-n}$ satisfy $t^1_{-n}\leq t^2_{-n}$, $a^1_n,a^2_n\in A_n$ satisfy $a^1_n\leq a^2_n$, and $a^1_{-n},a^2_{-n}\in A_{-n}$ satisfy $a^1_{-n}\leq a^2_{-n}$. Then, we note the increasing trend of $q$ for
\begin{equation}\label{qq}
q(\tilde\omega)\equiv\tilde u_{n,t^2_n,t^2_{-n}}(a^2_n,a^2_{-n},\tilde\omega)-\tilde u_{n,t^1_n,t^1_{-n}}(a^1_n,a^1_{-n},\tilde\omega),
\end{equation}
due to the increasing differences between $(t_n,t_{-n},a_n,a_{-n})\in T_n\times T_{-n}\times A_n\times A_{-n}$ and $\tilde\omega\in\tilde\Omega$ as stated in Monotonic Assumption~\ref{as-utility}. Thus, for $\mu^1,\mu^2\in {\cal P}(\tilde\Omega)$ satisfying $\mu^1\leq \mu^2$,
\begin{equation}\label{b2}
\int_{\tilde\Omega}q(\tilde\omega)\cdot\mu^1(d\tilde\omega)\leq \int_{\tilde\Omega}q(\tilde\omega)\cdot\mu^2(d\tilde\omega).
\end{equation}
In view of~(\ref{v-def-new}) and~(\ref{qq}), this will amount to the  increasing differences that $\tilde v_{n,t_n,t_{-n}}(a_n,a_{-n},\mu)$ enjoys between $(t_n,t_{-n},a_n,a_{-n})\in T_n\times T_{-n}\times A_n\times A_{-n}$ and $\mu\in {\cal P}(\tilde{\Omega})$. \qed

\noindent{\bf Proof of Proposition~\ref{magazj}: }First, the average in~(\ref{w-def-new}) and the simultaneous supermodularity and submodularity stated in Proposition~\ref{common} will lead to the simultaneous supermodularity and submodularity of $\tilde w_{n,t_n}(a_n,a_{-n},\nu)$ at every component $\nu_{t_{-n}}$. Due to the component-wise nature of the partial order assigned to ${\cal M}(T_{-n},{\cal P}(\tilde{\Omega}))$, this also means that $\tilde w_{n,t_n}(a_n,a_{-n},\nu)$ is both supermodular and submodular in $\nu\equiv (\nu_{t_{-n}})_{t_{-n}\in T_{-n}}\in {\cal M}(T_{-n},{\cal P}(\tilde{\Omega}))$.

Now suppose $t_n\in T_n$ is fixed. Then, due to the average in~(\ref{w-def-new}) and the increasing differences stated in Proposition~\ref{common}, $\tilde w_{n,t_n}(a_n,a_{-n},\nu)$ will have increasing differences between $a_n\in A_n$ and $(a_{-n},\nu)\in \prod_{m\neq n}{\cal M}(T_m,A_m)\times {\cal M}(T_{-n},{\cal P}(\tilde{\Omega}))$, as well as between $a_{-n}\in \prod_{m\neq n}{\cal M}(T_m,A_m)$ and $\nu\in {\cal M}(T_{-n},{\cal P}(\tilde{\Omega}))$.

Next, suppose $a_{-n}\in \prod_{m\neq n}{\cal M}(T_m,A_m)$ and $\nu\in {\cal M}(T_{-n},{\cal P}(\tilde{\Omega}))$ are fixed, while $a^1_n,a^2_n\in A_n$ satisfy $a^1_n\leq a^2_n$. Define $f\equiv(f_{t_n,t_{-n}})_{t_n\in T_n,t_{-n}\in T_{-n}}$ so that
\begin{equation}\label{ff}
f_{t_n,t_{-n}}\equiv\tilde v_{n,t_n,t_{-n}}(a^2_n,a_{-n,t_{-n}},\nu_{t_{-n}})-\tilde v_{n,t_n,t_{-n}}(a^1_n,a_{-n,t_{-n}},\nu_{t_{-n}}).
\end{equation}
Due to the increasing differences between $a_n\in A_n$ and $(t_n,t_{-n},a_{-n,t_{-n}},\nu_{-n,t_{-n}})\in T_n\times T_{-n}\times A_{-n}\times {\cal P}(\tilde{\Omega})$ as stated in Proposition~\ref{common}, as well as the memberships of $a_{-n}$ in $\prod_{m\neq n}{\cal M}(T_m,A_m)$ and $\nu$ in ${\cal M}(T_{-n},{\cal P}(\tilde{\Omega}))$, we know that $f_{t_n,t_{-n}}$ is increasing in both $t_n$ and $t_{-n}$. But by Monotonic  Assumption~\ref{as-mono-probability}, this will translate into
\begin{equation}
\sum_{t_{-n}\in T_{-n}}p_{n,t^1_n|t_{-n}}\cdot f_{t^1_n,t_{-n}}\leq \sum_{t_{-n}\in T_{-n}}p_{n,t^1_n|t_{-n}}\cdot f_{t^2_n,t_{-n}}\leq  \sum_{t_{-n}\in T_{-n}}p_{n,t^2_n|t_{-n}}\cdot f_{t^2_n,t_{-n}},
\end{equation}
whenever $t^1_n\leq t^2_n$. In view of~(\ref{w-def-new}) and~(\ref{ff}), this will amount to the increasing differences that $\tilde w_{n,t_n}(a_n,a_{-n},\nu)$ enjoys between $a_n\in A_n$ and $t_n\in T_n$.

Suppose $a_n\in A_n$ and $a_{-n}\in \prod_{m\neq n}{\cal M}(T_m,A_m)$ are fixed, while $\nu^1,\nu^2\in {\cal M}(T_{-n},{\cal P}(\tilde{\Omega}))$ satisfy $\nu^1\leq \nu^2$. Define $g\equiv (g_{t_n,t_{-n}})_{t_n\in T_n,t_{-n}\in T_{-n}}$ so that
\begin{equation}\label{gg}
g_{t_n,t_{-n}}\equiv\tilde v_{n,t_n,t_{-n}}(a_n,a_{-n,t_{-n}},\nu^2_{t_{-n}})-\tilde v_{n,t_n,t_{-n}}(a_n,a_{-n,t_{-n}},\nu^1_{t_{-n}}).
\end{equation}
Due to the increasing differences between $t_n\in T_n$ and $\nu_{t_{-n}}\in {\cal P}(\tilde{\Omega})$ as stated in Proposition~\ref{common}, we know that $g_{t_n,t_{-n}}$ is increasing in $t_n$. But by averaging, this will translate into
\begin{equation}
\sum_{t_{-n}\in T_{-n}}p^A_{n|t_{-n}}\cdot g_{t^1_n,t_{-n}}\leq \sum_{t_{-n}\in T_{-n}}p^A_{n|t_{-n}}\cdot g_{t^2_n,t_{-n}},
\end{equation}
whenever $t^1_n\leq t^2_n$. In view of~(\ref{w-def-new}) and~(\ref{gg}), this will amount to the increasing differences that $\tilde w_{n,t_n}(a_n,a_{-n},\nu)$ enjoys between $t_n\in T_n$ and $\nu\in {\cal M}(T_{-n},{\cal P}(\tilde{\Omega}))$ in scenario A.

In addition, suppose $a_n\in A_n$ and $a_{-n}\in \prod_{m\neq n}{\cal M}(T_m,A_m)$ are fixed, while $\nu^1,\nu^2\in 1(T_{-n},{\cal P}(\tilde{\Omega}))$ satisfy $\nu^1\leq \nu^2$. Define $h\equiv (h_{t_n,t_{-n}})_{t_n\in T_n,t_{-n}\in T_{-n}}$ so that
\begin{equation}\label{hh}
h_{t_n,t_{-n}}\equiv\tilde v_{n,t_n,t_{-n}}(a_n,a_{-n,t_{-n}},\nu^2_{t_{-n}})-\tilde v_{n,t_n,t_{-n}}(a_n,a_{-n,t_{-n}},\nu^1_{t_{-n}}).
\end{equation}
Due to $\nu^1$ and $\nu^2$'s memberships in $1(T_{-n},{\cal P}(\tilde\Omega))$, neither $\nu^1_{t_{-n}}$ nor $\nu^2_{t_{-n}}$ depends on $t_{-n}$. So by the increasing differences between $(t_n,t_{-n},a_{-n,t_{-n}})\in T_n\times T_{-n}\times A_{-n}$ and $\nu_{t_{-n}}\in {\cal P}(\tilde{\Omega})$ as stated in Proposition~\ref{common}, as well as the membership of $a_{-n}$ in $\prod_{m\neq n}{\cal M}(T_m,A_m)$, we know that $h_{t_n,t_{-n}}$ is increasing in both $t_n$ and $t_{-n}$. But by Monotonic  Assumption~\ref{as-mono-probability},
\begin{equation}
\sum_{t_{-n}\in T_{-n}}p^B_{n,t^1_n|t_{-n}}\cdot h_{t^1_n,t_{-n}}\leq \sum_{t_{-n}\in T_{-n}}p^B_{n,t^1_n|t_{-n}}\cdot h_{t^2_n,t_{-n}}\leq \sum_{t_{-n}\in T_{-n}}p^B_{n,t^2_n|t_{-n}}\cdot h_{t^2_n,t_{-n}},
\end{equation}
whenever $t^1_n\leq t^2_n$. In view of~(\ref{w-def-new}) and~(\ref{hh}), this will amount to the increasing differences that $\tilde w_{n,t_n}(a_n,a_{-n},\nu)$ enjoys between $t_n\in T_n$ and $\nu\in 1(T_{-n},{\cal P}(\tilde{\Omega}))$ in scenario B.\qed

\noindent{\bf Proof of Proposition~\ref{boring}: }For scenario A, we first show that ${\cal Q}^A_{n,t_n}$ as defined through~(\ref{p-def-new-a}) is a sublattice of $({\cal P}(\tilde{\Omega}))^{T_{-n}}$. Suppose  $\nu^1,\nu^2\in {\cal Q}^A_{n,t_n}$. Then, according to~(\ref{p-def-new-a}), $\nu^1_{t_{-n}},\nu^2_{t_{-n}}\in \tilde P^A_{n,t_n,t_{-n}}$ for any $t_{-n}\in T_{-n}$. So by Monotonic Assumption~\ref{as-prior},
\begin{equation}
\nu^1_{t_{-n}}\wedge \nu^2_{t_{-n}},\hspace*{.5in}\nu^1_{t_{-n}}\vee \nu^2_{t_{-n}}\in\tilde P^A_{n,t_n,t_{-n}},\hspace*{.8in}\forall t_{-n}\in T_{-n}.
\end{equation}
In addition, both $\nu^1_{t_{-n}}\wedge\nu^2_{t_{-n}}$ and  $\nu^1_{t_{-n}}\vee\nu^2_{t_{-n}}$ will continue to be increasing in $t_{-n}$ just because both $\nu^1_{t_{-n}}$ and $\nu^2_{t_{-n}}$ are. But by~(\ref{p-def-new-a}), this leads back to
\begin{equation}\label{pat-a}
\nu^1\wedge \nu^2,\hspace*{.8in}\nu^1\vee \nu^2\in {\cal Q}^A_{n,t_n}.
\end{equation}
Hence, ${\cal Q}^A_{n,t_n}$ is a sublattice of $({\cal P}(\tilde\Omega))^{T_{-n}}$.

We next prove that ${\cal Q}^A_{n,t_n}$ is nonempty. Since $\tilde P^A_{n,t_n,t_{-n}}$ for each $t_{-n}\in T_{-n}$ is nonempty, we can pick one $\nu_{t_{-n}}$ from every $\tilde P^A_{n,t_n,t_{-n}}$. Let us go through every $m\neq n$ in the order of $m=1,...,n-1,n+1,...,\bar n$. Suppose we are at a particular $m\neq n$. Then, for every $t_{-(n,m)}\in T_{-(n,m)}\equiv \prod_{l\neq n,m}T_l$, we go through the procedure of
\begin{equation}
\nu_{t_m,t_{-(n,m)}}=\bigwedge_{\tau_m=t_m}^{\bar t_m}\nu'_{\tau_m,t_{-(n,m)}},\hspace*{.8in}\forall t_m=1,2,...,\bar t_m-1,
\end{equation}
and
\begin{equation}
\nu_{t_m,t_{-(n,m)}}=\nu'_{t_m,t_{-(n,m)}},\hspace*{.8in}\forall t_m=1,2,...,\bar t_m-1.
\end{equation}
We now show that the $\nu$ assembled from all the $\nu_{t_{-n}}$'s after the procedure is a member of ${\cal Q}^A_{n,t_n}$ as defined through~(\ref{p-def-new-a}). First, due to Monotonic Assumption~\ref{as-mono-prior}, we can iteratively show that during the procedure,
\begin{equation}
\nu_{t_m,t_{-(n,m)}}\in \tilde P^A_{n,t_m,t_{-(n,m)}},\;\;\;\mbox{ in the order of }t_m=\bar t_m,\bar t_m-1,...1,
\end{equation}
for every $t_{-(n,m)}\in T_{-(n,m)}$. Second, after the procedure, for every $m\neq n$ and every $t_{-(n,m)}\in T_{-(n,m)}$, the new $\nu_{t_m,t_{-(n,m)}}$ is certainly increasing in $t_m$. For any $t^1_{-n},t^2_{-n}\in T_{-n}$ satisfying $t^1_{-n}\leq t^2_{-n}$, note that $t^1_m\leq t^2_m$ for any $m\neq n$. We can thus traverse from $t^1_{-n}$ to $t^2_{-n}$ in the order of $t^1_{-n}\equiv (t^1_1,t^1_{-(n,1)})$, $(t^1_1+1,t^1_{-(n,1)})$, $...$, $(t^2_1,t^1_{-(n,1)})\equiv (t^2_1,t^1_2,t^1_{-(n,1,2)})$, $(t^2_1,t^1_2+1,t^1_{-(n,1,2)})$, $...$, $(t^2_{(1,2)},t^1_{-(n,1,2)})\equiv (t^2_{(1,2)},t^1_3,t^1_{-(n,1,2,3)})$, $......$, $(t^2_{-(n,\bar n)},t^2_{\bar n}-1)$, $(t^2_{-(n,\bar n)},t^2_{\bar n})\equiv t^2_{-n}$. Along the path, the $t_{-n}$ encountered keeps rising. This implies that the associated $\nu_{t_{-n}}$ would keep rising as well. Consequently, we can reach $\nu_{t^1_{-n}}\leq \nu_{t^2_{-n}}$. That is, the assembled $\nu$ is a member of ${\cal M}(T_{-n},{\cal P}(\tilde{\Omega}))$. Taking both points together, we can see that $\nu$ is a member of ${\cal Q}^A_{n,t_n}$ as it is defined at~(\ref{p-def-new-a}). So ${\cal Q}^A_{n,t_n}\neq\emptyset$.

Finally, we verify the increasing trend of ${\cal Q}^A_{n,t_n}$ in $t_n$. Suppose $t^1_n,t^2_n\in T_n$ satisfy $t^1_n\leq t^2_n$; also,  $\nu^1\in {\cal Q}^A_{n,t^1_n}$ and $\nu^2\in {\cal Q}^A_{n,t^2_n}$. Then, according to~(\ref{p-def-new-a}), $\nu^1_{t_{-n}}\in \tilde P^A_{n,t^1_n,t_{-n}}$ and $\nu^2_{t_{-n}}\in \tilde P^A_{n,t^2_n,t_{-n}}$ for any $t_{-n}\in T_{-n}$, and both $\nu^1_{t_{-n}}$ and $\nu^2_{t_{-n}}$ are increasing in $t_{-n}$. So by Monotonic Assumption~\ref{as-mono-prior},
\begin{equation}\label{b19}
\nu^1_{t_{-n}}\wedge \nu^2_{t_{-n}}\in \tilde P^A_{n,t^1_n,t_{-n}},\hspace*{.5in}\nu^1_{t_{-n}}\vee \nu^2_{t_{-n}}\in \tilde P^A_{n,t^2_n,t_{-n}},\hspace*{.8in}\forall t_{-n}\in T_{-n}.
\end{equation}
Both $\nu^1_{t_{-n}}\wedge\nu^2_{t_{-n}}$ and  $\nu^1_{t_{-n}}\vee\nu^2_{t_{-n}}$ are still increasing in $t_{-n}$.
But by~(\ref{p-def-new-a}), this leads back to
\begin{equation}\label{b20}
\nu^1\wedge \nu^2\in \tilde P_{n,t^1_n},\hspace*{.8in}\nu^1\vee \nu^2\in \tilde P_{n,t^2_n}.
\end{equation}
Hence, ${\cal Q}^A_{n,t_n}$ is increasing in $t_n$.

For scenario B, we know that ${\cal Q}^B_{n,t_n}$ as defined through~(\ref{p-def-new-b}) is nonempty just because $\tilde P^B_{n,t_n}$ is nonempty. It is a sublattice of $({\cal P}(\tilde{\Omega}))^{T_{-n}}$ just because, due to Monotonic Assumption~\ref{as-prior}, $\tilde P^B_{n,t_n}$ is a sublattice of ${\cal P}(\tilde\Omega)$. Also, it is increasing in $t_n$ just because, due to Monotonic Assumption~\ref{as-mono-prior}, $\tilde P^B_{n,t_n}$ is increasing in $t_n$. \qed

\noindent{\bf Proof of Lemma~\ref{l-import}: }Suppose $x^1,x^2\in X$ satisfy $x^1\leq x^2$ and $y^1,y^2\in Y$ satisfy $y^1\leq y^2$. For any $\epsilon>0$, we can choose $z^{12}\in \tilde Z(y^2)$ so that
\begin{equation}\label{z12-def}
f(x^1,y^2,z^{12})\geq \sup_{z\in\tilde Z(y^2)}f(x^1,y^2,z)-\epsilon=g(x^1,y^2)-\epsilon,
\end{equation}
and $z^{21}\in \tilde Z(y^1)$ so that
\begin{equation}\label{z21-def}
f(x^2,y^1,z^{21})\geq \sup_{z\in\tilde Z(y^1)}f(x^2,y^1,z)-\epsilon=g(x^2,y^1)-\epsilon.
\end{equation}
Since $\tilde Z(y)$'s are sublattices that increase with $y$,
\begin{equation}\label{belonging}
z^{12}\wedge z^{21}\in\tilde Z(y^1),\hspace*{.8in}z^{12}\vee z^{21}\in \tilde Z(y^2).
\end{equation}
Now, by $f$'s increasing differences between $x\in X$ and $y\in Y$,
\begin{equation}\label{term1}
f(x^1,y^1,z^{12})-f(x^2,y^1,z^{12})+f(x^2,y^2,z^{12})-f(x^1,y^2,z^{12})\geq 0;
\end{equation}
by $f$'s increasing differences between $x\in X$ and $z\in Z$,
\begin{equation}\label{term2}
f(x^1,y^1,z^{21})-f(x^2,y^1,z^{21})-f(x^1,y^1,z^{12}\vee z^{21})+f(x^2,y^1,z^{12}\vee z^{21})\geq 0;
\end{equation}
by $f$'s supermodularity in $z\in Z$,
\begin{equation}\label{term3}
f(x^1,y^1,z^{12}\wedge z^{21})-f(x^1,y^1,z^{12})-f(x^1,y^1,z^{21})+f(x^1,y^1,z^{12}\vee z^{21})\geq 0;
\end{equation}
in addition, by $f$'s increasing differences between $y\in Y$ and $z\in Z$,
\begin{equation}\label{term4}
f(x^2,y^1,z^{12})-f(x^2,y^2,z^{12})-f(x^2,y^1,z^{12}\vee z^{21})+f(x^2,y^2,z^{12}\vee z^{21})\geq 0.
\end{equation}
Adding up~(\ref{term1}) to~(\ref{term4}) together, we obtain
\begin{equation}\label{together}
f(x^1,y^1,z^{12}\wedge z^{21})-f(x^1,y^2,z^{12})-f(x^2,y^1,z^{21})+f(x^2,y^2,z^{12}\vee z^{21})\geq 0.
\end{equation}
When this is combined with~(\ref{z12-def}) to~(\ref{belonging}), we have
\begin{equation}\begin{array}{l}
g(x^1,y^1)+g(x^2,y^2)\geq f(x^1,y^1,z^{12}\wedge z^{21})+f(x^2,y^2,z^{12}\vee z^{21})\\
\;\;\;\;\;\;\;\;\;\;\;\;\geq f(x^1,y^2,z^{12})+f(x^2,y^1,z^{21})\geq g(x^1,y^2)+g(x^2,y^1)-2\epsilon.
\end{array}\end{equation}
Since $\epsilon>0$ can be made arbitrarily small, it follows that
\begin{equation}
g(x^1,y^1)+g(x^2,y^2)\geq g(x^1,y^2)+g(x^2,y^1).
\end{equation}
That is, $g$ has increasing differences between $x\in X$ and $y\in Y$. \qed

\noindent{\bf Proof of Proposition~\ref{challenging}: }At any fixed $n\in N$, we can identify $a_{n,t_n}\in A_n$ with $x\in X$, $(t_n,a_{-n})\in T_n\times \prod_{m\neq n}{\cal M}(T_m,A_m)$ with $y\in Y$, and $\nu\in ({\cal P}(\tilde{\Omega}))^{T_{-n}}$
with $z\in Z$. Also, we can identify $\tilde w_{n,t_n}(a_{n,t_n},a_{-n},\nu)$ with $f(x,y,z)$, ${\cal Q}_{n,t_n}$ with $\tilde Z(y)$, and $\tilde s_{n,t_n}(a_{n,t_n},a_{-n})$ with $g(x,y)$. From Proposition~\ref{magazj}, we know that (I), (II), and (V) of Lemma~\ref{l-import} are true. From Proposition~\ref{boring}, we know that (III) and (IV) of Lemma~\ref{l-import} are also true. Now~(\ref{s-def-new}) dictates that the relationship between $f$, $\tilde Z$, and $g$ in Lemma~\ref{l-import} also applies here. So by that lemma, we can derive that $\tilde s_{n,t_n}(a_{n,t_n},a_{-n})$ has increasing differences between $a_{n,t_n}\in A_n$ and $(t_n,a_{-n})\in T_n\times \prod_{m\neq n}{\cal M}(T_m,A_m)$.\qed

\noindent{\bf Proof of Proposition~\ref{significant0}: }The proof has similarities to that for Theorem 1 of Yang and Qi \cite{YQ13}. Define $\tilde b_{n,t_n}(a_{-n})$ as player $n$'s highest best response to the given other-player monotone type-to-action profile $a_{-n}\in \prod_{m\neq n}{\cal M}(T_m,A_m)$ when his own type is $t_n\in T_n$:
\begin{equation}
\tilde b_{n,t_n}(a_{-n})=\sup \tilde B_{n,t_n}(a_{-n}),
\end{equation}
where the latter set is defined at~(\ref{best-def}). The properties of $\tilde B_{n,t_n}(a_{-n})$ will guarantee that $\tilde b_{n,t_n}(a_{-n})$ is both well defined and monotone in $(t_n,a_{-n})$. So given $a_{-n}\in\prod_{m\neq n}{\cal M}(T_m,A_m)$, the set $\tilde{\cal B}_n(a_{-n})$ contains the element $(\tilde b_{n,t_n}(a_{-n}))_{t_n\in T_n}\in {\cal M}(T_n,A_n)$ and hence is nonempty.

For an arbitrary nonempty subset $B$ of $\tilde{\cal B}_n(a_{-n})$, we show that $\sup B\in \tilde{\cal B}_n(a_{-n})$. Let
\begin{equation}\label{uu-def}
B|_{t_n}=\{b'\in A_n|b'=b_{t_n}\mbox{ for some }b\equiv (b_{t_n})_{t_n\in T_n}\in B\},\hspace*{.8in}\forall t_n\in T_n.
\end{equation}
Due to~(\ref{m-def}), $B|_{t_n}$ must be a subset of $\tilde B_{n,t_n}(a_{-n})$. But as the latter is a nonempty complete sublattice of $A_n$, we know that
\begin{equation}
\sup B|_{t_n}\in \tilde B_{n,t_n}(a_{-n}).
\end{equation}
Since the partial order on $A_n^{\;T_n}$ is defined in the component-wise fashion, we have from~(\ref{uu-def})
\begin{equation}
\sup B=(\sup B|_{t_n})_{t_n\in T_n}.
\end{equation}
From the fact that  $B\subseteq \tilde B_n(a_{-n})\subseteq {\cal M}(T_n,A_n)$, we also know that $\sup B\in {\cal M}(T_n,A_n)$. Therefore, according to~(\ref{m-def}), $\sup B\in \tilde{\cal B}_n(a_{-n})$. Symmetrically, we can also show that $\inf B\in \tilde{\cal B}_n(a_{-n})$. Thus the latter is a complete sublattice of ${\cal M}(T_n,A_n)$.

To show that $\tilde{\cal B}_n$ is a monotone correspondence from $\prod_{m\neq n}{\cal M}(T_m,A_m)$ to ${\cal M}(A_n,T_n)$, suppose $a^1_{-n},a^2_{-n}\in \prod_{m\neq n}{\cal M}(T_m,A_m)$ with $a^1_{-n}\leq a^2_{-n}$. Since $\tilde B_{n,t_n}$ is monotone in $a_{-n}$,
\begin{equation}\label{oops}
\tilde B_{n,t_n}(a^1_{-n})\leq \tilde B_{n,t_n}(a^2_{-n}),\hspace*{.8in}\forall t_n\in T_n.
\end{equation}
For $b^1\equiv (b^1_{t_n})_{t_n\in T_n}\in \tilde{\cal B}_n(a^1_{-n})$ and $b^2\equiv (b^2_{t_n})_{t_n\in T_n}\in \tilde{\cal B}_n(a^2_{-n})$, we have from~(\ref{m-def}) that
\begin{equation}
b^1_{t_n}\in \tilde B_{n,t_n}(a^1_{-n})\;\;\;\mbox{and}\;\;\; b^2_{t_n}\in \tilde B_{n,t_n}(a^2_{-n}),\hspace*{.8in}\forall t_n\in T_n.
\end{equation}
But due to~(\ref{oops}), we will have
\begin{equation}
b^1_{t_n}\wedge b^2_{t_n}\in \tilde B_{n,t_n}(a^1_{-n})\;\;\;\mbox{and}\;\;\;b^1_{t_n}\vee b^2_{t_n}\in \tilde B_{n,t_n}(a^2_{-n}),\hspace*{.8in}\forall t_n\in T_n.
\end{equation}
Note that $b^1\wedge b^2$ is merely $(b^1_{t_n}\wedge b^2_{t_n})_{t_n\in T_n}$ and $b^1\vee b^2$ is merely $(b^1_{t_n}\vee b^2_{t_n})_{t_n\in T_n}$, and they are within ${\cal M}(T_n,A_n)$ because $b^1$ and $b^2$ are. Hence, we have from~(\ref{m-def}) that
\begin{equation}
b^1\wedge b^2\in \tilde{\cal B}_n(a^1_{-n})\;\;\;\mbox{and}\;\;\;b^1\vee b^2\in \tilde{\cal B}_n(a^2_{-n}).
\end{equation}
This will translate into the monotonicity of the correspondence $\tilde{\cal B}_n$.\qed


\noindent{\bf Proof of Proposition~\ref{common-p}: }First, due to the average in the $\lambda$-dependent version of~(\ref{v-def-new}) and the increasing differences stated in Parametric Assumption~\ref{as-utility-p}, $\tilde v_{n,t_n,t_{-n}}(a_n,a_{-n},\mu|\lambda)$ will have increasing differences between $a_n\in A_n$ and $\lambda\in\Lambda$. Next, suppose $\lambda^1,\lambda^2\in \Lambda$ satisfy $\lambda^1\leq \lambda^2$. Then, we note the increasing trend of $q$ for
\begin{equation}\label{qq-p}
q(\tilde\omega)\equiv\tilde u_{n,t_n,t_{-n}}(a_n,a_{-n},\tilde\omega|\lambda^2)-\tilde u_{n,t_n,t_{-n}}(a_n,a_{-n},\tilde\omega|\lambda^1),
\end{equation}
due to the increasing differences between $\tilde\omega\in\tilde\Omega$ and $\lambda\in\Lambda$ as stated in Parametric Assumption~\ref{as-utility-p}. Thus, for $\mu^1,\mu^2\in {\cal P}(\tilde\Omega)$ satisfying $\mu^1\leq \mu^2$, we have the same relation as that stated in~(\ref{b2}).
In view of the $\lambda$-dependent version of~(\ref{v-def-new}) and~(\ref{qq-p}), this will amount to the increasing differences that $\tilde v_{n,t_n,t_{-n}}(a_n,a_{-n},\mu|\lambda)$ enjoys between $\mu\in {\cal P}(\tilde{\Omega})$ and $\lambda\in\Lambda$. \qed

\noindent{\bf Proof of Proposition~\ref{magazj-p}: }Let $n\in N$, $t_n\in T_n$, and $a_{-n}\in \prod_{m\neq n}{\cal M}(T_m,A_m)$ be fixed. First, suppose $\nu\in {\cal M}(T_{-n},{\cal P}(\tilde{\Omega}))$ is fixed, while $a^1_n,a^2_n\in A_n$ satisfy $a^1_n\leq a^2_n$. Define $f\equiv(f_{t_{-n}}(\lambda))_{t_{-n}\in T_{-n},\lambda\in\Lambda}$ so that
\begin{equation}\label{ff-p}
f_{t_{-n}}(\lambda)\equiv \tilde v_{n,t_n,t_{-n}}(a^2_n,a_{-n,t_{-n}},\nu_{t_{-n}}|\lambda)-\tilde v_{n,t_n,t_{-n}}(a^1_n,a_{-n,t_{-n}},\nu_{t_{-n}}|\lambda).
\end{equation}
Due to the increasing differences between $a_n\in A_n$ and $(t_{-n},a_{-n,t_{-n}},\nu_{-n,t_{-n}})\in T_{-n}\times A_{-n}\times {\cal P}(\tilde{\Omega})$ as stated in Proposition~\ref{common}, as well as the memberships of $a_{-n}$ in $\prod_{m\neq n}{\cal M}(T_m,A_m)$ and $\nu$ in ${\cal M}(T_{-n},{\cal P}(\tilde{\Omega}))$, we know that $f_{t_{-n}}(\lambda)$ is increasing in $t_{-n}$. By the increasing differences between $a_n\in A_n$ and $\lambda\in\Lambda$ as stated in  Proposition~\ref{common-p}, we know that $f_{t_{-n}}(\lambda)$ is increasing in $\lambda$. But by Parametric Assumption~\ref{as-mono-probability-p}, these will translate into
\begin{equation}\begin{array}{ll}
\sum_{t_{-n}\in T_{-n}}p_{n,t_n|t_{-n}}(\lambda^1)\cdot f_{t_{-n}}(\lambda^1)&\leq \sum_{t_{-n}\in T_{-n}}p_{n,t_n|t_{-n}}(\lambda^1)\cdot f_{t_{-n}}(\lambda^2)\\
&\leq  \sum_{t_{-n}\in T_{-n}}p_{n,t_n|t_{-n}}(\lambda^2)\cdot f_{t_{-n}}(\lambda^2),
\end{array}\end{equation}
whenever $\lambda^1\leq \lambda^2$. In view of the $\lambda$-dependent version of~(\ref{w-def-new}) and~(\ref{ff-p}), this will amount to the increasing differences that $\tilde w_{n,t_n}(a_n,a_{-n},\nu|\lambda)$ enjoys between $a_n\in A_n$ and $\lambda\in \Lambda$.

Now, suppose $a_n\in A_n$ is fixed, while $\nu^1,\nu^2\in {\cal M}(T_{-n},{\cal P}(\tilde{\Omega}))$ satisfy $\nu^1\leq \nu^2$. Define $g\equiv (g_{t_{-n}}(\lambda))_{t_{-n}\in T_{-n},\lambda\in\Lambda}$ so that
\begin{equation}\label{gg-p}
g_{t_{-n}}(\lambda)\equiv\tilde v_{n,t_n,t_{-n}}(a_n,a_{-n,t_{-n}},\nu^2_{t_{-n}}|\lambda)-\tilde v_{n,t_n,t_{-n}}(a_n,a_{-n,t_{-n}},\nu^1_{t_{-n}}|\lambda).
\end{equation}
Due to the increasing differences between $\nu_{t_{-n}}\in {\cal P}(\tilde{\Omega})$ and $\lambda\in \Lambda$ as stated in Proposition~\ref{common-p}, we know that $g_{t_{-n}}(\lambda)$ is increasing in $\lambda$. But by averaging over the probability $p^A_{n}$ which is invariant in $\lambda$ according to Parametric Assumption~\ref{as-mono-probability-p}, this will translate into
\begin{equation}
\sum_{t_{-n}\in T_{-n}}p^A_{n|t_{-n}}\cdot g_{t_{-n}}(\lambda^1)\leq \sum_{t_{-n}\in T_{-n}}p^A_{n|t_{-n}}\cdot g_{t_{-n}}(\lambda^2),
\end{equation}
whenever $\lambda^1\leq \lambda^2$. In view of the $\lambda$-dependent version of~(\ref{w-def-new}) and~(\ref{gg-p}), this will amount to the increasing differences that $\tilde w_{n,t_n}(a_n,a_{-n},\nu|\lambda)$ enjoys between $\nu\in {\cal M}(T_{-n},{\cal P}(\tilde{\Omega}))$ and $\lambda\in\Lambda$ in scenario A.

In addition, suppose $a_n\in A_n$ is fixed, while $\nu^1,\nu^2\in 1(T_{-n},{\cal P}(\tilde{\Omega}))$ satisfy $\nu^1\leq \nu^2$. Define $h\equiv (h_{t_{-n}}(\lambda))_{t_{-n}\in T_{-n},\lambda\in\Lambda}$ so that
\begin{equation}\label{hh-p}
h_{t_{-n}}(\lambda)\equiv\tilde v_{n,t_n,t_{-n}}(a_n,a_{-n,t_{-n}},\nu^2_{t_{-n}}|\lambda)-\tilde v_{n,t_n,t_{-n}}(a_n,a_{-n,t_{-n}},\nu^1_{t_{-n}}|\lambda).
\end{equation}
Due to $\nu^1$ and $\nu^2$'s memberships in $1(T_{-n},{\cal P}(\tilde\Omega))$, neither $\nu^1_{t_{-n}}$ nor $\nu^2_{t_{-n}}$ depends on $t_{-n}$. So by the increasing differences between $(t_{-n},a_{-n,t_{-n}})\in T_{-n}\times A_{-n}$ and $\nu_{t_{-n}}\in {\cal P}(\tilde{\Omega})$ as stated in Proposition~\ref{common}, as well as the membership of $a_{-n}$ in $\prod_{m\neq n}{\cal M}(T_m,A_m)$, we know that $h_{t_{-n}}(\lambda)$ is increasing in $t_{-n}$. By the increasing differences between $\nu_{t_{-n}}\in {\cal P}(\tilde\Omega)$ and $\lambda\in\Lambda$ as stated in  Proposition~\ref{common-p}, we know that $h_{t_{-n}}(\lambda)$ is increasing in $\lambda$. But by Parametric Assumption~\ref{as-mono-probability-p},
\begin{equation}\begin{array}{ll}
\sum_{t_{-n}\in T_{-n}}p^B_{n,t_n|t_{-n}}(\lambda^1)\cdot h_{t_{-n}}(\lambda^1)&\leq \sum_{t_{-n}\in T_{-n}}p^B_{n,t_n|t_{-n}}(\lambda^1)\cdot h_{t_{-n}}(\lambda^2)\\
&\leq \sum_{t_{-n}\in T_{-n}}p^B_{n,t_n|t_{-n}}(\lambda^2)\cdot h_{t_{-n}}(\lambda^2),
\end{array}\end{equation}
whenever $\lambda^1\leq\lambda^2$. In view of the $\lambda$-dependent version of~(\ref{w-def-new}) and~(\ref{hh-p}), this will amount to the increasing differences that $\tilde w_{n,t_n}(a_n,a_{-n},\nu|\lambda)$ enjoys between $\nu\in 1(T_{-n},{\cal P}(\tilde{\Omega}))$ and $\lambda\in\Lambda$ in scenario B.\qed

\noindent{\bf Proof of Proposition~\ref{boring-p}: }For scenario A, suppose $\lambda^1,\lambda^2\in \Lambda$ satisfy $\lambda^1\leq \lambda^2$; also,  $\nu^1\in {\cal Q}^A_{n,t_n}(\lambda^1)$ and $\nu^2\in {\cal Q}^A_{n,t_n}(\lambda^2)$. Then, according to the $\lambda$-dependent version of~(\ref{p-def-new-a}), $\nu^1_{t_{-n}}\in \tilde P^A_{n,t_n,t_{-n}}(\lambda^1)$ and $\nu^2_{t_{-n}}\in \tilde P^A_{n,t_n,t_{-n}}(\lambda^2)$ for any $t_{-n}\in T_{-n}$, and both $\nu^1_{t_{-n}}$ and $\nu^2_{t_{-n}}$ are increasing in $t_{-n}$. So by Parametric Assumption~\ref{as-mono-prior-p},
\begin{equation}\label{b19-p}
\nu^1_{t_{-n}}\wedge \nu^2_{t_{-n}}\in \tilde P^A_{n,t_n,t_{-n}}(\lambda^1),\hspace*{.5in}\nu^1_{t_{-n}}\vee \nu^2_{t_{-n}}\in \tilde P^A_{n,t_n,t_{-n}}(\lambda^2),\hspace*{.8in}\forall t_{-n}\in T_{-n}.
\end{equation}
Both $\nu^1_{t_{-n}}\wedge\nu^2_{t_{-n}}$ and  $\nu^1_{t_{-n}}\vee\nu^2_{t_{-n}}$ are still increasing in $t_{-n}$.
But by the $\lambda$-dependent version of~(\ref{p-def-new-a}), this will lead back to
\begin{equation}\label{b20-p}
\nu^1\wedge \nu^2\in {\cal Q}^A_{n,t_n}(\lambda^1),\hspace*{.5in}\nu^1\vee \nu^2\in {\cal Q}^A_{n,t_n}(\lambda^2).
\end{equation}
Hence, ${\cal Q}^A_{n,t_n}(\lambda)$ is increasing in $\lambda$.

For scenario B, we know that ${\cal Q}^B_{n,t_n}(\lambda)$ defined at the $\lambda$-dependent version of~(\ref{p-def-new-b}) is increasing in $\lambda$ just because, due to Parametric Assumption~\ref{as-mono-prior-p}, $\tilde P^B_{n,t_n}(\lambda)$ is increasing in $\lambda$. . \qed

\noindent{\bf Proof of Proposition~\ref{challenging-p}: }At any fixed $n\in N$, $t_n\in T_n$, and $a_{-n}\in\prod_{m\neq n}{\cal M}(T_m,A_m)$, we can identify $a_{n,t_n}\in A_n$ with $x\in X$, $\lambda\in\Lambda$ with $y\in Y$, and $\nu\in {\cal M}(T_{-n},{\cal P}(\tilde{\Omega}))$ in scenario A or $\nu\in 1(T_{-n},{\cal P}(\tilde{\Omega}))$ in scenario B with $z\in Z$. Also, we can identify $\tilde w_{n,t_n}(a_{n,t_n},a_{-n},\nu|\lambda)$ with $f(x,y,z)$, ${\cal Q}_{n,t_n}(\lambda)$ with $\tilde Z(y)$, and $\tilde s_{n,t_n}(a_{n,t_n},a_{-n}|\lambda)$ with $g(x,y)$. From Propositions~\ref{magazj} and~\ref{magazj-p}, we know that (I), (II), and (V) of Lemma~\ref{l-import} are true. We have the validity of Lemma~\ref{l-import}'s (III) from Proposition~\ref{boring} and that of the lemma's (IV) from Proposition~\ref{boring-p}. Now the $\lambda$-dependent version of~(\ref{s-def-new}) dictates that the relationship between $f$, $\tilde Z$, and $g$ in Lemma~\ref{l-import} also applies here. So by that lemma, we can derive that $\tilde s_{n,t_n}(a_{n,t_n},a_{-n}|\lambda)$ has increasing differences between $a_{n,t_n}\in A_n$ and $\lambda\in\Lambda$.\qed

\noindent{\bf Proof of Proposition~\ref{significant0-p}: }The proof has similarities to that for Theorem 2 of Yang and Qi \cite{YQ13}. We first show that $\tilde B_{n,t_n}(a_{-n}|\lambda)$ defined at the $\lambda$-dependent version of~(\ref{best-def}) is monotonically increasing in $\lambda\in \Lambda$ at every $n\in N$, $t_n\in T_n$, and $a_{-n}\in\prod_{m\neq n}{\cal M}(T_m,A_m)$. For that purpose, let $\lambda^1,\lambda^2\in \Lambda$ satisfying $\lambda^1\leq \lambda^2$, $a^1_{n,t_n}\in \tilde B_{n,t_n}(a_{-n}|\lambda^1)$, and $a^2_{n,t_n}\in \tilde B_{n,t_n}(a_{-n}|\lambda^2)$ be given. It can be checked that
\begin{equation}\begin{array}{ll}
0&\leq \tilde s_{n,t_n}(a^1_{n,t_n},a_{-n}|\lambda^1)-\tilde s_{n,t_n}(a^1_{n,t_n}\wedge a^2_{n,t_n},a_{-n}|\lambda^1)\\
&\leq \tilde s_{n,t_n}(a^1_{n,t_n},a_{-n}|\lambda^2)-\tilde s_{n,t_n}(a^1_{n,t_n}\wedge a^2_{n,t_n},a_{-n}|\lambda^2)\\
&=\tilde s_{n,t_n}(a^1_{n,t_n}\vee a^2_{n,t_n},a_{-n}|\lambda^2)-\tilde s_{n,t_n}(a^2_{n,t_n},a_{-n}|\lambda^2)\leq 0,
\end{array}\end{equation}
where the first inequality is due to the optimality of $a^1_{n,t_n}$ for $\tilde s_{n,t_n}(\cdot,a_{-n}|\lambda^1)$ as demanded by $\tilde B_{n,t_n}(a_{-n}|\lambda^1)$'s definition at the $\lambda$-dependent version of~(\ref{best-def}), the second inequality comes from Proposition~\ref{challenging-p}, the only equality is attributable to the fact that $A_n$ is totally ordered, and the last inequality is due to the optimality of $a^2_{n,t_n}$ for $\tilde s_{n,t_n}(\cdot,a_{-n}|\lambda^2)$ as demanded by $\tilde B_{n,t_n}(a_{-n}|\lambda^2)$'s definition at the $\lambda$-dependent version of~(\ref{best-def}). The only possibility is for all inequalities to be equalities. Thus, we must have $a^1_{n,t_n}\wedge a^2_{n,t_n}\in \tilde B_{n,t_n}(a_{-n}|\lambda^1)$ and $a^1_{n,t_n}\vee a^2_{n,t_n}\in \tilde B_{n,t_n}(a_{-n}|\lambda^2)$. Hence, the correspondence $\tilde B_{n,t_n}(a_{-n}|\lambda)$ is increasing in $\lambda$.

In view of the definition of $\tilde{\cal B}_n(a_{-n}|\lambda)$ from $\tilde B_{n,t_n}(a_{-n}|\lambda)$ through the $\lambda$-dependent version of~(\ref{m-def}), it is clear that $\tilde{\cal B}_n(a_{-n}|\lambda)$ will be increasing in $\lambda$ as well. \qed



\begin{thebibliography}{99}
	
\bibitem{A07} Ahn, D.S. 2007. Hierarchies of Ambiguous Beliefs. {\em Journal of Economic Theory}, {\bf 136}, pp. 286-301.

\bibitem{A53} Allais, M. 1953. Le Comportement de l'Homme Rationnel devant le Risque: Critique des Postulats et Axiomes de l'Ecole Americaine. {\em Econometrica}, {\bf 21}, pp. 503-546.

\bibitem{AA63}
Anscombe, F.J. and R.J. Aumann. 1963. A Definition of Subjective Probability. {\em The Annals of Mathematical Statistics}, {\bf 34}, pp. 199-205.



\bibitem{A76}
Aumann, R.J. 1976. Agreeing to Disagree. {\em Annals of Statistics}, {\bf 4}, pp. 1236-1239.

\bibitem{AT11} Azrieli, Y. and R. Teper. 2011. Uncertainty Aversion and Equilibrium Existence in Games with Incomplete Information. {\em Games and Economic Behavior}, {\bf 73}, pp. 310-317.

\bibitem{Bd11} Bade, S. 2011. Ambiguous Act Equilibria. {\em Games and Economic Behavior}, {\bf 71}, pp. 246-260.



\bibitem{BOP06} Bose, S., E. Ozdenoren, and A. Pape. 2006. Optimal Auctions with Ambiguity. {\em Theoretical Economics}, {\bf 1}, pp. 411-438.

\bibitem{BR14} Bose, S. and L. Renou. 2014. Mechanism Design with Ambiguous Communication Devices. {\em Econometrica}, {\bf 82}, pp. 1853-1872.


\bibitem{CH94} Camerer C.F. and T.H. Ho. 1994. Violations of the Betweenness Axiom and Nonlinearity in Probability. {\em Journal of Risk and Uncertainty}, {\bf 8}, pp. 167–196.

\bibitem{CKL13} Charness, G., E. Karni, and D. Levin. 2013. Ambiguity Attitudes and Social Interactions: An Experimental Investigation. {\em Journal of Risk and Uncertainty}, {\bf 46}, pp. 1-25.


\bibitem{CY89} Curley, S.P. and J.F. Yates. 1989. An Empirical Evaluation of Descriptive Models of Ambiguity Reactions in Choice Situations. {\em
Journal of Mathematical Psychology}, {\bf 33}, pp. 397-427.

\bibitem{D64} Debreu, G. 1964. Continuity Properties of Paretian Utility. {\em International Economic Review}, {\bf 5}, pp. 285-293.


\bibitem{D67} Dempster, A.P. 1967. Upper and Lower Probabilities Induced by a Multivalued Mapping. {\em Annals of Mathematical Studies}, {\bf 38}, pp. 325-339.

\bibitem{D08} Di Tillio. 2008. Subjective Expected Utility in Games. {\em Theoretical Economics}, {\bf 3}, pp. 287-323.


\bibitem{DW94} Dow, J. and S. Werlang. 1994. Nash Equilibrium under Knightian Uncertainty: Breaking Down Backward Induction. {\em Journal of Economic Theory}, {\bf 64}, pp. 305-324.

\bibitem{EK00} Eichberger, J. and D. Kelsey. 2000. Non-additive Beliefs and Strategic Equilibria. {\em Games and Economic Behavior}, {\bf 30}, pp. 183-215.

\bibitem{E61}
Ellsberg, D. 1961. Risk, Ambiguity and Savage Axioms. {\em Quarterly Journal of Economics}, {\bf 75}, pp. 643-669.

\bibitem{E97} Epstein, L.G. 1997. Preference, Rationalizability, and Equilibrium. {\em Journal of Economic Theory}, {\bf 73}, pp. 1-29.

\bibitem{EW96} Epstein, L. and T. Wang. 1996. Beliefs about Beliefs without Probabilities. {\em Econometrica}, {\bf 64}, pp. 1343-1373.

\bibitem{FH90} Fagin, R. and J. Halpern. 1990. A New Approach to Updating Beliefs. {\em Proceedings of the 6th Conference on Uncertainty and AI}, pp. 317-325, Elsevier, New York.


\bibitem{GM13}
Gilboa, I. and M. Marinacci. 2013. Ambiguity and the Bayesian Paradigm. In D. Acemoglu, M. Arellano, and E. Dekel (Eds.), {\em Advances in Economics and Econometrics: Theory and Applications, Tenth World Congress of the Econometric Society}. Cambridge University Press. 

\bibitem{GS89}
Gilboa, I. and D. Schmeidler. 1989. Maxmin Expected Utility with Non-unique Prior. {\em Journal of Mathematical Economics}, {\bf 18}, pp. 141-153.

\bibitem{GMT16} Grant, S., I. Meneghel, and R. Tourky. 2016. Savage Games. {\em Theoretical Economics}, {\bf 11}, pp. 641-682.


\bibitem{H67-8} Harsanyi, J.C. 1967-68. Games with Incomplete Information Played by Bayesian Players. {\em Management Science}, {\bf 14}, pp. 159-182, 320-334, 486-502.


\bibitem{H74} Hildenbrand, W. 1974. {\em Core and Equilibria of a Large Economy}. Princeton University Press, Princeton, NJ.



\bibitem{KU05} Kajii, A. and T. Ui. 2005. Incomplelte Information Games with Multiple Priors. {\em Japanese Economic Review}, {\bf 56}, pp. 332-351.


\bibitem{K96} Klibanoff, P. 1996. Uncertainty, Decision, and Normal Form Games. Mimeo, Northwestern University.


\bibitem{KS90} Khan, M.A. and Y.N. Sun. 1990. On a Reformulation of Cournot-Nash Equilibria. {\em Journal of Mathematical Analysis and Applications}, {\bf 146}, pp. 442-460.

\bibitem{KT84} Klein, E. and A.C. Thompson. 1984. {\em Theory of Correspondences}. John Wiley \& Sons, New York.


\bibitem{L96}
Lo, K.-C. 1996. Equilibrium in Beliefs under Uncertainty. {\em Journal of Economic Theory}, {\bf 71}, pp. 443-484.

\bibitem{L98} Lo, K.-C. 1998. Sealed Bid Auctions with Uncertain Averse Bidders. {\em Economic Theory}, {\bf 12}, pp. 1-20.


\bibitem{M00} Marinacci, M. 2000. Ambiguous Games. {\em Games and Economic Behavior}, {\bf 31}, pp. 191-219.




\bibitem{M74} Mas-Colell, A. 1974. An Equilibrium Existence Theorem without Complete or Transitive Preferences. {\em Journal of Mathmatical Economics}, {\bf 1}, pp. 237-246.

\bibitem{MZ85} Mertens, J.F. and S. Zamir. 1985. Formulation of Bayesian Analysis for Games and Incomplete Information. {\em International Journal of Game Theory}, {\bf 14}, pp. 1-29.

\bibitem{MR90} Milgrom, P.R. and J. Roberts. 1990. Rationalizability, Learning, and Equilibrium in Games with Strategic Complementarities. {\em Econometrica}, {\bf 58}, pp. 1255–1277.

\bibitem{MS94} Milgrom, P.R. and C. Shannon. 1994. Monotone Comparative Statics. {\em Econometrica}, {\bf 62}, pp. 157-180.

\bibitem{MW82} Milgrom, P.R. and R.J. Weber. 1982. A Theory of Auctions and Competitive Bidding. {\em Econometrica}, {\bf 50}, pp. 1089-1122.


\bibitem{MK00} Munkres, J.R. 2000. {\em Topology, 2nd Edition}, Prentice Hall, New Jersey.

\bibitem{N50} Nash, J.F. 1950. Equilibrium Points in $n$-person Games. {\em Proceedings of the National Academy of Sciences}, {\bf 36}, pp. 48-49.

\bibitem{N51} Nash, J.F. 1951. Non-cooperative Games. {\em Annals of Mathematics}, {\bf 54}, pp. 286-295.

\bibitem{VNM44}
von Neumann, J. and O. Morgenstern. 1944. {\em Theory of Games and Economic Behaviour}. Princeton University Press, Princeton, NJ.


\bibitem{RS13} Riedel, F. and L. Sass. 2014. Ellsberg Games. {\em Theory and Decision}, {\bf 76}, pp. 469-509.




\bibitem{S54} Savage, L.J. 1972. {\em The Foundation of Statistics, 2nd Revised Edition}. Dover Publications, New York.


\bibitem{S69} Schmeidler, D. 1969. Competitive Equilibria in Markets with a Continuum of Traders and Incomplete Preferences. {\em Econometrica}, {\bf 37}, pp. 578-585.

\bibitem{S89} Schmeidler, D. 1989. Subjective Probability and Expected Utility without Additivity. {\em Econometrica}, {\bf 57}, pp. 571-587.

\bibitem{SS75} Shafer, W. and H. Sonnenschein. 1975. Equilibrium in Abstract Economies without Ordered Preferences. {\em Journal of Mathematical Economics}, {\bf 2}, pp. 345-348.

\bibitem{SS07} Shaked, M. and J.G. Shanthikumar. 2007. {\em Stochastic Orders}. Springer, Berlin.


\bibitem{T55} Tarski, A. 1955. A Lattice-theoretical Fixpoint Theorem and Its Applications. {\em Pacific Journal of Mathematics}, {\bf 5}, pp. 285-309.

\bibitem{T79} Topkis, D.M. 1979. Equilibrium Points in Nonzero-sum $n$-person Submodular Games. {\em SIAM Journal on Control and Optimization}, {\bf 17}, pp. 773-787.

\bibitem{T98} Topkis, D.M. 1998. {\em Supermodularity and Complementarity}. Princeton University Press, Princeton, NJ.



\bibitem{V90} Vives, X. 1990. Nash Equilibrium with Strategic Complementarities. {\em Journal of Mathematical Economics}, {\bf 19}, pp. 305-321.

\bibitem{WG96} Wu, G. and R. Gonzalez. 1996. Curvature of the Probability Weighing Function. {\em Management Science}, {\bf 42}, pp. 1676-1690.


\bibitem{YQ13} Yang, J. and X. Qi. 2013. The Nonatomic Supermodular Game. {\em Games and Economic Behavior}, {\bf 82}, pp. 609-620.

\bibitem{VZV07} van Zandt, T. and X. Vives. 2007. Monotone Equilibrium in Bayesian Games of Strategic Complementarities. {\em Journal of Economic Theory}, {\bf 134}, pp. 330-360.

\bibitem{Z94} Zhou, L. 1994. The Set of Nash Equilibria of a Supermodular Game is a Complete Lattice. {\em Games and Economic Behavior}, {\bf 7}, pp. 295-300.

\end{thebibliography}
\end{document}